\documentclass[twoside,a4paper,  11pt]{article}
\usepackage[pdftex]{graphicx}
\usepackage{calc}
\usepackage{amsmath}
\usepackage{amssymb}

\usepackage{textcomp}

\usepackage{sectsty}
\chaptertitlefont{\Large}
\sectionfont{\large}

\begin{document}

\noindent {\large {\bf Longitudinal Bunch Diagnostics using Coherent Transition Radiation Spectroscopy}\vspace{1mm}

\noindent  {\it Physical Principles, Multichannel Spectrometer, Experimental Results, Mathematical Methods} }\vspace{5mm}

\noindent  Bernhard Schmidt$^{1}$,  Stephan Wesch$^{1}$,  Toke K\"ovener$^{2,3}$, Christopher Behrens$^{1}$, Eugen Hass$^{2}$,\\ Sara Casalbuoni$^{4}$,  and Peter Schm\"user$^{1,2}$. 
\vspace{2mm}

\noindent \small{1.\,Deutsches Elektronen-Synchrotron DESY Hamburg,~2.\,Universit\"at Hamburg,~3.\,CERN, Geneva,\\
\noindent 4.\,Institute for Beam Physics and Technology, Karlsruhe Institute of Technology.}\vspace{1mm}

\noindent \small{Corresponding authors: Bernhard.Schmidt@desy.de\,, Peter.Schmueser@desy.de}\\
{\small This work is licensed under a Creative Commons Attribution 4.0 International License}
\vspace{2mm}

\vspace{3mm}

\noindent {\bf Abstract}\\
This report summarizes the work on electron bunch diagnostics using coherent transition radiation spectroscopy which our group has carried out over the past 13 years and which is still ongoing.\\
The generation  and properties of transition radiation (TR) are thoroughly treated. The  spectral energy density, as described by the  Ginzburg-Frank formula,  is computed  analytically, and the modifications caused by the finite size of the TR screen  and  near-field diffraction effects are carefully analyzed.  The principles of electron bunch shape reconstruction using coherent transition radiation (CTR) are outlined. The three-dimensional form factor is defined and its separation into a transverse and a longitudinal part. Spectroscopic measurements  yield only the absolute magnitude  of the form factor but not its phase, which however  is needed for computing the bunch shape via the inverse Fourier transformation. Two phase retrieval methods  are  investigated and illustrated with  model calculations:   analytic phase computation by means of the Kramers-Kronig dispersion relation, and iterative phase retrieval. Particular attention is paid to the ambiguities which are unavoidable in the reconstruction of  longitudinal charge density profiles from spectroscopic measurements.   The origin of these ambiguities has been identified and a thorough mathematical analysis is presented. 
The experimental part of the paper comprises a description  of our multichannel infrared and THz spectrometer and a selection of measurements at FLASH (Free-electron LASer in Hamburg),   comparing the bunch profiles derived from the spectroscopic data with the profiles determined with a transversely deflecting microwave structure.\\
 
\noindent   The appendices are devoted to the mathematical methods. A rigorous derivation of the Kramers-Kronig phase formula is presented in Appendix~A.  Numerous analytic model calculations can be found in Appendix B. The differences between normal and truncated Gaussians are discussed in Appendix C.  Finally, Appendix D contains a short  description of the propagation of an electromagnetic wave front  by two-dimensional fast Fourier transformation.  This is the basis of a powerful numerical Mathematica\texttrademark ~code {\it THzTransport}, developed in our group, which permits the propagation of electromagnetic wave fronts (visible light, infrared or THz radiation) through an optical beam line consisting of drift spaces, lenses,   mirrors and apertures.

\section{Introduction}
\label{intro}
 The electron bunches in  the high-gain free-electron laser FLASH\footnote{The physics  of high-gain free-electron lasers and the technology of the soft X-ray FEL FLASH  is described in \cite{Schmueser-FEL}.} are longitudinally compressed to achieve peak currents in the  kA range which are necessary to drive the high-gain FEL process in the undulator magnets. Bunch compression is accomplished by a two-stage process: first an {\it energy chirp} (energy-position relationship) is imprinted onto the typically 10\,ps long bunches emerging from the electron gun, and then the chirped bunches are passed through magnetic chicanes where the length is reduced to about 100\,fs or less. A linearization of the accelerating voltage is achieved by superimposing the 1.3\,GHz accelerating field with its third harmonic. A superconducting 3.9\,GHz cavity \cite{Helen-Edwards} permits optimization of the bunch compression process.

Magnetic compression of intense electron bunches is strongly affected by collective effects in the chicanes and cannot  be adequately described by linear beam transfer theory. Space charge forces,  coherent synchrotron radiation and wake fields have a profound influence on the time profile and internal energy distribution of the compressed bunches. The collective effects have been  studied by various numerical simulations (see \cite{Dohlus} and the references quoted therein) but the parameter uncertainties are large and experimental data are  indispensable for determining the  length and the longitudinal density profile of the bunches before they enter the undulator. \vspace{2mm}

\noindent Our group has applied 
two time-domain techniques permitting a direct visualization of longitudinal electron bunch profiles with very high resolution: (1) a transversely deflecting microwave  structure TDS, and (2)  electro-optic  (EO) detection systems (see \cite{steffen-PRSTAB} and the references quoted therein), which will not be discussed here.  In addition, a high-resolution frequency-domain technique has been developed based on  a multichannel single-shot spectrometer for recording  coherent transition radiation in the infrared  and THz regime.\vspace{2mm}

\noindent {\bf Transversely deflecting microwave structure TDS} \\
In the TDS  the temporal profile of the electron bunch is transferred to a spatial profile on a view screen by a rapidly varying electromagnetic field \cite{LOLA,TDS1,TDS2}.  The TDS used at FLASH  is a 3.6~m long traveling-wave structure operating at 2.856~GHz in which a combination of electric and magnetic fields produces a transverse force for the electrons. 
\begin{figure}[ht!]
\begin{center}
\includegraphics[width=14cm]{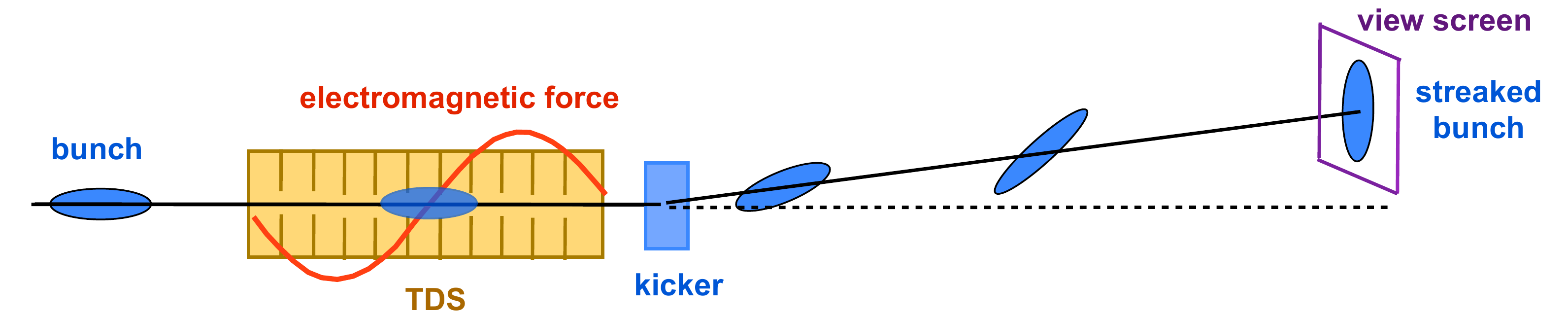}
\end{center}
\caption{ \small{Principle of longitudinal charge density measurement using a transversely deflecting microwave structure. For optimum resolution the radio-frequency (RF) phase is chosen such that the bunch center coincides with the zero-crossing of the RF wave. This condition holds  along the entire axis of the traveling-wave structure since electron bunch and RF wave move synchronously with  a speed very close to $c$ (light velocity in vacuum). }}
\label{TDS-schematic}
\end{figure} 

\begin{figure}[htb!]
\begin{center}
\includegraphics[width=8cm]{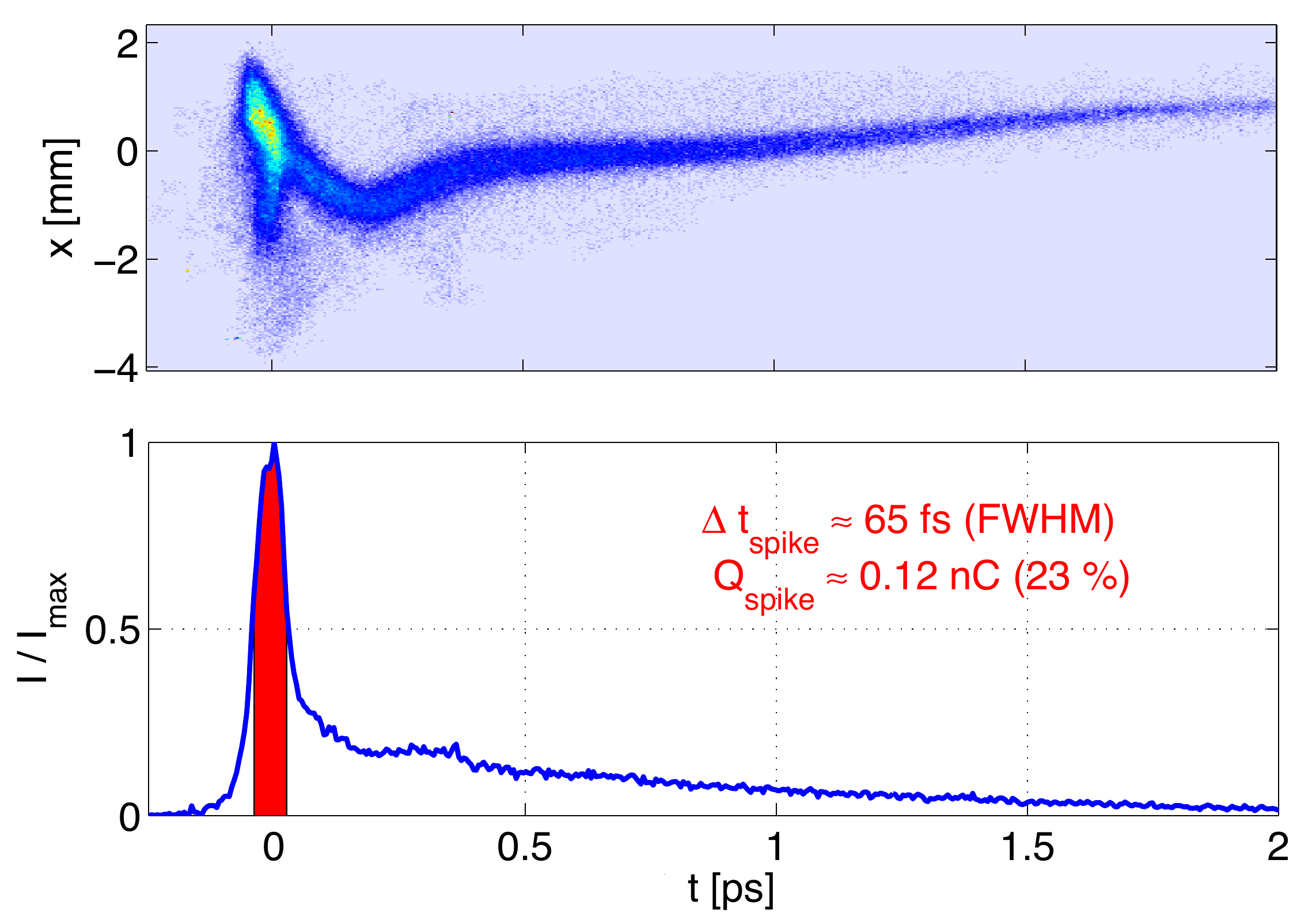}
\end{center}
\caption{ \small{{\it Top\,}: Two-dimensional image of a single electron
bunch whose time profile is translated into a spatial
profile on an observation screen. The bunch head is at the  left side.
{\it Bottom\,}: Current as a function of time. The maximum current
is $I_{\rm max}=1.8$\,kA in this measurement \cite{roehrs-thesis}.}}
\label{TDS-Spike-tail}
\end{figure}

 The bunches pass the TDS near zero crossing of the RF field, and the electrons receive a vertical kick which depends on their longitudinal position inside the bunch. The longitudinal bunch profile is thereby transformed into a streak image  on the observation screen.  A single bunch out of a train can be streaked.
With a fast kicker magnet, this bunch is deflected towards the view   screen  and recorded by a digital camera.
The principle of the TDS is explained in Fig.\,\ref{TDS-schematic}.
The time resolution of the TDS installed at FLASH may be as good as 10\,fs (rms),   depending on the beam optics chosen.  An essential prerequisite for good resolution is a large beta function at the position of the TDS.
An image of a streaked electron bunch is shown in Fig.~\ref{TDS-Spike-tail}. 
  An important application was time-resolved phase space tomography \cite{roehrs-PRSTAB, roehrs-thesis} to determine the so-called {\it slice emittance}. 
  The TDS is  routinely used as a diagnostic tool  at FLASH,  the XFEL and other accelerators.\vspace{20mm}

\noindent {\bf Infrared and THz spectroscopy} \\
Complementary to the above-mentioned  time-domain techniques is spectroscopy in the frequency domain. In particular, coherent transition radiation (CTR) in the infrared and far-infrared (THz) regime has a long tradition as a tool  for the longitudinal  diagnostics of short electron bunches.  The intensity of coherent radiation is proportional to $N^2|F(\omega|^2$, where $N$ is the number of electrons in the bunch and $F(\omega)$ is the longitudinal form factor  (written here as a function of circular frequency $\omega=2 \pi c/\lambda$). We will discuss in  detail various methods for determining the bunch length and the  internal bunch structure  from the measured CTR spectra.

In the past, our group has carried out numerous CTR autocorrelation studies with Martin-Puplett interferometers (see e.g. \cite{Geitz}). 
In the interferometer the optical delay between the two arms is varied in small steps by moving a mirror.  Each bunch makes just one entry in the autocorrelation plot, hence  many successive bunches are needed to obtain an average longitudinal shape. 
To open the way to CTR spectroscopy on single electron bunches  
 a novel multichannel infrared and THz spectrometer with fast readout was jointly developed at DESY and the University of Hamburg~\cite{Wesch-NIM}. This spectrometer, named  CRISP (Coherent transition Radiation Intensity Spectrometer),  and its experimental applications are described in this report. 

 \vspace{5mm}

\section{Production and Properties of Transition Radiation}\label{TransRad}
When a relativistic electron crosses the boundary between two media of different permittivity, the electromagnetic field carried by the particle changes abruptly upon the transition from one medium to the other.  To satisfy the boundary conditions for the electric and magnetic field vectors one has add  two radiation fields, one propagating in  forward direction, the other in backward direction. This radiation is called transition radiation. The boundary-condition method is straightforward for an infinite planar boundary, it can be generalized to describe the radiation from screens of simple other shapes:  a circular disc, a circular hole in an infinite plane, or a semi-infinite half plane \cite{Ter-Mik, Bol}.
 This will not be discussed here because the mathematical effort is considerable and the results apply only for the far-field diffraction regime. Radiation from a screen of arbitrary shape cannot be calculated analytically. \\
An alternative approach to compute the radiation by a relativistic charged particle at the transition from vacuum into a metal is based on the Weizs\"äcker-Williams method of virtual quanta, see e.g. \cite{Jackson}. The assumption is made that the virtual photons, constituting the self-field of the particle, are converted into real photons by reflection at the metallic interface. Effectively this means that the Fourier components of the transverse electric field of the electron are reflected at the metal surface. Then the Huygens-Fresnel principle is applied to compute the outgoing 
electromagnetic wave. An important prerequisite is the fact that the electromagnetic field of an ultrarelativistic electron is concentrated in a flat disc perpendicular to the direction of motion and is thus essentially transverse.

\subsection{Electromagnetic field of a relativistic point charge}
\begin{figure}[ht!]
\centering
\includegraphics[angle=0,width=12cm]{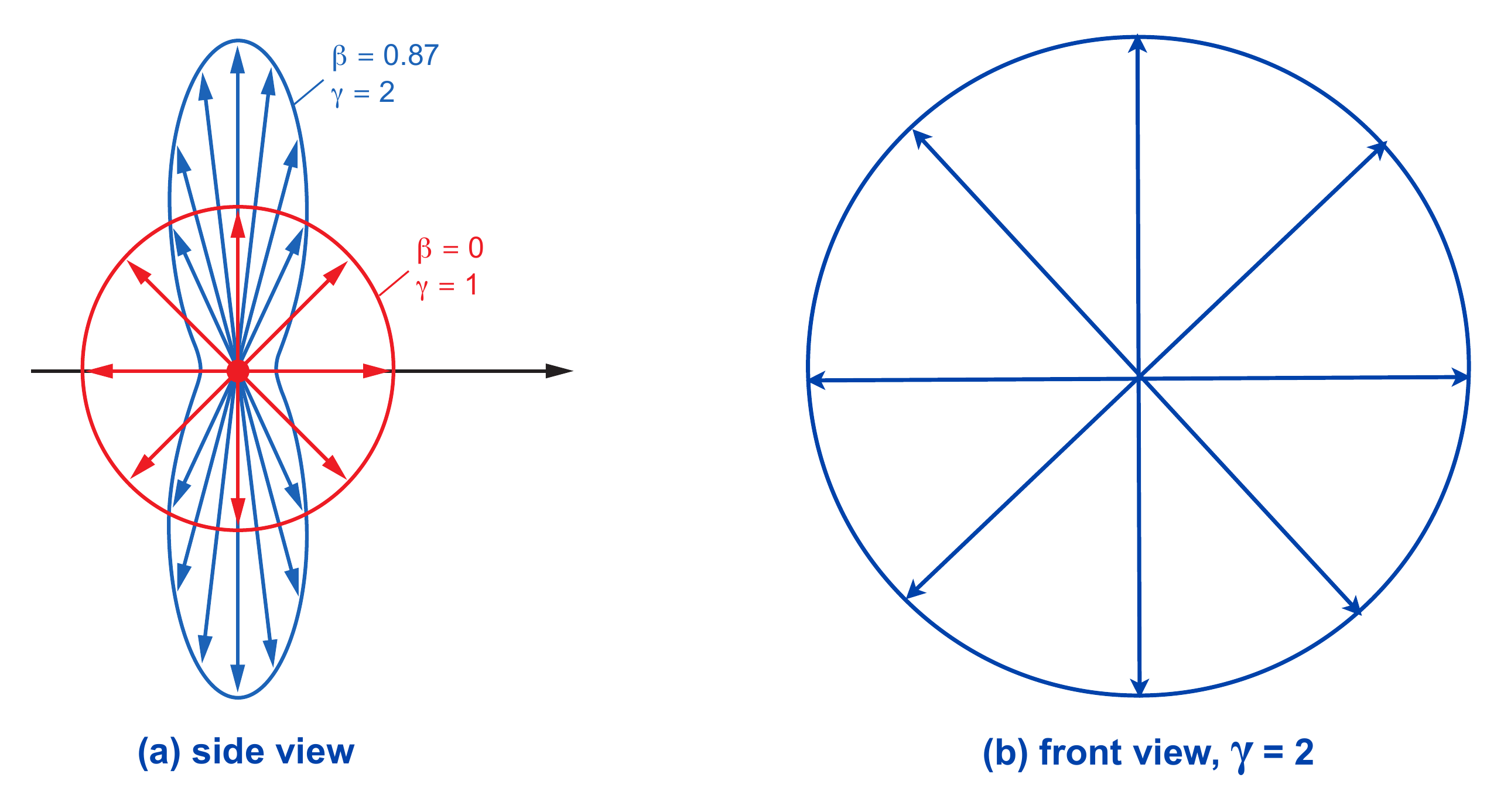}
  \caption[]{\small{{\bf (a)} Polar angle distributions of the electric field of a positron at rest (red) and a positron moving with $v=0.87\,c$ (blue).  {\bf (b)} Field line pattern viewed along the direction of motion for the case $v=0.87\,c$,   showing the radial polarization.} }
\label{rel-pointcharge}
\end{figure}

The  electromagnetic field of a point charge $q$ moving with constant speed $v=\beta\,c$ along the $z$ direction can be determined by starting with  the 4-vector potential in the particle rest frame
$$A'^\mu=\left(\frac{\Phi'}{c},\,\boldsymbol{A'}\right).$$
Here $\Phi'$ is the scalar potential and $\boldsymbol{A'}$ is the vector potential. In the rest frame there is only a scalar potential, the vector potential vanishes
\begin{equation}
\Phi'=\frac{q}{4 \pi \varepsilon_0 r'}~,~~~~\boldsymbol{A}'=(0,\,0,\,0)~~\Rightarrow~
A'^{ \mu}=\left(\frac{\Phi'}{c},\,0,\,0,\,0\right).
\end{equation}
Now we carry out a  Lorentz transformation  into the laboratory system.  
\begin{equation}
\Phi=\gamma(\Phi'+v\,A'_z)=\gamma \Phi' \,,~~~~A_z=\gamma \left(A'_z+\frac{ v}{c^2} \Phi' \right)
=\frac{\gamma \, v}{c^2}\Phi' ~~~\mathrm{with}~~~\gamma=\frac{1}{\sqrt{1-v^2/c^2}}\,.
\end{equation}
The transverse components remain invariant:
$A_x=A_x'=0$, $A_y=A_y'=0 \,.$
For a charged particle, moving with constant speed on a straight line,  there is a simple connection between scalar and vector potential
\begin{equation}
\boldsymbol{A}=\frac{\boldsymbol{v}}{c^2}\,\Phi \,.
\end{equation}
The electric and magnetic fields are computed using the Maxwell equations.
\begin{equation}\label{E-bewegtesTeilchen-Kap7}
\boldsymbol{E}(r,\theta)=\frac{q}{4 \pi \varepsilon_0 r^2}\cdot\frac{(1-\beta^2)}{(1-\beta^2\sin^2{\theta})^{3/2}}
\cdot \frac{\boldsymbol{r}}{r},~~~\boldsymbol{B}(r,\theta)=\frac{1}{c^2}\,\boldsymbol{v \times E}(r,\theta)
\end{equation}
where $\theta$ is the angle between the  $z$ axis and the vector $\boldsymbol{r}=(x,y,z)$ and $\beta=v/c$.  Details of the computation can be found in Refs.\,\cite{Casalbuoni-I, Theorie-II}. The polar angle distributions of the electric field vector of a positron  which is either at rest or is moving with $v=0.87\,c$ ($\gamma=2$) are plotted in Fig.\,\ref{rel-pointcharge}. (For convenience we plot here the electric field lines of a positive charge. For  an electron the  field vectors point inwards and the  arrow heads would be hardly visible).

  Specifically, the field components parallel and perpendicular to the particle velocity $\boldsymbol{v}$ are
\begin{eqnarray}\label{Epar-Eperp}
\left|\boldsymbol{E}_\parallel \right|&=&\frac{q}{4 \pi \varepsilon_0 r^2}\cdot(1-\beta^2)=\frac{q}{4 \pi \varepsilon_0 r^2}\cdot \frac{1}{\gamma^2}~~~~\mbox{for}~~ 
(\theta=0) \,,\nonumber \\
\left|\boldsymbol{E}_\perp\right|&=&\frac{q}{4 \pi \varepsilon_0 r^2}\cdot \frac{1}{\sqrt{1-\beta^2}}=\frac{q}{4 \pi \varepsilon_0 r^2}\cdot \gamma
~~~~\mbox{for}~~(\theta=\pi/2)\,.
\label{Eperp}
\end{eqnarray}
With increasing Lorentz factor $\gamma$ there is a rapidly  increasing anisotropy, the transverse field component grows with $\gamma$ while  the longitudinal component drops as $1/\gamma^2$. For electron energies in the GeV range the field is almost completely transverse.

\subsection{Spectral energy of transition radiation in the backward hemisphere }
\begin{figure}[ht!]
\centering
\includegraphics[angle=0,width=16cm]{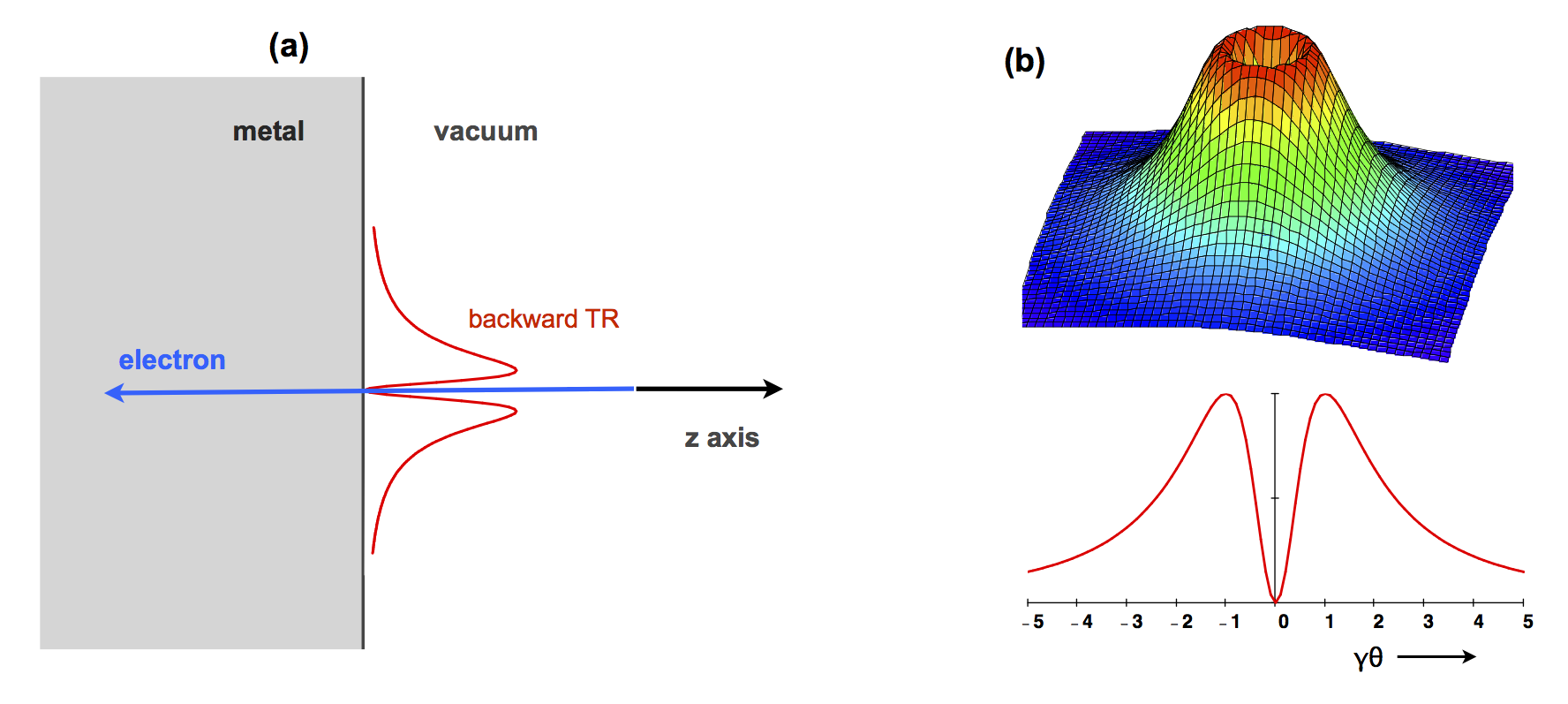}
  \caption[]{\small{{\bf (a)} Schematic view of the generation of backward transition radiation. A relativistic electron moves in negative $z$ direction and crosses the interface between vacuum and a metal. Transition radiation is emitted into the  positive $z$ hemisphere.  {\bf (b)}
  The spectral energy density of  transition radiation as a function of the scaled angle $\gamma \,\theta$ according to the Ginzburg-Frank formula. The maximum occurs at $\gamma \,\theta =1$. Note that the intensity vanishes at $\theta=0$ (exact backward direction with respect to the electron velocity). This is due to the radial polarization of TR, depicted in Fig.\,\ref{rel-pointcharge}.}}
\label{Ginz-Frank}
\end{figure}

 For the special case that an electron passes from vacuum into a  metal, only backward radiation is emitted at the interface since electromagnetic waves cannot propagate inside the metal. The generation of backward TR is shown schematically in 
Fig.\,\ref{Ginz-Frank}a.  A characteristic feature of transition radiation  is its radial polarization (see Fig.\,\ref{rel-pointcharge}b) which is very different from the well-known  linear, circular or elliptic polarization of a laser beam.  \\
In case of an infinite planar boundary, the spectral energy density of   transition radiation, emitted into the backward hemisphere,  is given by the Ginzburg-Frank formula
\begin{equation} \label{GF}
\left[\frac{d^2U}{d\omega d\Omega}\right]_{\rm GF} =\frac{e^2}{4 \pi^3 \varepsilon_0 c}\cdot 
\frac{\beta^2 \sin^2 \theta}{(1-\beta^2 \cos^2 \theta)^2}
\end{equation}
with $\beta=v/c$ and $\theta$ the angle against the backward direction. For a derivation of this formula we refer to Landau-Lifshitz \cite{Landau-Lifshitz}, see also \cite{Ter-Mik}.  Note that the Ginzburg-Frank formula is only valid if the radiation is observed in the far-field (Fraunhofer) diffraction regime. 
The angular distribution is shown in Fig.\,\ref{Ginz-Frank}b. The intensity vanishes in the exact backward direction at $\theta=0$, which is a  consequence of the radial polarization. 
        
\noindent The angular distribution has its maximum  at the angle
\begin{equation}
\theta_{\rm m}=\frac{1}{\gamma}=\sqrt{1-\beta^2}=\frac{m_ec^2 }{E_e} 
\end{equation}
 but extends to significantly larger angles.
For an infinite screen, the spectral TR energy density (\ref{GF}) does not depend on the circular frequency\footnote{In the following $\omega$ is called ``frequency'' for short.} 
$\omega=2\pi f$, provided one measures in the far-field and stays well below the plasma frequency $\omega_p=\sqrt{n_e e^2/(\varepsilon_0 m_e)}$ of the metal, which is in the ultraviolet. In the next section  we will show that for a finite TR screen the radiation energy acquires an $\omega$ dependence and its angular distribution is widened.

\subsection{Generalizations of the Ginzburg-Frank formula}
The Ginzburg-Frank  formula  is not applicable in most practical cases because two basic conditions of the analytic derivation may  not be fulfilled: (a) the radiation screens used in an accelerator are of limited size, and (b) the radiation is usually observed in the near-field and not in the far-field diffraction regime. To construct a generalization of the Ginzburg-Frank  formula  we make use of the fact that the  electric
 field of an ultrarelativistic  particle  is predominantly  perpendicular to the direction of motion, see Eq.\,(\ref{Epar-Eperp}). The field resembles closely  that of an  electromagnetic wave propagating in vacuum. This is the reason why  the Weizs\"acker-Williams method of virtual quanta can be utilized for computing backward TR from a metallic screen: by reflection at the screen, the virtual photons of the particle's self-field are converted to the real photons of a backward-moving electromagnetic wave.  The virtual-photon method would completely fail for non-relativistic particles which have large longitudinal field components. 
\vspace{1mm}

\noindent The virtual-photon method has been used by us \cite{Casalbuoni-I} to compute the wave propagating in backward direction for radiation screens of arbitrary  size, both in the far-field and in the near-field diffraction regimes.
The transverse electric field component of a highly relativistic electron ($q=-e$) moving along the $z$ axis  is \cite{Casalbuoni-I}, \cite{Theorie-II}
\begin{equation}
E_\perp(\rho,z,t)=-\gamma \,\frac{e}{4 \pi \varepsilon_0}\,
\frac{\rho}{(\rho^2+\gamma^2(z-\beta c t)^2)^{3/2}}    ~~~\mathrm{with}~~   \rho=\sqrt{x^2+y^2}\,.
\end{equation}
 The field depends on the distance $\rho$ from the axis but not on the azimuthal angle. 
When the relativistic electron passes   by an observer at a small distance its transverse electric field appears as a very short time pulse. The Fourier transform of the transient field is derived in \cite{Casalbuoni-I}, see also Jackson~\cite{Jackson}:
\begin{equation}\label{Fourier-Eperp-A}
\tilde{E}_\perp(\rho,\omega)= \frac{-e \,\omega}{(2 \pi)^{3/2} \varepsilon_0 
  \gamma \beta^2 c^2}\,K_1\left(\frac{\omega\, \rho}{ \gamma \beta c } \right) .
\end{equation}
The function $K_1$ is a modified Bessel function. \vspace{2mm}

\noindent To preserve cylindrical symmetry the  TR screen  is chosen to be a circular disc of radius $a$ which is centered with respect to the $z$ axis. On this screen we use  cylindrical coordinates 
$(\rho,\varphi)$. The radiation is detected on a remote observation screen. 
\begin{figure}[ht!]
\centering
\includegraphics[angle=0,width=12cm]{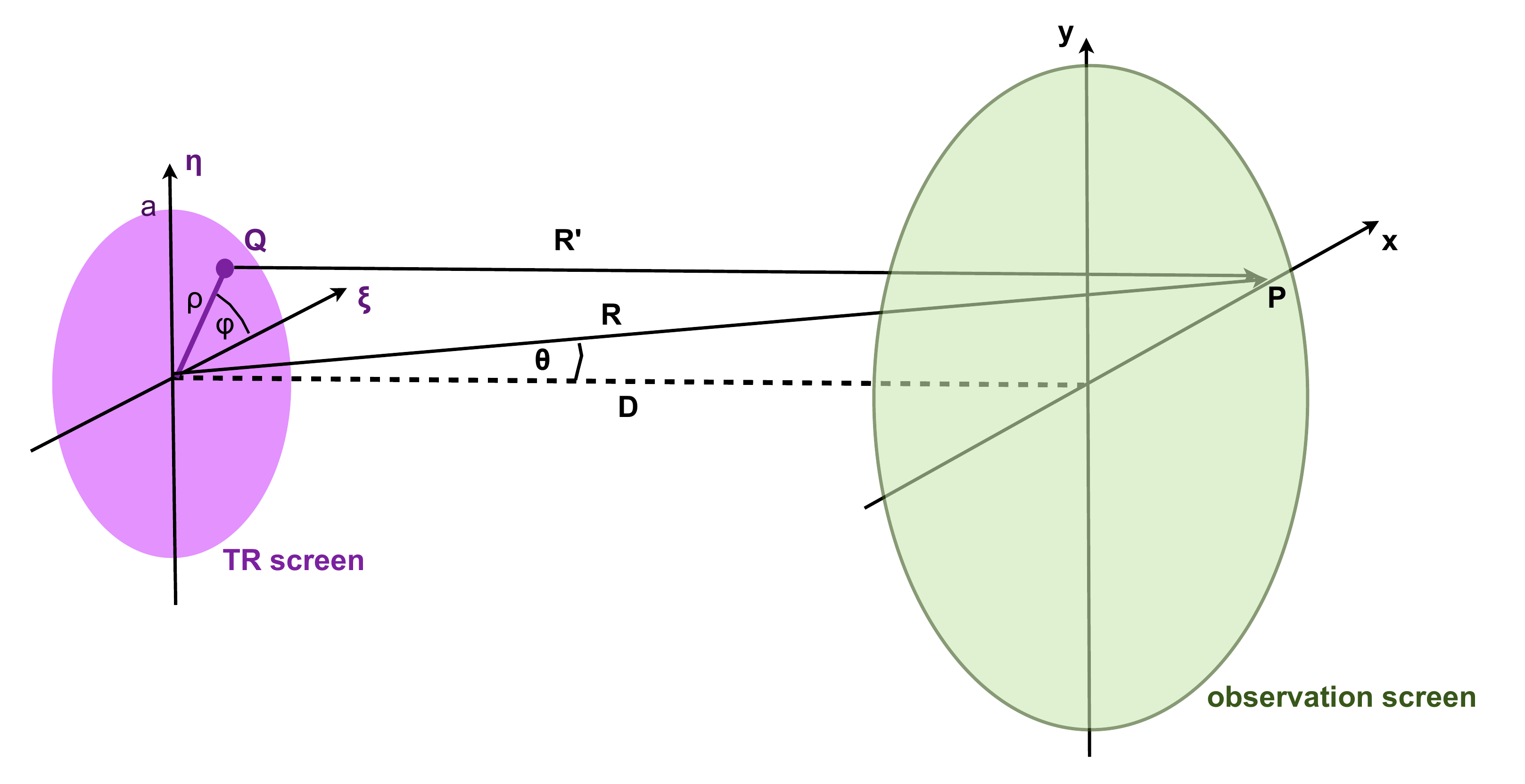}
  \caption[]{ \small{Diffraction geometry for a circular TR screen of radius $a$ and an observation  screen at large distance $D\gg a$.}}
\label{diff-geom}
\end{figure}   
Because of the cylindrical symmetry of the emitted transition radiation we can restrict ourselves to points on the $x$ axis of the observation  screen, hence the coordinates of our observation  point are $P=(x,0,D)$.
The distance between $P$ and the center of the TR screen  is $R=\sqrt{D^2+x^2}$, while the distance between $P$ and an  arbitrary point $Q=(\rho,\phi,0)$ on the TR screen  is $R'=\sqrt{D^2 + (x-\rho \cos \phi)^2+(\rho \sin \phi)^2}$.
The radius $a$ of the  TR screen is assumed to be much smaller than $R$, hence $\rho \le a \ll R$ and the square root can be expanded into a Taylor series:
\begin{equation} \label{distance-R2}
R'=\sqrt{D^2 + (x-\rho \cos \phi)^2+(\rho \sin \phi)^2}\,\approx \, R\,\left(1  -\left(\frac{\rho }{R}\right) \,\frac{x \,\cos \phi }{R}
+\frac{1}{2}\,\left(\frac{\rho }{R}\right)^2 \right)~~~~\mathrm{with}~~R=\sqrt{D^2+x^2}\,.
\end{equation}
The   term which is linear in $(\rho/R)$   describes  far-field diffraction,  the quadratic  term $(\rho/R)^2$  accounts  for the additional   near-field diffraction effects.  \vspace{2mm}

\noindent {\bf Far-field diffraction}\\
In the far-field regime, the  Fourier-transformed electric field on the   observation screen  can be computed analytically  \cite{Casalbuoni-I}. We express it here as a function of the wave number $k=\omega/c$.
\begin{equation}\label{Fourier-Eperp-B}
\left[\tilde{E}(\theta,k)\right]_{\rm far} =\frac{e}{(2\pi)^{3/2} \varepsilon_0\, c}\frac{\exp(i k R)}{R}
\frac{\beta \sin{\theta}}{1-\beta^2 \cos^2{\theta} }[1-T(\theta,k)]~
\end{equation}
with a correction term that accounts for the finite  TR screen radius $a$:
\begin{equation}
T(\theta,k)=\frac{k a  }{ \beta \gamma}\,J_0(k a \sin \theta) 
K_1\left(\frac{k a}{\beta  \gamma} \right)+\frac{k a }{\beta^2\gamma^2\sin{\theta}}J_1(k a \sin{\theta}) \,
K_0\left(\frac{k a}{ \beta \gamma} \right).
\label{T(theta,k)}
\end{equation}
The resulting spectral energy as a function  of the angle $\theta$ is \cite{Casalbuoni-I}
\begin{equation}
\left[\frac{d^2U}{d\omega d\Omega}\right]_{\rm far} =\frac{e^2 k^4}{4\pi^3 \varepsilon_0 c  \beta^4\gamma^2}\left|\int_0^a    J_1(k \rho \sin{\theta})K_1\left(\frac{k \rho }{\beta \gamma} \right)\, 
        \rho \,d\rho \right|^2.
        \label{GF-farfield-int}
\end{equation}
The integration can be done analytically and yields the  far-field generalization of the Ginzburg-Frank formula for a circular radiation screen of finite radius $a$
\begin{equation}\label{GF-farfield}
\left[\frac{d^2U}{d\omega d\Omega}\right]_{\rm far} =\left[\frac{d^2U}{d\omega d\Omega}\right]_{\rm GF} \cdot  \left[1- T(\theta,k)\right]^2~~~~~~(\omega=k\,c).
\end{equation}
The correction term $T(\theta,k)$ vanishes for $a \rightarrow \infty$, so this formula reduces to the  standard  Ginzburg-Frank formula (\ref{GF}) for a sufficiently large TR screen.  Rectangular or other screen shapes can be treated with the Fourier-transform algorithm explained in Appendix D.\vspace{2mm}

\noindent {\bf Near-field diffraction}\\
In order to cover also the near-field we have to include the second-order term in Eq.\,(\ref{distance-R2}).
The angular dependence of the spectral energy is then given by 
\begin{equation}\label{GF-nearfield}
\left[\frac{d^2U}{d\omega d\Omega}\right]_{\rm near} =\frac{e^2 k^4}{4\pi^3 \varepsilon_0 c  \beta^4\gamma^2}\left|\int_0^a    J_1(k \rho \sin{\theta})K_1\left(\frac{k \rho }{\beta \gamma} \right)\, \exp\left( \frac{i k\,\rho^2}{2 R}\right) \, \rho \,d\rho \right|^2
\end{equation}
This is the near-field generalization of the Ginzburg-Frank formula for a circular radiation screen.
The difference to (\ref{GF-farfield-int}) is the extra phase factor $\exp(i k\rho^2/(2R))$. 
The integral in (\ref{GF-nearfield}) must be evaluated numerically. \vspace{10mm}

\begin{figure}[ht!]
\centering
\includegraphics[angle=0,width=10cm]{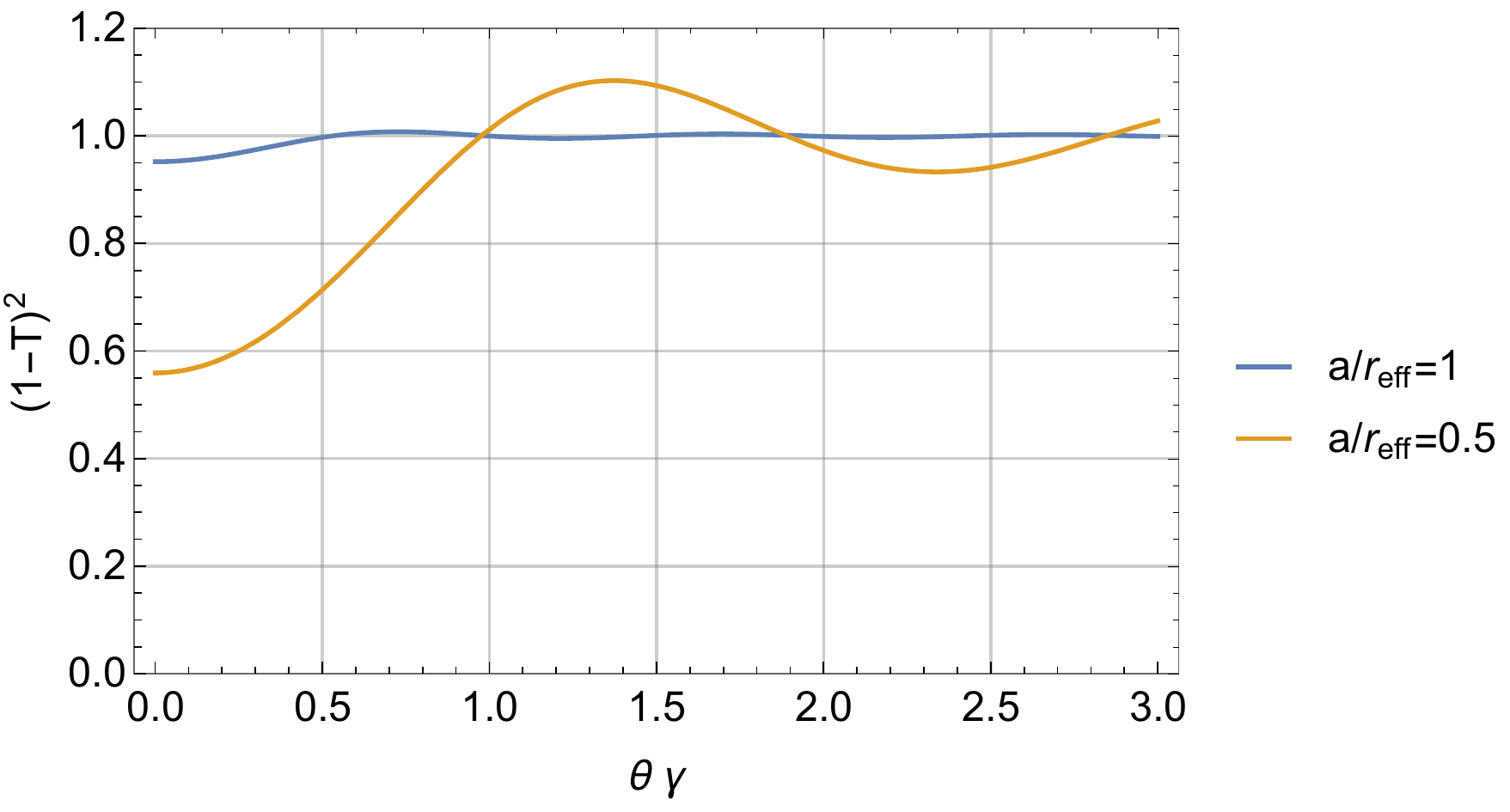}
  \caption[]{ \small{The factor $(1-T(\theta_s,\xi))^2$, plotted  as a function of the scaled angle $\theta_s=\gamma\,\theta$,  for two values of the scaled disc radius $\xi=a/r_{\rm eff}$.}}
\label{eff-radius}
\end{figure}

\noindent {\bf Effective source size and far-field condition}\\
When the disc radius $a$ is  large and the observation screen very far away one should expect that the Ginzburg-Frank angular distribution is recovered. This is indeed the case. The question is, how large the radius has to be. It turns out that the answer depends on the wavelength and the Lorentz factor. Following Castellano et al. \cite{Castellano} we define an {\it effective source radius} by 
\begin{equation} \label{TR-reff}
r_{\rm eff}=\gamma \lambda\,.
\end{equation}
The first condition for obtaining the Ginzburg-Frank angular distribution is that the TR source radius has to exceed the effective source radius
\begin{equation}\label{eff-size}
 r_{\rm source} \equiv a \ge r_{\rm eff}=\gamma \lambda \,.
\end{equation} 
Quantitatively, we can understand the effective source  size condition as follows.
We rewrite  the correction term  (\ref{T(theta,k)}) using scaled variables $\xi=a/r_{\rm eff}$ and $\theta_s=\gamma\,\theta$ and restricting ourselves to small angles: 
\begin{equation}
T(\theta_s,\xi)=2\pi\,\xi \, \left(\frac{1}{\theta_s}\,K_0(2\pi\,\xi)J_1(2\pi\,\xi\,\theta_s)+K_1(2\pi\,\xi)J_0(2\pi\,\xi\,\theta_s)\right)
\end{equation}
The factor $(1-T(\theta_s,\xi))^2$ is plotted in Fig.\,\ref{eff-radius} as a function of the scaled angle $\theta_s=\gamma\,\theta$ for two screen radii:
$a/r_{\rm eff}=1$ and $a/r_{\rm eff}=0.5.$
It is obvious that the reduction of the spectral energy  caused by  the finite TR screen size is small for $a/ r_{\rm eff}\ge 1$ but becomes very significant for 
$a/r_{\rm eff}\le 0.5$.

\begin{figure}[ht!]
\centering
\includegraphics[angle=0,width=14cm]{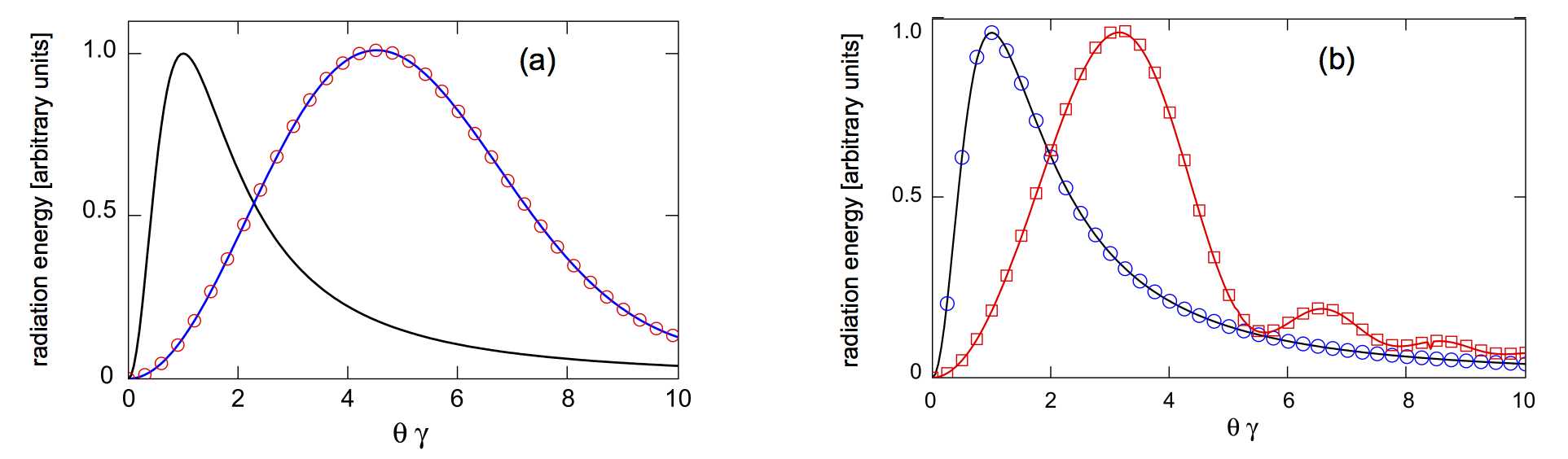}
  \caption[]{ \small{{\bf (a)} Far-field TR from a circular disc of radius $a=3\,$mm for $\lambda=0.3~$mm,
  $\gamma=100$, $r_{\rm eff}=30\,$mm, $D=4~$m. The far-field criterion   (\ref{farfield-cond}) is satisfied:
  $D> \gamma r_{\rm eff}=3\,$m.   However, the effective source-size criterion is badly violated, 
  $a \ll r_{\rm eff}$. Black curve: Ginzburg-Frank formula; blue curve: far-field computation using Eq.\,(\ref{GF-farfield}); red circles: numerical near-field computation  using Eq.\,(\ref{GF-nearfield}).  The distributions are individually normalized to a maximum value of 1. \newline
 {\bf (b)} Near-field  TR from a circular disc of radius $a=30\,$mm for $\lambda=0.3~$mm,
  $\gamma=100$, $r_{\rm eff}=30\,$mm, $D=0.2\,$m.  The effective source-size criterion is satisfied but the far-field condition is strongly violated, $D \ll \gamma r_{\rm eff}=3\,$m. Black curve: Ginzburg-Frank formula; blue circles: far-field prediction (\ref{GF-farfield}); red curve: near-field prediction (\ref{GF-nearfield}); red squares: numerical calculation using the exact square root expression for $R'$.}}
\label{sourcesize}
\end{figure}  

\vspace{4mm}

 \noindent The second  condition for obtaining the Ginzburg-Frank angular distribution is to have far-field diffraction, which requires
\begin{equation} \label{farfield-cond}
D \gg \gamma \,r_{\rm eff}=\gamma^2 \lambda \,.
\end{equation}
This inequality follows from formula (\ref{distance-R2}). In the far-field, the quadratic term $(\rho/R)^2$  must be much smaller than the  term which is linear in $(\rho/R)$. Assuming that the source size criterion is fulfilled, then $\rho_{\rm max}=a=r_{\rm eff}=\gamma \lambda$, and with $R\approx D$ and  $x \approx  D/\gamma$ for a typical point on the observation screen one gets the inequality (\ref{farfield-cond}).
In  fact, when (\ref{farfield-cond}) is  fulfilled,  a numerical evaluation yields an almost perfect agreement between the formulas  (\ref{GF-farfield}) and (\ref{GF-nearfield}). 
\vspace{2mm}

\noindent Significant differences arise, however, when one of these conditions is violated. In Fig.\,\ref{sourcesize}  we show several examples. When the far-field condition is satisfied but the source-size condition is violated, the formulas  (\ref{GF-farfield}) and (\ref{GF-nearfield}) are in agreement but both predict a wider angular distribution than the Ginzburg-Frank formula (\ref{GF}). When the source-size condition is satisfied but the far-field condition is violated, the far-field formula~(\ref{GF-farfield})  yields the same angular distribution as the Ginzburg-Frank formula~(\ref{GF}) but the near-field formula (\ref{GF-nearfield})  yields a wider distribution. \vspace{2mm}

\noindent For the typical electron Lorentz factors at FLASH of $\gamma >1000$    the effective source-size and the far-field conditions are both violated except at very small wavelengths in the  few $\mu$m range. Hence Eq.\,(\ref{GF-nearfield}) must be used to compute the spectral energy
 entering   a detector with aperture angle $\theta_{\rm ap}$
\begin{equation}
U_{\rm det}(\omega) =
 \int_0^{\theta_{\rm ap}} \left[\frac{d^2U}{d\omega d\Omega}\right]_{\rm near} 
\,2\pi\,\sin{\theta}\,d\theta\,.
\end{equation}
It is very important to realize that only  a small fraction of the TR energy is emitted at very small angles, $\theta \le \theta_{\rm m}$. Hence  for intensity reasons the aperture angle of a  TR detector is  often chosen to be much larger than  $\theta_{\rm m}=1/\gamma$. Thereby one accepts  near-field transition radiation which has a wide tail towards larger angles and, more importantly, the intensity observed at these larger angles is enhanced by the growing solid angle $d\Omega=2\pi \sin{\theta}\,d\theta$. The radiation energy entering the detector continues to grow with increasing  aperture angle even beyond $\theta_{\rm ap}  \ge 100 \,\theta_{\rm m}$.

 \clearpage

\section{Electron Bunch Shape Reconstruction using Coherent  Transition Radiation }
Consider an electron bunch as sketched in Fig.\,\ref{3D-Bunch}. We want to determine the transition radiation produced by the $N \gg 1$ particles in the bunch as a  function of frequency $\omega$ and emission angle $\theta$. 
The spectral energy receives  contributions from incoherent and coherent radiation. \vspace{1mm}

\noindent {\it Incoherence} means that interference terms average to zero because of  random phase relations, and hence it is  permitted to add probabilities. The  incoherent spectral energy density produced by a bunch of $N$ electrons is simply $N$ times the spectral energy density produced by one electron:
\begin{equation}
\left[\frac{d^2U}{d\omega d\Omega}\right]_{\rm incoh} =N \,\left[\frac{d^2U}{d\omega d\Omega}\right]_{1}\,. 
\end{equation}
Incoherent transition radiation in the visible range is very useful for transverse beam diagnostics (e.g. emittance measurements) but is not suited for determining  details of the longitudinal bunch structure. \vspace{1mm}

\noindent {\it Coherence} means that one  has to add complex amplitudes. The absolute square of the sum amplitude yields the probability, and when multiplying this amplitude with its complex conjugate, interference terms come in.    
Full coherence means that the radiation fields of all $N$ electrons add constructively. The intensity $I_N$ is then $N^2$ times the intensity $I_1$ emitted by a single electron. Usually, however, there is partly constructive and partly destructive interference, and in that case $I_N < N^2I_1$.  The form factor, or more accurately its absolute square,  is a measure  of the degree of constructive interference.

\subsection{Form factor}\label{formfactor}
\subsubsection{Three-dimensional form factor}

\begin{figure}[ht!]
\centering
\includegraphics[angle=0,width=10cm]{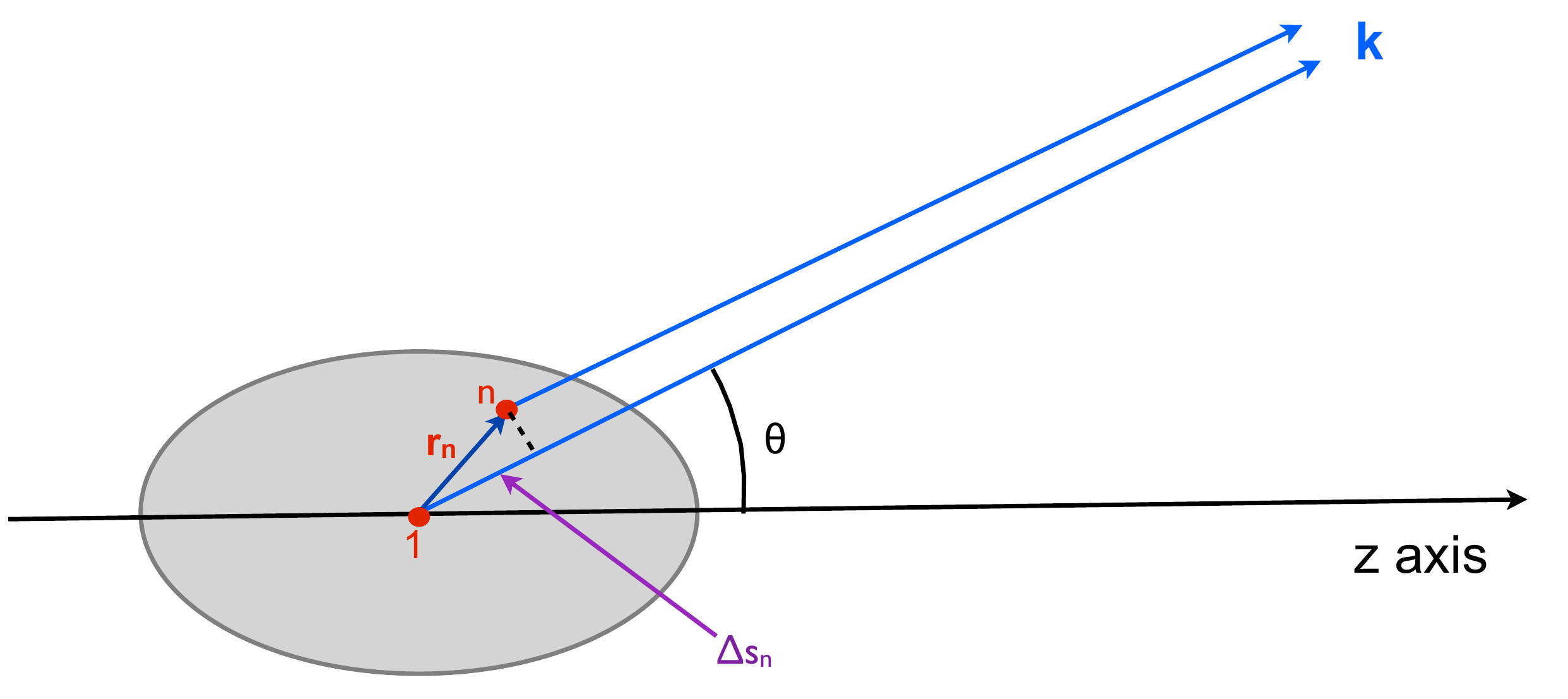}
  \caption[]{\small{Computation of the phase shift between the waves produced by an arbitrary electron at $\boldsymbol{r}_n$ and the reference electron at $\boldsymbol{r}_1=0$. The distances to a far remote  observation point differ by 
  $\Delta s_n=s_1-s_n=\boldsymbol{k\cdot r}_n/k$, the phases differ by
  $\Delta \varphi_n=\Delta s_n \,(2\pi/\lambda)=\boldsymbol{k\cdot r}_n\,$. }
   }
\label{3D-Bunch}
\end{figure}

As said above,  coherence means that complex amplitudes have to be added. This will be done now.
To compute coherent transition radiation (CTR)  we place a ``reference electron'' at the bunch center and label it with the index ``1''. The radiation field of this single electron is given by Eq.\,(\ref{Fourier-Eperp-B}), we rewrite it in the form
\begin{equation}\label{Fourier-Eperp-C}
\tilde{E}_1(\boldsymbol{k})=\frac{e }{(2\pi)^{3/2} \varepsilon_0 c}\frac{\exp(i k R)}{R}
\frac{\beta \sin{\theta}}{1-\beta^2 \cos^2{\theta} }[1-T(\theta,k)]
\end{equation}
with   the  wave  vector
\begin{equation}
\boldsymbol{k}=\frac{\omega}{c}(\sin{\theta},0,\cos{\theta})\,,~~~~k=|\boldsymbol{k}|=\omega\,c=\frac{2\pi}{\lambda}\,.
\end{equation}
An arbitrary electron ``$n$'' at a position $\boldsymbol{r}_n$ inside the bunch produces a field of the same mathematical form, but since it crosses the TR screen at a different time and a different position, there will be a phase shift $\Delta \varphi_n=\boldsymbol{k\cdot r}_n$ with respect to the reference electron, see Fig.\,\ref{3D-Bunch}.
The total field in $\boldsymbol{k} $ direction is obtained by summing over all electrons in the bunch
$$\tilde{E}_{\rm tot}(\boldsymbol{k})=\tilde{E}_1(\boldsymbol{k}) \,\sum_{n=1}^N \exp(i\,\boldsymbol{k \cdot r}_n)\,.$$

A typical electron bunch consists of some $10^9$ electrons, and it is useful to replace the discrete distribution of $N$ point particles by a continuous particle density $\rho(\boldsymbol{r} )$ which we normalize to 1.
Hence the total field can be written as
\begin{equation}
\tilde{E}_{\rm tot}(\boldsymbol{k})=N\,\tilde{E}_1(\boldsymbol{k}) \,\int  \rho(\boldsymbol{r} )\exp(\,i\,\boldsymbol{k \cdot r})\,d^3r \equiv N\,\tilde{E}_1(\boldsymbol{k}) \tilde{F}_{\rm 3D}(\boldsymbol{k} )
\end{equation}
where we have defined 
the three-dimensional bunch form factor
\begin{equation}
\tilde{F}_{\rm 3D}(\boldsymbol{k} )=\int  \rho(\boldsymbol{r} )\exp(\,i\,\boldsymbol{k \cdot r})\,d^3r
~~~\mathrm{with}  ~~~\int  \rho(\boldsymbol{r} )\,d^3r=1\,.
\end{equation}
From this equation follows immediately 
$$\tilde{F}_{\rm 3D}(0)=1\,.$$
The Fourier back transformation reads
\begin{equation}
\rho(\boldsymbol{r} )=\frac{1}{(2\pi)^3}\int  \tilde{F}_{3D}(\boldsymbol{k} )\exp(-i\,\boldsymbol{k \cdot r})\,d^3k\,.
\end{equation}
The coherent spectral energy density produced by a bunch of $N$ electrons is the product of   the spectral energy density produced by a single electron, the number of electrons squared and the absolute square of the form factor: 
\begin{equation}
\left[\frac{d^2U}{d\omega d\Omega}\right]_{\rm coh} =N^2 \,\left[\frac{d^2U}{d\omega d\Omega}\right]_{1} \,
|\tilde{F}_{\rm 3D}(\boldsymbol{k} )|^2.
\end{equation}

In the following treatment  we make the simplifying assumption that the transverse density distribution  of the electron bunch is independent of the longitudinal position in the bunch. In other words, we assume that the slice emittance and other beam parameters are constant along the bunch. Then the 3D particle density can be factorized:
\begin{equation}
\rho(x,y,z)=\rho_{\rm trans}(x,y)\,\rho_{\rm long}(z)
\end{equation}
and the 
3D form factor is the product of the  transverse  and the longitudinal form factors:
\begin{equation}
\tilde{F}_{\rm 3D}(\boldsymbol{k} )=\tilde{F}_{\rm trans}(k_x,k_y) \,\tilde{F}_{\rm long}(k_z)\,.
\label{3DFF}
\end{equation}
In reality,  the  electron bunches  in FLASH  are  affected by nonlinear effects in the magnetic bunch compressor and acquire   a slice emittance that varies along the bunch axis. The considerations in the next section show that a variable slice emittance has a rather  small impact on the observable CTR spectra. The effect  will be ignored here.

\subsubsection{Transverse form factor}
 The transverse charge density distribution is assumed to be a cylindrically symmetric Gaussian.  The normalized distribution is written in the form
$$\rho_{\rm trans}(x,y)=\frac{1}{\sqrt{2 \pi} \,\sigma}\exp\left(-\frac{x^2}{2\,\sigma^2}\right)\,\frac{1}{\sqrt{2 \pi}\, \sigma}\exp\left(-\frac{y^2}{2\,\sigma^2}\right).$$
The transverse form factor of the two-dimensional charge distribution is defined by the equation
$$\tilde{F}_{\rm trans}(k_x,k_y)=\iint \rho_{\rm trans}(x,y)\, \exp( i\,[k_x x +k_y y])\,dxdy\,.$$
Because of the cylindrical symmetry we can assume without loss of generality  that the wave vector $\boldsymbol{k}$ is located in the $(x,z)$ plane. Then
$$\boldsymbol{k}=k(\sin \theta, 0,\cos \theta)\,,~~k_x x +k_y y=k \,x\,\sin \theta\,.$$
Using these relations  the transverse form factor can be written as a function of wave number $k$ and emission angle $\theta$
\begin{equation}
\tilde{F}_{\rm trans}(k,\theta)=\exp \left(-\frac{k^{2} \sigma^2 \sin^2 \theta}{2}\right).
\label{FFtrans-Gauss}
\end{equation}

Backward TR is confined to a fairly narrow cone around the $z$ axis (a typical aperture angle of the detector is 100\,mrad).  CTR spectroscopy is therefore not suited for determining the transverse density distribution in the bunch. At large wavelengths (small wave numbers) the product $k \,\sigma \sin \theta =2\pi \sin \theta \,\sigma/\lambda$ is close to zero, and the transverse form factor  is close to 1. 
However,  at small wavelengths ($\lambda < \sigma$) there is considerable destructive interference and the  transverse form factor may drop to small values. This reduction depends  on the aperture angle of the spectrometer. To get an impression we use the 
following typical electron beam parameters at the position of the CRISP spectrometer in the FLASH linac: \vspace{1mm}

\noindent Lorentz factor $\gamma=1500$, normalized emittance
$\varepsilon_n \approx 2\,\mu$m, beta functions $\beta_x=\beta_y=7\,$m, $\sigma\approx 100\,\mu$m.\vspace{1mm}

 The  square of the transverse form factor must be averaged over the aperture  of the spectrometer, with a weight factor given by the angular-dependent radiation energy density.   This average depends on the wave number $k=2\pi/\lambda$,  the 
rms electron beam radius $\sigma$, and the aperture angle $\theta_{\rm ap}$:
\begin{equation}
\langle |\tilde{F}_{\rm trans}|^2 \rangle(k, \sigma,\theta_{\rm ap})
=\frac{\int_0^{\theta_{\rm ap}} {\cal U}(\theta,k)\, \exp \left(-k^{2} \sigma^2 \sin^2 \theta\right)
\,\sin{ \theta} \,d\theta } { \int_0^{\theta_{\rm ap}} {\cal U}(\theta,k)
\,\sin{ \theta} \,d\theta} 
\label{FF2av}
\end{equation}
where  ${\cal U}(\theta,k)$ stands for the near-field radiation energy density\,(\ref{GF-nearfield}), which is evaluated here for the case  of a single  aperture  at a distance of $D=1\,$m. A more accurate treatment,  based  on  a {\it THzTransport} simulation of the radiation transport from the TR screen through the CTR beamline to the multichannel spectrometer,  will be presented in Section \ref{spectrometer}.\vspace{2mm}

\begin{figure}[ht!]
\centering
\includegraphics[angle=0,width=17.5cm]{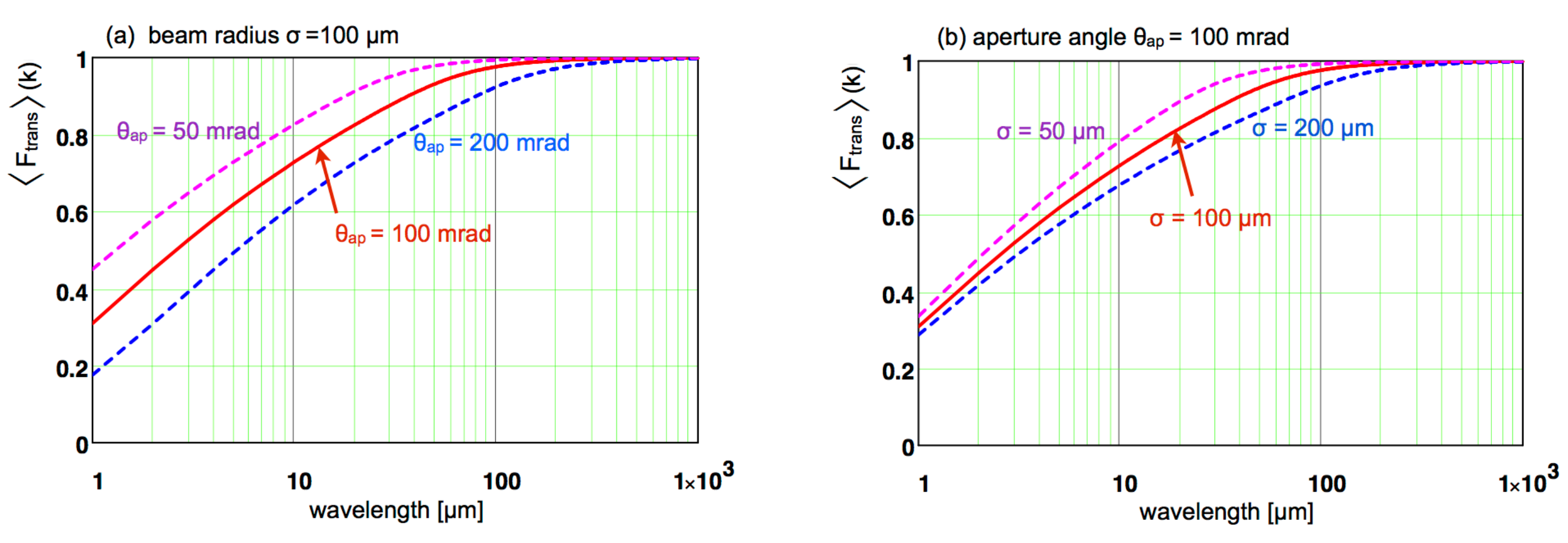}
  \caption[]{ \small{{\bf (a)} The rms transverse form factor  $\langle \tilde{F}_{\rm trans} \rangle(k)$,  plotted as a function of wavelength $\lambda=2\pi/k$, for  aperture angles of $\theta_{\rm ap}=50,\, 100,\,200\,$mrad and a fixed rms electron beam radius $\sigma=100\,\mu$m. \newline {\bf (b)} The rms transverse form factor, plotted versus $\lambda=2\pi/k$, for a fixed aperture angle of $\theta_{\rm ap}=100\,$mrad and  rms electron beam radii of $\sigma=50,\,100, \,200\,\mu$m. }
   }
\label{Fig-F2trans}
\end{figure}

\noindent  The root-mean-square value of the transverse form factor is the square root of expression\,(\ref{FF2av}):
\begin{equation}
\langle \tilde{F}_{\rm trans}\rangle(k)=\sqrt{\langle |\tilde{F}_{\rm trans}|^2 \rangle(k, \sigma,\theta_{\rm ap})} ~.
\label{FFav}
\end{equation}
The rms transverse form factor $\langle \tilde{F}_{\rm trans}\rangle(k)$ depends implicitly on the parameters $\sigma$ and $\theta_{\rm ap}$, which are fixed quantities in a given experimental setup,  but these dependencies  are not written down here to simplify the notation. \\
The impact of the  spectrometer aperture angle $\theta_{\rm ap}$    is depicted in Fig.\,\ref{Fig-F2trans}a where $\langle \tilde{F}_{\rm trans}\rangle(k)$ is plotted versus $\lambda=2\pi/k$  for a typical rms beam radius of  $\sigma=100\,\mu$m and various aperture angles. 
The impact of the rms beam radius is shown in Fig.\,\ref{Fig-F2trans}b for an aperture angle of 100\,mrad (this is roughly the aperture   of  our  spectrometer setup).
The suppression of small wavelengths by the transverse form factor is substantial. The measured spectra have to  be corrected for these losses. This will be done with the help of the response function defined in Section~\ref{spectrometer}.

\subsubsection{Longitudinal form factor} \label{FFlong}
In the following we assume that the measured spectra have been corrected properly for the above-mentioned suppression effects at small wavelengths. 
 The spectrometer aperture angle $\theta_{\rm ap}\approx 100\,$mrad  is so  small that $\cos \theta$ is almost equal to 1 in the angular range $0 \le \theta \le \theta_{\rm ap}$, hence $k_z=k\cos \theta$ can be replaced by $k=\omega/c$.  
 
 When an electron bunch crosses the  TR source screen it generates a radiation pulse whose time duration $\tau$ is related to the bunch length $\ell$ by
 $\ell=v \tau=\beta c \tau$. 
For comparison with time-domain diagnostic instruments such as the TDS it is advantageous to express  the longitudinal density distribution $\rho_{\rm long}(z)$ of the  bunch as a function of time:
$$\rho(t)=\rho_{\rm long}(\beta c\, t)\,.$$
The longitudinal form factor can be written as a function  of $\omega$
$${\cal F}(\omega)=\tilde{F}_{\rm long}(\omega/c)\,.$$
The subscript ``long''  is not  needed anymore and will be dropped in the following.

The Fourier transformation relations  between normalized longitudinal density distribution and longitudinal form factor are\footnote{The factors ``1'' in front of the  Fourier integral and ``$1/(2\pi)$''  in front  of the  inverse Fourier integral are  chosen such that the basic requirement ${\cal F}(0)=1$ is fulfilled:  at very low frequency  (very large wavelength) the  bunch acts as a  point charge whose form factor must be unity. }
\begin{equation}
{\cal F}(\omega)=\int_{-\infty}^{\infty} \rho(t) \exp(i\,\omega\,t) dt\,,~~~
\rho(t)=\frac{1}{2\pi}\,\int_{-\infty}^{\infty} {\cal F}(\omega)\exp(-i\,\omega\,t) d\omega ~~~\mathrm{with}~~
\int_{-\infty}^{\infty} \rho(t) dt=1 \,.
\label{Fourier-formfactor}
\end{equation}
At   low frequencies, namely when the wavelength of the radiation is long compared to the bunch length,  the form factor of a bunch centered at $t=0$ is a real number close to 1.    All electrons radiate coherently which means that there is constructive interference among  their radiation fields. Spectral measurements in this range yield  no information  on the internal charge distribution in the bunch.  To gain such information,  measurements at wavelengths significantly shorter than the bunch length have to be carried out. In that case  the form factor becomes a complex-valued function 
\begin{equation}
{\cal F}(\omega)=F(\omega)\,e^{i \,\Phi(\omega)}
\label{FFlong-amp-phase}
\end{equation}
whose magnitude $F(\omega)=|{\cal F}(\omega)|$ is  generally less than 1. 
If both $F(\omega)$ and $\Phi(\omega)$ were known, a unique reconstruction of the charge distribution $\rho(t)$ could be achieved by the inverse Fourier transformation. 
Unfortunately only the spectral intensity is accessible in  spectroscopic experiments at  accelerators. Hence the modulus $F(\omega)=|{\cal F}(\omega)|$ of the longitudinal form factor can be   determined  while the phase  $\Phi(\omega)$ remains unknown.

Some limited information  is provided by the autocorrelation function which is the inverse Fourier transform of 
$|{\cal F}(\omega)|^2=|F(\omega)|^2$ and thus a measurable quantity:
\begin{equation}
A(t)=\frac{1}{2\pi}\,\int_{-\infty}^{\infty} |{\cal F}(\omega)|^2\exp(-i\,\omega\,t) d\omega\,.
\label{autocorrelation-function}
\end{equation}
The autocorrelation function  provides no information on the internal structure.  

\subsection{Ambiguities in bunch shape reconstruction from spectroscopic data}\label{ambiguity}
The normalized particle density $\rho(t)$ is a real function. Decomposing the form factor and its complex conjugate into real and imaginary part: 
$${\cal F}(\omega)=\int \rho(t) \cos(\omega t)\, dt +i\int \rho(t) \sin(\omega t)\, dt \,,~~~
{\cal F}^*(\omega)=\int \rho(t)  \cos(\omega t)\, dt -i\int \rho(t) \sin(\omega t)\, dt  $$
reveals two important properties.\\
a) There exists a  relation between the form factor and its complex conjugate:
 ${\cal F}(\omega)={\cal F}^*(-\omega)$\,.
 In the next section we will introduce  complex frequencies $\hat{\omega}=\omega_{\rm r}+i\,\omega_{\rm i}$. Then the relation between ${\cal F}$ and ${\cal F}^*$ can be written as
 \begin{equation}
{\cal F}(\hat{\omega})={\cal F}^*(-\hat{\omega}^*) \,.
 \label{FF-FF*}
 \end{equation}
b) The form factor is a real function  if $\rho(t)$ is symmetric with respect to $t=0$.\vspace{3mm}

\noindent The fact that only the magnitude $F(\omega)=|{\cal F}(\omega)|$ of the form factor is measured but not its phase  has undesirable (but unavoidable)  consequences.
\begin{itemize}
\item{Time reversal does not change $F(\omega)$. The time profiles $\rho(t)$ and $\rho(-t)$ yield the same 
spectrum: \\
{\it Head and tail of the bunch cannot be distinguished}. }
\item{A time shift $\rho(t)\rightarrow \rho(t+t_0)$ results in an extra phase factor $e^{i\,\omega t_0}$ but leaves $F(\omega)$ invariant:\\
 {\it The arrival time of the bunch at the TR screen cannot be measured}.}
\item{The magnitude $F(\omega)$ does not uniquely specify the internal bunch structure (see below):\\
{\it There exist many different bunch structures yielding exactly the same spectrum}. }
\end{itemize}
\begin{figure}[htb]
		\centering
		\includegraphics[width=15cm]{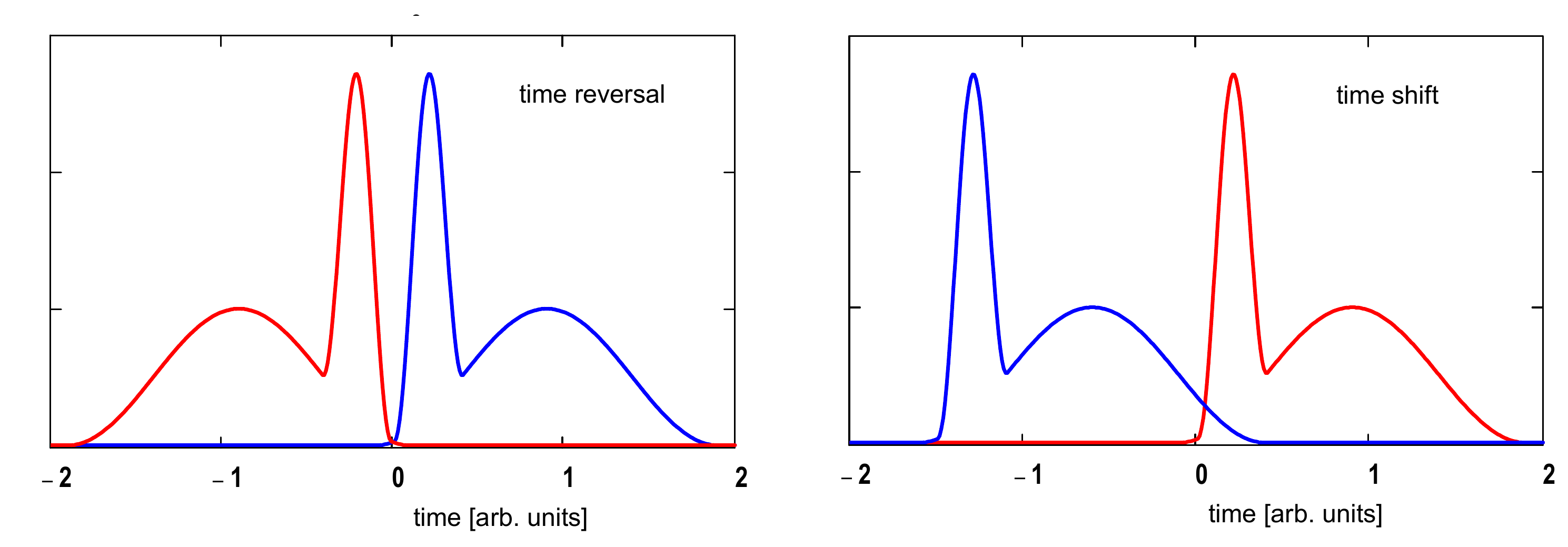} 
		\caption{\small{ Time reversal or time shift of a bunch leave $F(\omega)=|{\cal F}(\omega)|$ invariant. }}		
		\label{time-revers-shift}			
		\end{figure} 	
The determination of the phase  $\Phi(\omega)$ is obviously of great importance.  Two phase retrieval methods will be discussed in the following sections, but one has to be aware of a fundamental limitation: the unique reconstruction of a function from the magnitude of its Fourier transform is mathematically impossible in the one-dimensional case\footnote{In  two or more dimensions the reconstruction is unique in the sense that the set of false reconstructions is  a set of measure zero. For a proof see M.H. Hayes \cite{Hayes}.}.  This was nicely demonstrated by Akutowicz \cite{Akutowicz}. He considered  two functions $f_1(t)$ and $f_2(t)$  which vanish for $t <0$ and which for $t \ge 0$ are given by
\begin{eqnarray}
f_1(t)&=&e^{-\beta t}  \nonumber \\
f_2(t)&=&e^{-\beta t} \left(1+\frac{4 \beta^2 (1-\cos(\alpha t))}{\alpha^2}-\frac{4 \beta \sin(\alpha t)}{\alpha} \right)
\end{eqnarray}
with real parameters $\alpha, \beta >0$.  
\begin{figure}[htb]
		\centering
		\includegraphics[width=7cm]{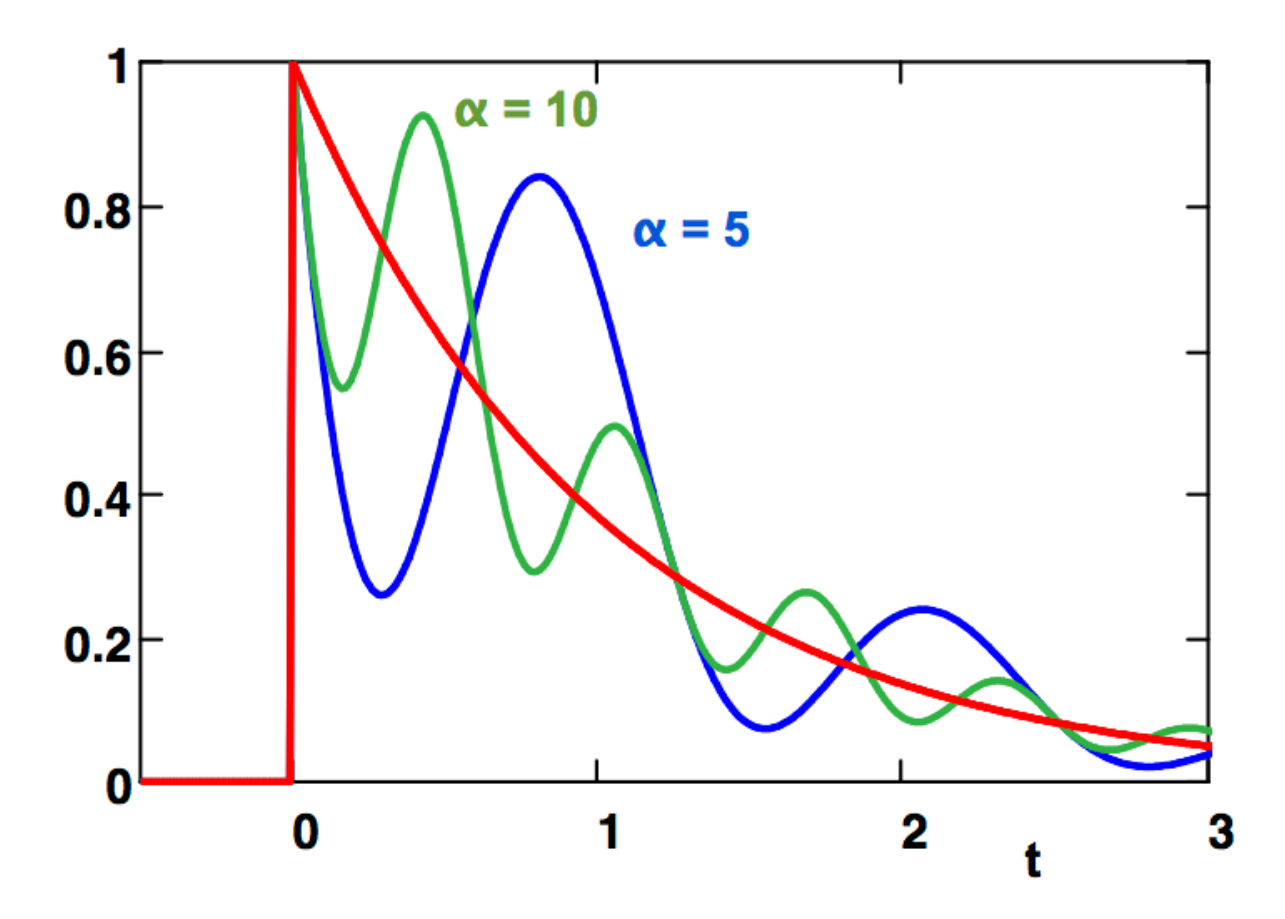} 
		\caption{\small{ The  functions $f_1(t)$ (red) and  $f_2(t)$ for $\beta=1$ and $\alpha=5$ (blue) resp. $\alpha=10$ (green). }}		
		\label{Akuto-functions}			
		\end{figure} 	
The first function has an infinitely  steep rise at $t=0$, followed by an exponential decay.  The second function has the same steep rise but the decay is superimposed with an oscillatory pattern, see Fig.~\ref{Akuto-functions}.  The complex Fourier transforms ${\cal F}_1(\hat{\omega})$ and ${\cal F}_2(\hat{\omega})$ differ considerably
\begin{equation}
{\cal F}_1(\hat{\omega})=\frac{1}{\beta-i \hat{\omega}} \,,~~~~{\cal F}_2(\hat{\omega})=\frac{\alpha^2+(\beta +  i\hat{\omega})^2}{[\alpha^2+(\beta -  i\hat{\omega})^2](\beta -  i\hat{\omega})}
\label{FF-Akuto}
\end{equation}
but  their absolute magnitudes are identical on the real $\omega$ axis:
\begin{equation}
|{\cal F}_1(\omega)|=|{\cal F}_2(\omega)|=\frac{1}{\sqrt{\beta^2+\omega^2}} ~~~\mathrm{for~real~}\omega\,.
\end{equation}
Therefore all three curves in Fig.\,\ref{Akuto-functions} are permitted reconstructions.
This example shows very clearly that non-trivial ambiguities in the bunch shape reconstruction are unavoidable and will occur in any phase retrieval method.

\subsection{Analytic phase retrieval}\label{ana-phase-retrieve}

{\bf Kramers-Kronig phase}\\
 The problem that only the magnitude of a complex-valued quantity of interest is measurable but not its phase  arises also for the optical reflection properties of solids. As shown in \cite{Wooten}  it is possible  to compute  the phase of the complex reflectivity amplitude  from the measured reflectivity by making use of the powerful mathematical theory of analytic (or holomorphic) functions, in  particular  by applying the Kramers-Kronig dispersion relation.
 The method has been adopted for 
the phase reconstruction of the complex bunch form factor,  see  \cite{lai-happek-sievers94, Lai-Sievers} and the references quoted therein. The {\it Kramers-Kronig phase} 
$\Phi_{\rm KK}(\omega)$   can be computed from the  real function  $F(\omega)=|{\cal F}(\omega)|$ by means of the following principal-value integral\footnote{The principal value means that the singularity of the integrand at $\omega'=\omega$ is approached symmetrically from below and above, see Eq.\,(\ref{princ-val-int})
  in Appendix A.}
\begin{equation} \boxed{~
\Phi_{\rm KK}(\omega)=\frac{2 \omega}{\pi} \, \mathcal {P} \int_{0}^{\infty} 
\frac{\ln(| {\cal F}(\omega')|)-\ln( |{\cal F}(\omega)|) } {\omega^2-\omega'^2}\,d\omega' \,. ~}
\label{KK-phase-sect2}
\end{equation}

A rigorous mathematical derivation of formula (\ref{KK-phase-sect2}), based on the theory of analytic functions, especially  the Cauchy Integral Formula and the Residue Theorem,   can be found in Appendix A. Here we indicate only a few steps. 
In analogy to the standard notation of complex numbers, $z=x+i\,y$, one defines a complex frequency by 
$\hat{\omega}=\omega_{\rm r}+i\,\omega_{\rm i}$. The form factor ${\cal F}(\omega)$, which is {\it a priori} a function of the real variable $\omega$,  is continued into the complex $\hat{\omega}$ plane, and ${\cal F}(\hat{\omega})$ can be shown to be an analytic function. The standard Kramers-Kronig dispersion relation between  real and imaginary part (Eq.\,(\ref{disp-relation}) in Appendix A)  is not applicable here since only the magnitude $|{\cal F}(\omega)|$  is known from measurement.
To separate magnitude and phase we compute the logarithm of expression\,(\ref{FFlong-amp-phase}) and insert the complex frequency
$$\ln({\cal F}(\hat{\omega}))=\ln(F(\hat{\omega}))+i\,\Phi(\hat{\omega})\,.$$
Our aim is to find a  relation between $\ln(F(\hat{\omega}))$ and $\Phi(\hat{\omega})$. This  involves an integration around the closed loop depicted in Fig.\,\ref{Weg-log(omega)} (see Appendix A) and the application of the   Residue Theorem. A severe problem, however,   is that the form factor drops to zero at infinite frequency, hence $\ln({\cal F}(\hat{\omega}))$ diverges on the large semicircle $\Gamma_1$ in Fig.\,\ref{Weg-log(omega)}. To circumvent this difficulty,   an auxiliary function, containing $\ln({\cal F}(\hat{\omega}))$ as a factor, is constructed which can be integrated along the semicircle $\Gamma_1$.
After many computational steps one arrives at Eq.\,(\ref{KK-phase-sect2}). 
\vspace{2mm}

\begin{figure}[htb]
		\centering
		\includegraphics[width=14cm]{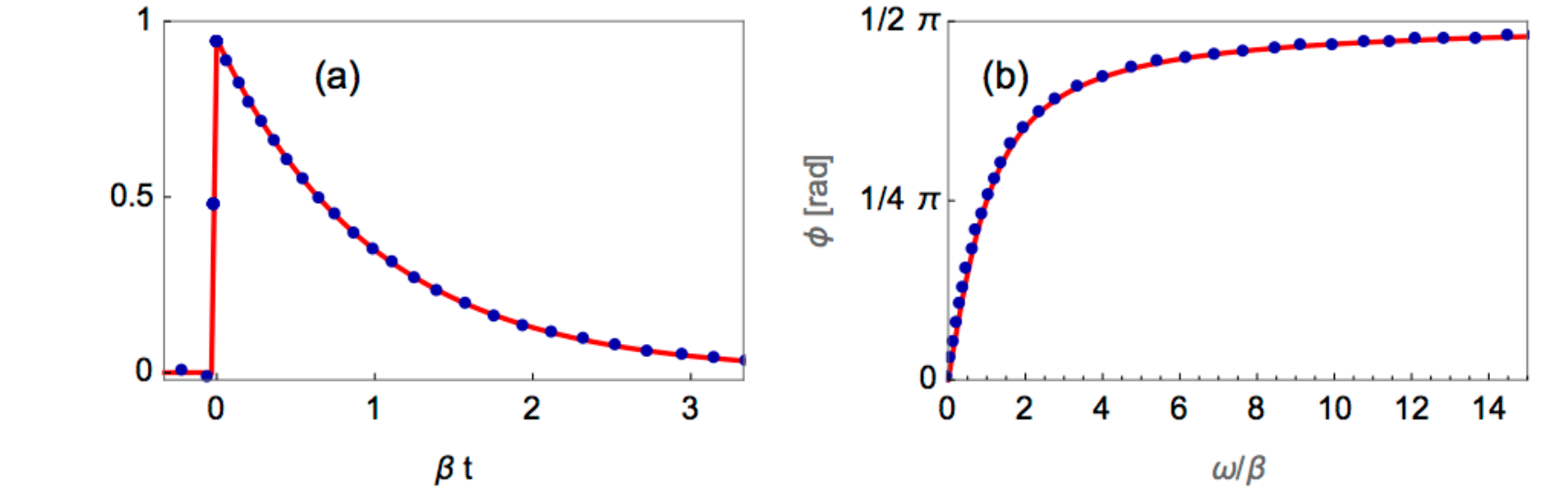} 
		\caption{\small{{\bf (a)} The function $f_1(t)$ (red curve) and its Kramers-Kronig reconstruction (blue dots). {\bf (b)} The phase $\Phi_1(\omega)$ of the  complex form factor ${\cal F}_1(\omega)$ (red) and the Kramers-Kronig phase $\Phi_{\rm KK}( \omega)$ (blue dots). }}		
		\label{f1-KK}
		\end{figure}

\noindent As a first application of formula (\ref{KK-phase-sect2}) we try to reconstruct the function  $f_1(t)$. The KK method reproduces   $f_1(t)$ indeed  accurately, see Fig.\,\ref{f1-KK}, and also the phase is reproduced. However, any attempt to reconstruct the oscillatory function $f_2(t)$ will fail, no matter what the  value of $\alpha$ is.  
 \vspace{3mm}

\noindent {\bf Blaschke phase}\\
An essential prerequisite for Eq.\,\,(\ref{KK-phase-sect2}) to hold  is that  the complex form factor does not have any zeros in the upper half of the complex  $\hat{\omega}$ plane  because otherwise $\ln({\cal F}(\hat{\omega}))$ would have  essential singularities at these points. In such a case the Residue Theorem is not applicable and  formula   (\ref{KK-phase-sect2}) cannot be used to compute $\Phi(\omega)$. 
It was proved by Blaschke \cite{Blaschke} that another phase has to be taken into consideration. Suppose ${\cal F}(\hat{\omega})$ has a zero at the point $\hat{\omega}_1=a_1+i\,b_1$ in the right upper quarter of the complex plane, i.e.  $\Re(\hat{\omega}_1)=a_1>0$ and $\Im(\hat{\omega}_1)=b_1>0$. 
Formula (\ref{FF-FF*}) shows that there is  another zero  in the left upper quarter at $\hat{\omega}_1'=-a_1+ib_1$. This pair of  zeros can be removed by modifying ${\cal F}(\hat{\omega})$
\begin{equation}
{\cal F}_{\rm mod}(\hat{\omega})={\cal F}(\hat{\omega}) {\cal B}(\hat{\omega})~~~\mathrm{with}~~
{\cal B}(\hat{\omega})= \frac{\hat{\omega}-(a_1-ib_1)}{\hat{\omega}-(a_1+ib_1)}\cdot \frac{\hat{\omega}-(-a_1-ib_1)}{\hat{\omega}-(-a_1+ib_1)} 
 \,.
 \label{Blaschke-factor}
\end{equation} 
Here ${\cal B}(\hat{\omega})$ is the so-called  Blaschke 
factor.  \\On the real $\omega$ axis
 the absolute magnitude of the Blaschke factor  is 1, hence
 \begin{equation}
 |{\cal F}_{\rm mod}(\omega)|=|{\cal F}(\omega)|~~~
\mathrm{for\, real}~~~\omega \,.
 \end{equation}
This is a very important equation. It means that the form factor ${\cal F}(\omega)$ and  the modified form factor ${\cal F}_{\rm mod}(\omega)$ describe  exactly the same  radiation spectrum. The phase of  ${\cal B}(\omega)$ is computed by the equation
\begin{equation}\label{Blaschke-phase}
\Phi_{\rm B}(\omega)=\arg({\cal B}(\omega))\,.
\end{equation}

 This procedure is repeated for every zero of ${\cal F}(\hat{\omega})$ until ${\cal F}_{\rm mod}(\hat{\omega})$ is free from any zeros. 
Then Eq.\,\,(\ref{KK-phase-sect2}) is valid for the modified form factor,  so 
$\Phi_{\rm mod}(\omega)=\Phi_{\rm KK}(\omega)$.
The reconstruction phase is given by the difference between 
 KK phase and Blaschke phase
\begin{equation}
\Phi_{\rm rec}(\omega)=\Phi_{\rm KK}(\omega)-\Phi_{\rm B}(\omega)\,.
\end{equation}

 \vspace{2mm}
		
\noindent Now we demonstrate that the Blaschke phase in combination with the KK phase enables a  faithful reconstruction	of $f_2(t)$. 		
\begin{figure}[htb!]
		\centering
		\includegraphics[width=15cm]{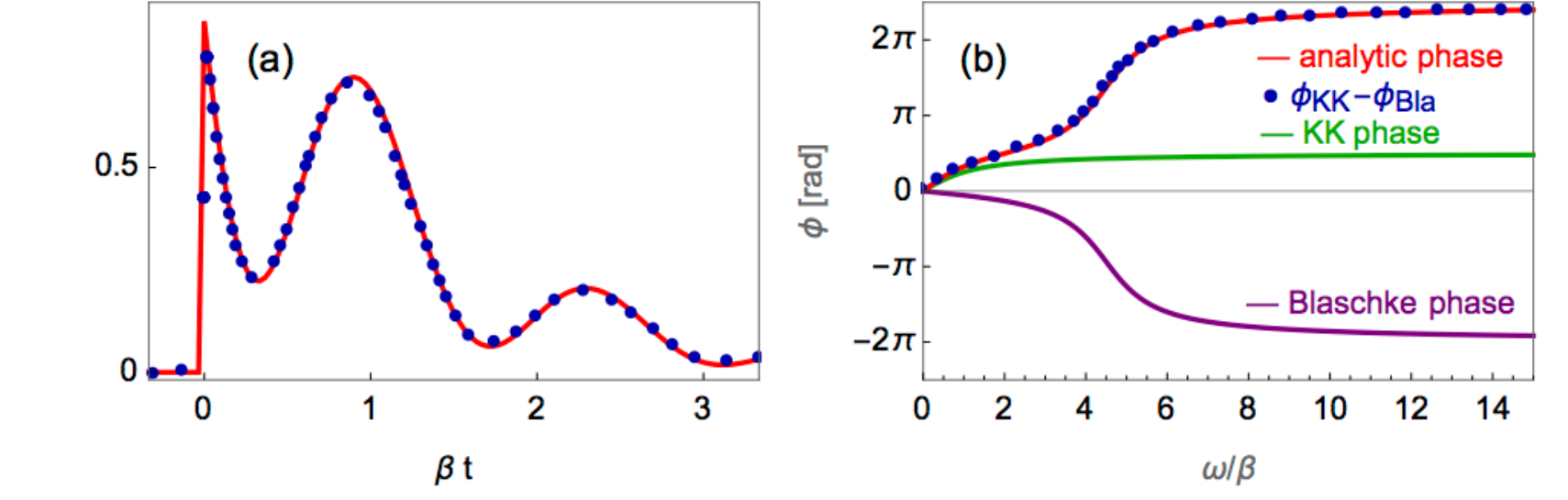} 
		\caption{\small{{\bf (a)} The function $f_2(t)$ (red curve) and its reconstruction (blue dots) using the  reconstruction phase $\Phi_{\rm rec}(\omega)=\Phi_{\rm KK}(\omega)-\Phi_{\rm B}(\omega)$. {\bf (b)} The analytic phase $\Phi_2(\omega)$ of the  complex form factor ${\cal F}_2(\omega)$ (red), the KK phase $\Phi_{\rm KK}(\omega)$ (green), the Blaschke phase $\Phi_{\rm B}(\omega)$ (purple) and the reconstruction phase $\Phi_{\rm rec}(\omega)=\Phi_{\rm KK}(\omega)-\Phi_{\rm B}(\omega)$ (blue dots).  }
		 }			
		\label{Akutow-f2-KKB}
		\end{figure}  
To this end we have to find the complex zeros  of the form factor 	${\cal F}_2(\hat{\omega})$, see Eq.\,(\ref{FF-Akuto}). This is easy: the numerator of ${\cal F}_2(\hat{\omega})$ must vanish if we insert the complex frequency $\hat{\omega}_1=a+ib\,$:
	$$
0=\alpha^2+(\beta +  i\hat{\omega}_1)^2=\alpha^2+(\beta +  i a-b)^2=
	\alpha^2+\beta^2-a^2+b^2-2b\beta +i\,2a(\beta-b) \,.$$
Both real and imaginary part of this equation must be zero. From this condition follows immediately 
$$b=\beta\,,~~~~a=\pm \alpha\,.$$
  The form factor  ${\cal F}_2(\hat{\omega})$ has just one pair of zeros in the upper half plane:  $\hat{\omega}_1=\alpha+i\beta$ in the right upper quarter of the complex $\hat{\omega}$ plane and its mirror image $\hat{\omega}'_1=-\alpha+i\beta$ in the left upper quarter. We consider the function $f_2(t)$ with the parameters $\alpha=5$ and $\beta=1$.
Using the Eqs.\,(\ref{Blaschke-factor}) and  (\ref{Blaschke-phase}) we 
compute the  Blaschke phase $\Phi_{\rm B}(\omega)$. The  reconstruction phase  
$
 \Phi_{\rm rec}(\omega)=\Phi_{\rm KK}(\omega)-\Phi_{\rm B}(\omega) 
$
is found to be identical with the analytic phase $\Phi_2(\omega)$ of the form factor ${\cal F}_2(\omega)$, see Fig.\,\ref{Akutow-f2-KKB}b.
When we use this reconstruction phase to compute the function $f_2(t)$ from the magnitude $|{\cal F}_2(\omega)|$ of the form factor we find perfect agreement with the original function 
 (Fig.\,\ref{Akutow-f2-KKB}a). This result is  a remarkable success of the dispersion relation theory. 
 \vspace{10mm}

\noindent {\bf Examples of analytic bunch shape reconstruction}\\
 Extensive  model calculations for bunch shape reconstruction  will be presented in Appendix B. Here we show only a few examples. \vspace{2mm}
 
\begin{figure}[htb]
		\centering
		\includegraphics[width=17cm]{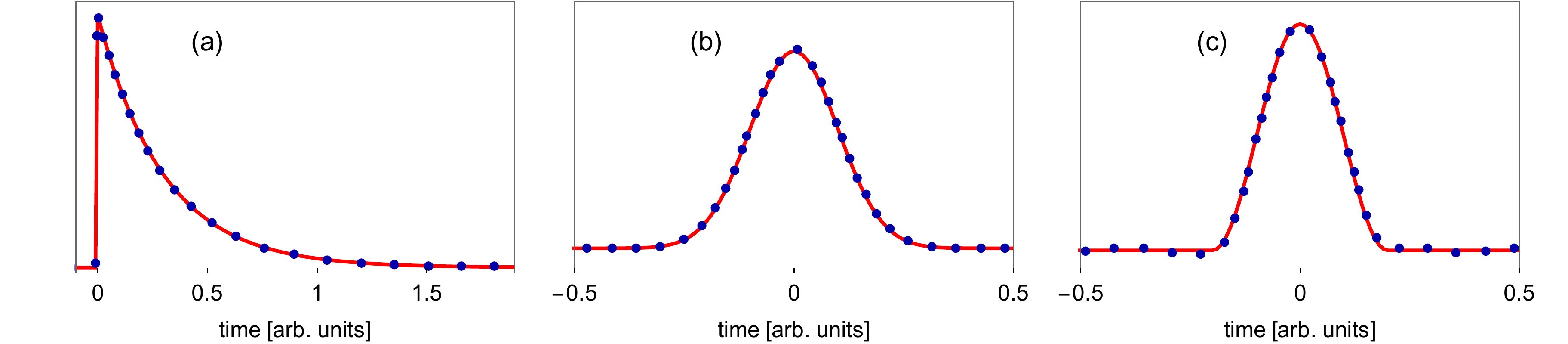} 	
\caption{\small{Reconstruction of bunch shape using the Kramers-Kronig phase for (a) the function $f_1(t)$ with an infinitely steep rise and an exponential decay, (b) a single truncated Gaussian, (c) one period of a cosine-squared wave.   Red curves: input $\rho(t)$, blue dots: reconstructed $\rho(t)$. }}
\label{Gauss-Cosine}
	\end{figure}

\noindent (1)  If the time profile $\rho(t)$ features a single peak,   such as the  function $f_1(t)$, a  Gaussian or one period of a cosine-squared wave,  the Kramers-Kronig (KK) phase permits a perfect reconstruction as demonstrated in Fig.\,\ref{Gauss-Cosine}. A  cosine-squared pulse of width $2b$, which is centered at $t_c$, is described by
\begin{equation}
\rho(t)=\frac{1}{b }\cos^2\left(\frac{\pi\,(t-t_c)}{2b}\right)~~\mathrm{for }~~~(t_c-b) \le t\le (t_c+b) \,,
~~~~~\rho(t)=0~~\mathrm{otherwise } \,.
\label{rho-trunc-cos-squared}
\end{equation}
The Fourier transform can be computed analytically:
\begin{equation}
{\cal F}(\omega) =\frac{\pi^2 \sin(\omega b)\exp( i \,\omega t_c)}{\omega b (\pi^2-\omega^2 b^2)} \,.
 \label{FF-cosine}
\end{equation}
 It is important to note  that subtle mathematical problems arise with bunches of truly Gaussian shape.
A Gaussian function violates causality because it extends over the full time range $-\infty <t < + \infty$. The unfortunate consequence is that the Gaussian form factor does not fulfill all requirements that are needed in the derivation of the Kramers-Kronig phase formula. A detailed study  will be presented  in Appendix A and Appendix C. 

\begin{figure}[ht]
		\centering
		\includegraphics[width=17cm]{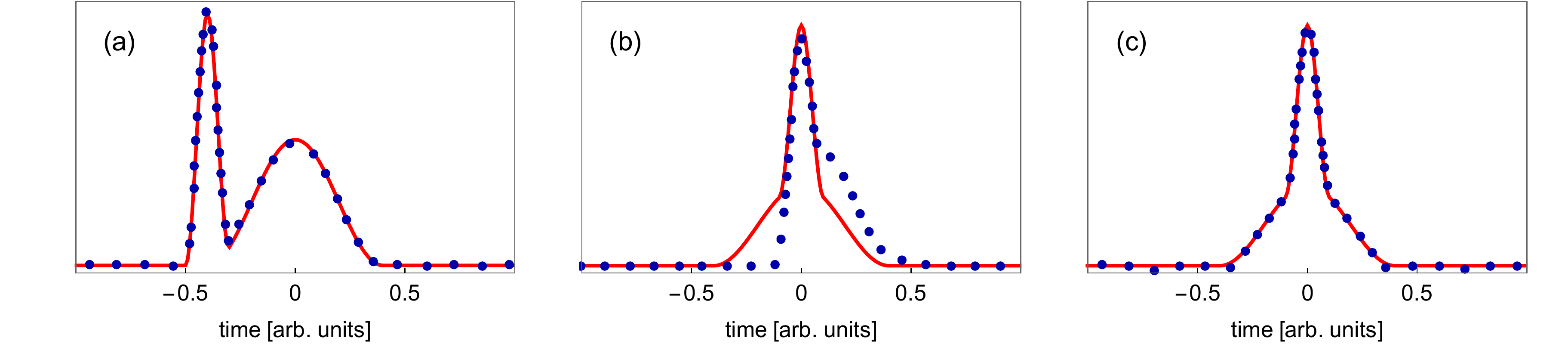} 	
\caption{\small{ Reconstruction of a bunch consisting of two cosine-squared pulses of different width. \newline {\bf (a)} Narrow peak  at the front: the Kramers-Kronig (KK) phase yields a perfect reproduction of the original bunch shape. {\bf (b)} Narrow peak  at the center: the KK phase yields a bad reproduction. {\bf (c)} Narrow peak  at the center: the KK phase combined with 
the Blaschke phase yields an excellent  reproduction of the original bunch shape. The mathematical details are presented in Appendix B.}}
\label{2cos-diffwidth}
	\end{figure} 
\vspace{2mm}
	 
\noindent (2) For bunch profiles  featuring several peaks the situation is confusing at first sight. In some cases the input charge distribution is faithfully reconstructed using the KK phase, in other cases significant differences are found.  We consider a bunch consisting of two  cosine-squared pulses  of different width.  The Kramers-Kronig method yields a precise reproduction of the original bunch shape when the narrow peak is at the front, but it completely fails when the narrow peak is centered with respect to the wide one (see Fig.\,\ref{2cos-diffwidth}). An excellent  reproduction of the original bunch shape is achieved if both  KK phase and  Blaschke phase are taken into account.\vspace{2mm}

\begin{figure}[htb!]
		\centering
		\includegraphics[width=18cm]{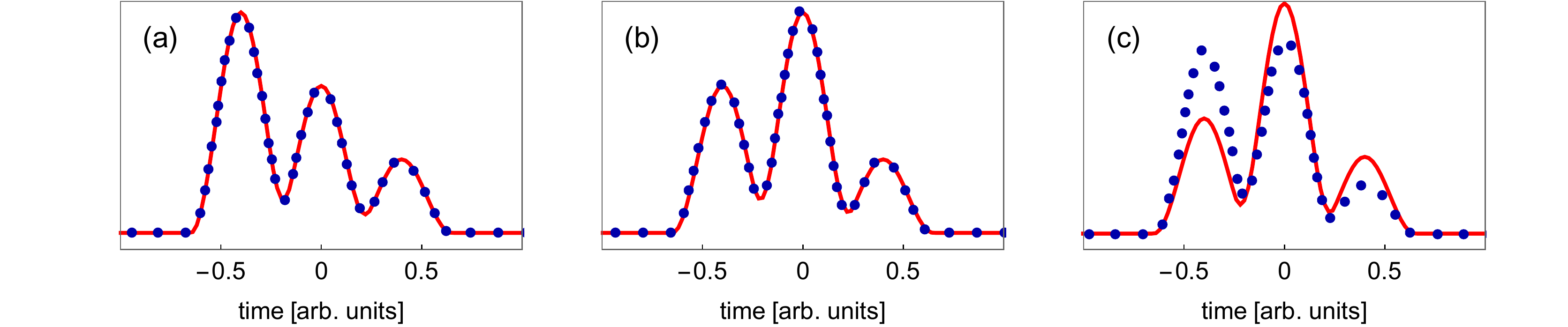} 	
\caption{\small{Three  cosine-squared pulses  of equal width and with uniform spacing. The amplitude ratios are $A_2/A_1=2/3, \, A_3/A_1=1/3$. {\bf (a)} The largest peak is at the front.	{\bf (b)} The largest peak is in the center.   The KK phase yields a perfect reconstruction in both cases. {\bf (c)}  For amplitude ratios of  $A_2/A_1=1/2, \, A_3/A_1=1/3$, the KK reconstruction  is perfect  when the highest peak is at the front but fails if it is in the center.}}
\label{3Cos2-KK}
	\end{figure}

\noindent (3) Our third  example is a bunch consisting of three  cosine-squared pulses of equal width and with uniform spacing. The amplitude ratios are $A_2/A_1=2/3, \, A_3/A_1=1/3$. The KK reconstruction agrees perfectly with the input distribution,   the highest peak may be at the front or in the center of the bunch, see 
Fig.\,\ref{3Cos2-KK}. But  a slight change of the parameters may lead  to  different  results. For example, when the amplitude ratios are  $A_2/A_1=0.5, \, A_3/A_1=0.3$, the KK reconstruction works if the highest peak is at the front but it fails if  it is in the center of the bunch. 

\begin{figure}[htb!]
		\centering
		\includegraphics[width=12cm]{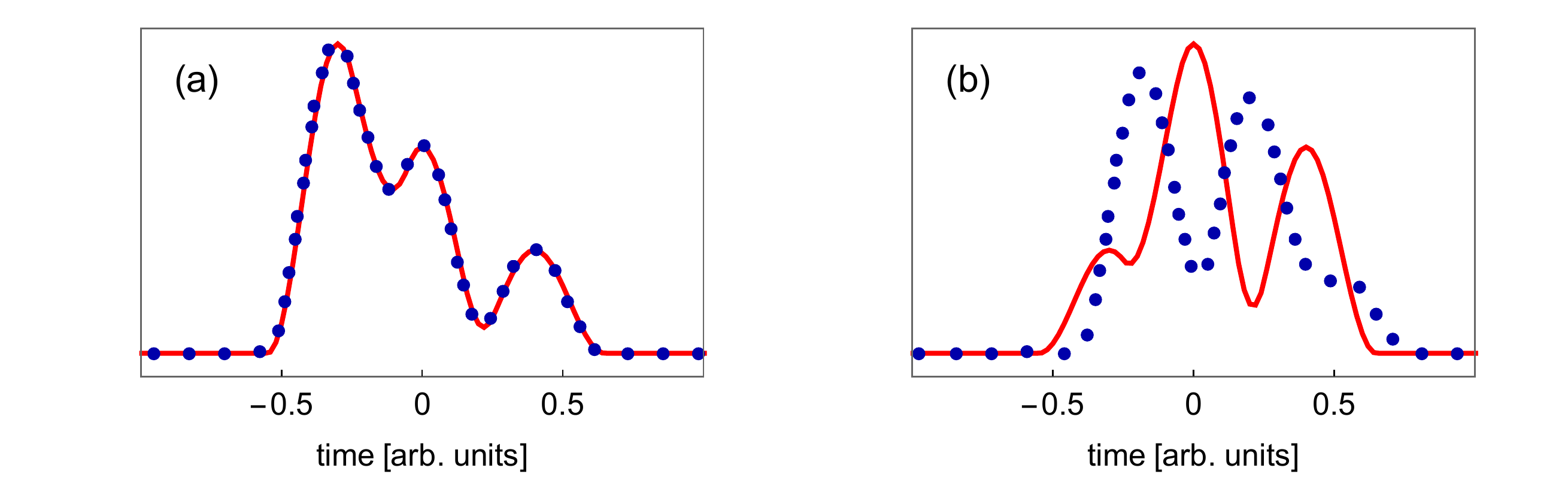} 	
\caption{\small{Three  cosine-squared pulses  of equal width and with non-uniform spacing.  The ratio of the distances is $d_1/d_2=4/3$. {\bf (a)} The largest peak is at the front.	The KK phase yields a perfect reconstruction. \newline {\bf (b)} The largest peak is in the center.   The KK reconstruction disagrees with the input distribution. A faithful reconstruction is achieved by taking the Blaschke phase into account, see Appendix B.}}
\label{3Cos2-KK-nonequi}
	\end{figure}

Another case are three  cosine-squared pulses of equal width but with   non-uniform spacing. Again the KK reconstruction agrees perfectly with the input distribution if the highest peak is at the head of the bunch but fails if it is in the center, see Fig.\,\ref{3Cos2-KK-nonequi}.
	\vspace{2mm}

\begin{figure}[htb!]
		\centering
		\includegraphics[width=13cm]{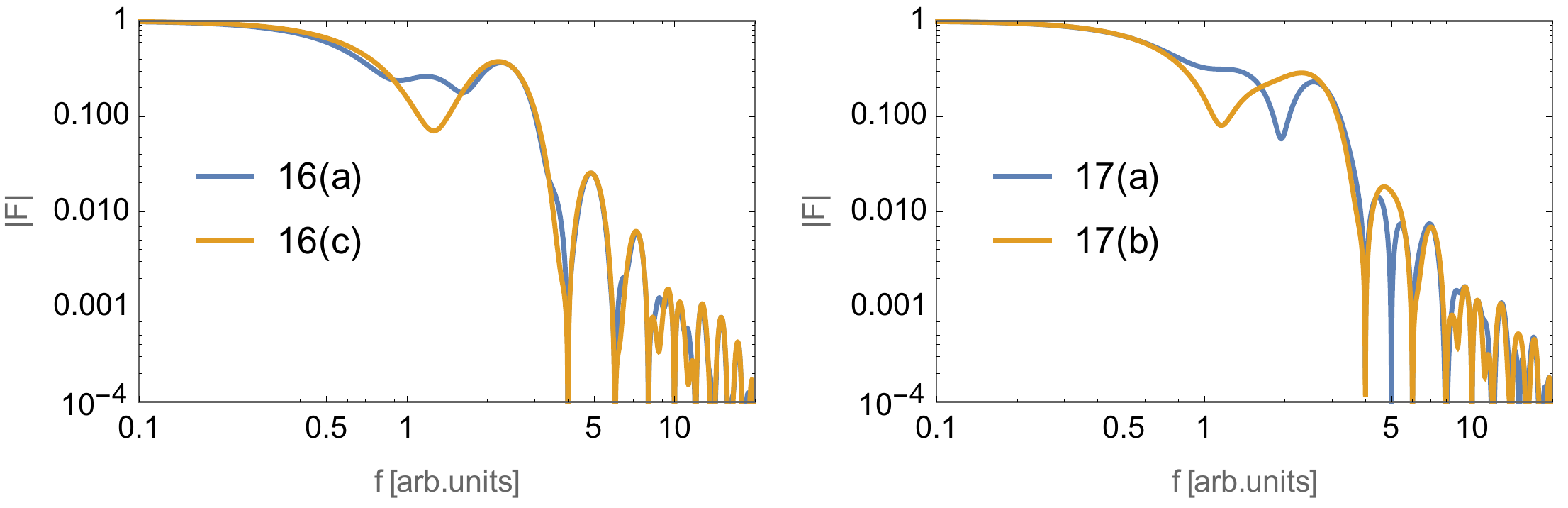} 	
\caption{\small{The computed form factor magnitudes $|{\cal F}(\omega)|$ of a superposition of three  cosine-squared pulses  of equal width, plotted versus $f=\omega/2\pi$. Left: uniform spacing, right: non-uniform spacing.  The blue curves refer to the cases where the KK reconstruction works while the yellow curves curves refer to the cases where the KK reconstruction fails. }}
\label{FF-models}
	\end{figure}

\noindent It is instructive to look at the form factors 	of the bunches composed of three  cosine-squared pulses. These are shown in  Fig.\,\ref{FF-models}. 
The blue curves (the KK reconstruction works) and the yellow curves (the KK reconstruction fails) are very similar, and there is no hint at all, why the KK reconstruction should work in one case but fail in the other.  \vspace{1mm}

\noindent The model profiles presented in this section demonstrate that a correct  reconstruction of time profile with the help of the Kramers-Kronig dispersion relation cannot be guaranteed. The KK phase reconstruction fails whenever the form factor has zeros in the upper half of the complex frequency plane. 
This is a specific  illustration of the more general theorem that a unique bunch shape reconstruction from the magnitude of the form factor is mathematically impossible.

\vspace{6mm}

\noindent {\bf The Blaschke correction does not work for real data}\\
The Blaschke phase is a known quantity in our model calculations where we choose a mathematically well-defined input distribution $\rho(t)$ and compute the complex form factor by Fourier transformation.  Whenever this form factor has one or more zeros in the upper half of the complex plane, the KK phase alone  is insufficient but the combination of  Kramers-Kronig phase and Blaschke phase enables a faithful reconstruction.\\
In spectroscopic experiments at accelerators, however,  the situation is much less favorable. There  exists simply  no information on such zeros of the form factor. The unfortunate consequence is that even the most precise determination of 
	$|{\cal F}(\omega)|=F(\omega)$ does not allow a unique bunch shape reconstruction, there will always be ambiguities. The only  phase which can be derived by analytical methods from the measured modulus of the form factor is the Kramers-Kronig phase, however the KK phase  leads to wrong reconstructions if an unknown Blaschke phase should be  present.\vspace{2mm}
	
	\noindent {\bf Criticism of the dispersion relation method}\\
Computing the phase of the form factor via the Kramers Kronig dispersion relation is only justified if the form factor is an analytic function of the 	
complex frequency. This requirement is fulfilled  in the model calculations presented in this section and in Appendix A, B and C, however it may not be the case for an experimentally determined form factor which is measured at a finite number of discrete frequencies $\omega_j$ and in a limited range. Extrapolations towards very small and very large frequencies are needed, and the data have errors. One has to make the implicit assumption that an analytic function exists whose magnitude agrees (within errors) with the measured values $F(\omega_j)$, and for this function the dispersion-theoretical approach can be applied. Obviously it is desirable to have an alternative phase retrieval method at hand which does not rely on such deep lying mathematical prerequisites. The iterative phase retrieval method offers this alternative.

\subsection{Iterative phase retrieval} \label{Iter-Kap3}
\subsubsection{Gerchberg-Saxton algorithm}
Iterative algorithms for phase retrieval from intensity data  are used in many research areas such as electron microscopy, X ray diffraction and astronomy. These are usually two-dimensional problems. An  overview can be found in \cite{Fienup}. 
Iterative  phase retrieval in the one-dimensional case of  longitudinal electron bunch reconstruction from spectroscopic data has been applied recently \cite{Bajlekov, Pelliccia}, 
and it has been claimed that thereby the restrictions of the KK method can be overcome,  the main argument being that the KK phase is computed by an  integral over all frequencies and thus depends on extrapolations into regimes where $F(\omega)$  has not been measured. This argument is misleading as it misses the main point, namely that  a unique reconstruction of a function from the magnitude of its Fourier transform is mathematically impossible in the one-dimensional case, see Fig.\,\ref{Akuto-functions}	and	the discussion   in Appendix A. It is obvious that {\it any} phase retrieval method will suffer from  this fundamental limitation. Our motivation to study the iterative method in parallel to the KK method is more of a practical nature: by comparison with time domain measurements  we want to explore which of the two methods yields the most likely bunch shape. \vspace{2mm}

\begin{figure}[htb]
		\centering
		\includegraphics[width=11cm]{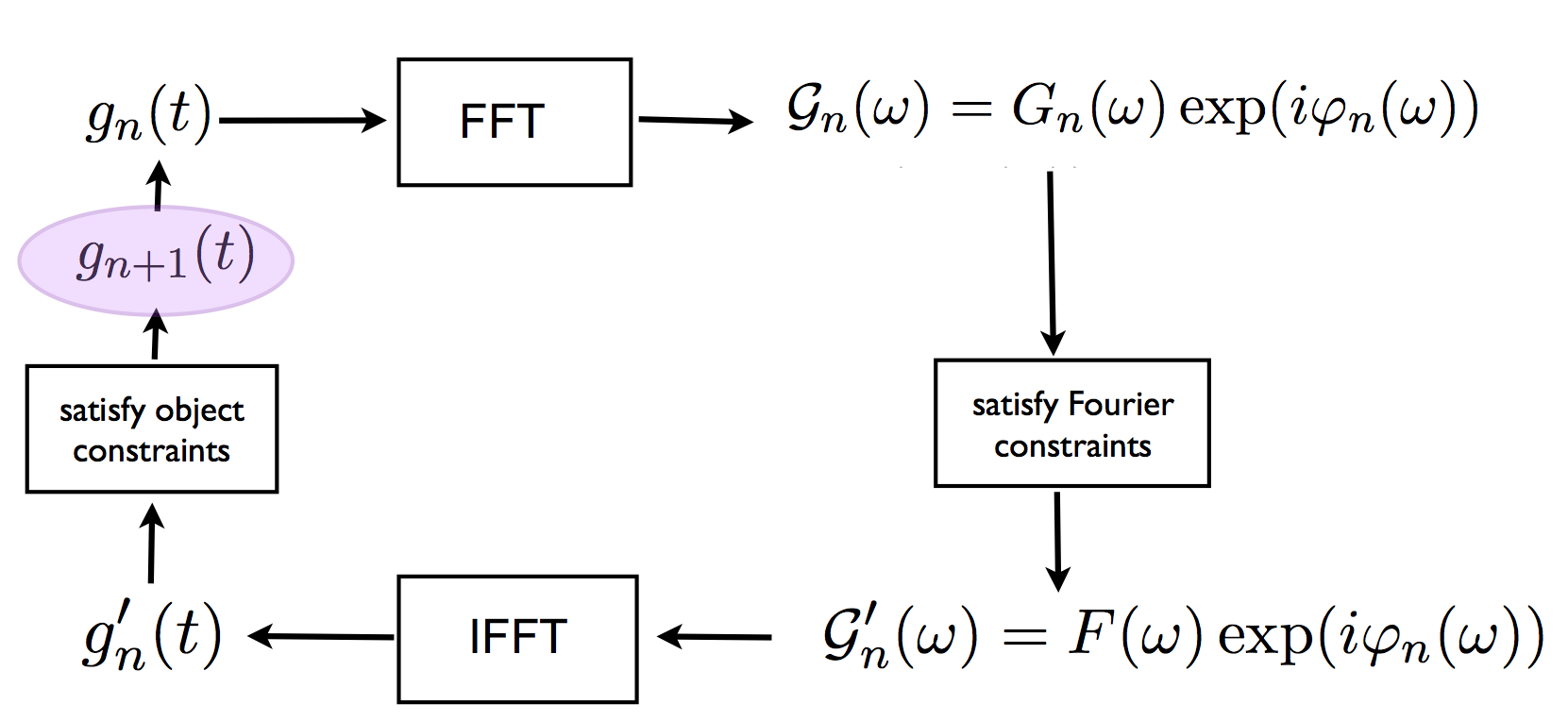} 	
\caption{\small{Block diagram  of the Gerchberg-Saxton algorithm. 
FFT stands for Fast Fourier Transformation, IFFT for Inverse Fast Fourier Transformation.}}
\label{Gerchberg-Saxton-algo}
	\end{figure}

\noindent For the iterative phase retrieval we use the Gerchberg-Saxton algorithm \cite{Gerchberg-Saxton}. A block diagram is shown in Fig.\,\ref{Gerchberg-Saxton-algo}. 
  The basic idea   is as follows. Let ${\cal F}(\omega)=F(\omega)\,\exp(i \,\Phi(\omega))$ be the form factor (Fourier transform) of the unknown longitudinal particle density distribution  $\rho(t)$. 
The magnitude of the form factor $F(\omega)=|{\cal F}(\omega)|$ is known from measurement   but  the phase $\Phi(\omega)$ is unknown. The iterative loop  may be started by making a first guess $g_1(t)$ of the particle density distribution and Fourier transforming it to obtain  a first estimate 
${\cal G}_1(\omega)=G_1(\omega)\,\exp(i \,\varphi_1(\omega))$ of the complex form factor. 
Then the computed  modulus $G_1(\omega)$ is replaced  with the measured modulus $F(\omega)$ but the computed phase $\varphi_1(\omega)$ is retained.
Next  an inverse Fourier transformation  is carried out leading to a modified   time profile $g'_1(t)$. This profile is subjected to 
 several constraints, the most important one being:  the particle density  is not allowed to assume negative values.
These constraints lead to a modified time profile $g_2(t)$ which is then used as starting distribution in the next iteration. 
Usually many iterations are needed until convergence is achieved, meaning that all constraints are fulfilled. Then one has arrived at a solution of the bunch shape reconstruction problem, but as stated above, this solution is not unique.\vspace{1mm}

 \noindent Alternatively the loop may be started with a guess of the complex form factor, taking the magnitude $F(\omega)$ from measurement and choosing the initial phase function either to be a constant, the Kramers-Kronig phase or a randomly varying function. In the following examples we choose random start phases.
 
 \subsubsection{Examples of iterative bunch shape reconstruction}
\begin{figure}[htb]
		\centering
		\includegraphics[width=17cm]{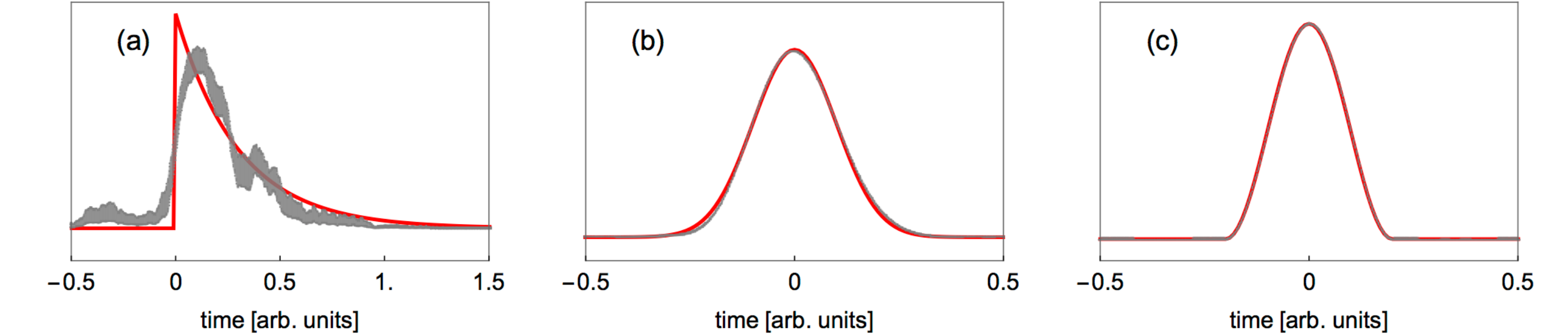} 	
\caption{\small{Iterative reconstruction of bunch shape  for (a) the function $f_1(t)$ with an infinitely steep rise and an exponential decay, (b) a single truncated Gaussian, (c) one period of a cosine-squared wave.   Red curves: input $\rho(t)$, shaded gray area: reconstructed $\rho(t)$. In the cases (b) and (c) the reconstruction is so good that it overlaps  the original curve.}}
\label{Aku-Gauss-Cos-iter}
	\end{figure}

\begin{figure}[htb]
		\centering
		\includegraphics[width=17cm]{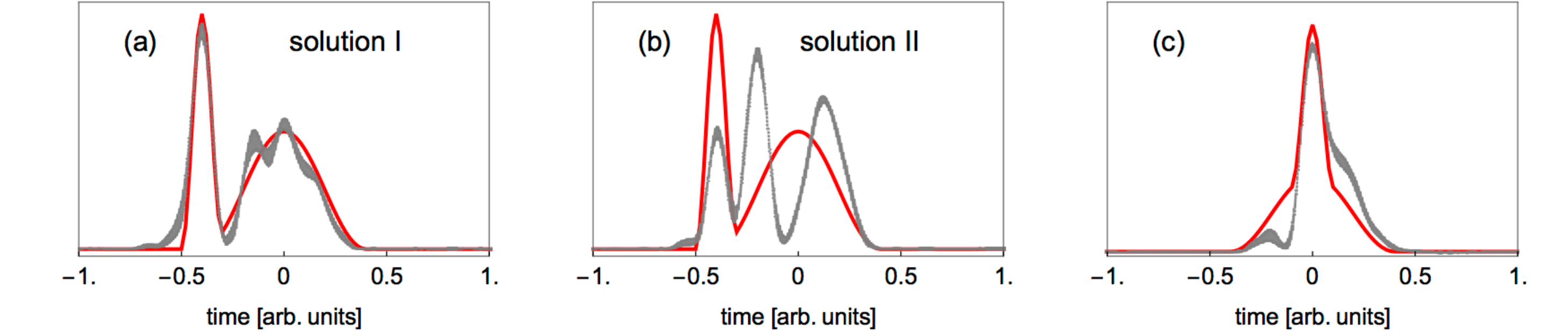} 	
\caption{\small{Iterative reconstruction of a bunch consisting of two superimposed cosine-squared  peaks of different width.
{\bf (a,b)} If the narrow peak is at the front there are two types of  solutions. Solution I is a time profile resembling the input profile,  solution II is a profile featuring three peaks. {\bf (c)}  If the narrow peak is in the center, the reconstruction yields always the same profile  which however is different from the original.  }}
\label{Two-Cos-IT}
	\end{figure}

\begin{figure}[htb]
		\centering
		\includegraphics[width=17cm]{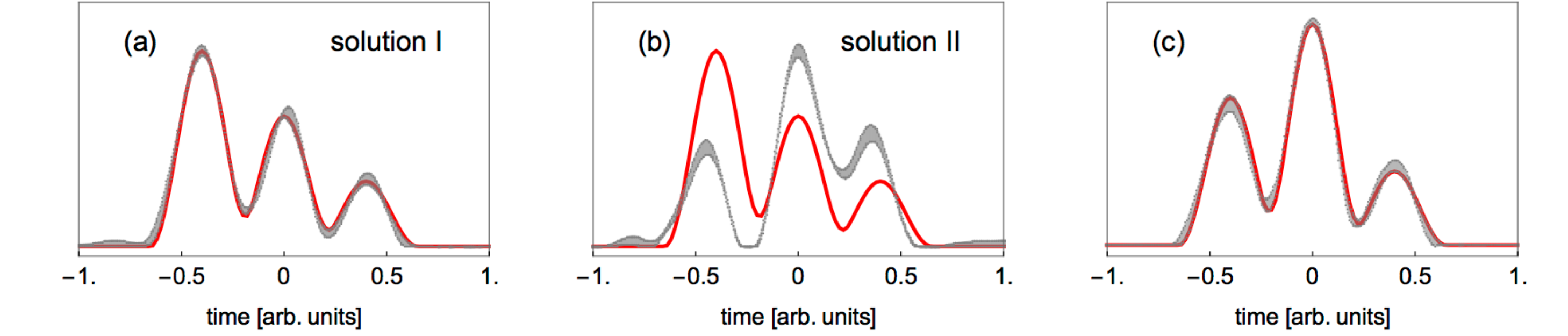} 	
\caption{\small{Iterative reconstruction of a bunch consisting of three cosine-squared pulses of equal width and with uniform spacing. The amplitude ratios are $A_2/A_1=2/3, \, A_3/A_1=1/3$. {\bf (a,\,b)}  If the largest peak is at the front there are two types of  solutions. Solution I: In about 2/3 of the 100 iteration cycles a faithful reconstruction is achieved. Solution II: In about 1/3 of the 100 iteration cycles  the Gerchberg-Saxton algorithm  converges to a different profile. {\bf (c)} If the largest peak is in the center there is a unique  solution: In all iteration cycles a faithful reconstruction is achieved.
 }}
\label{3Cos-eq-IT}
	\end{figure}

\begin{figure}[htb]
		\centering
		\includegraphics[width=17cm]{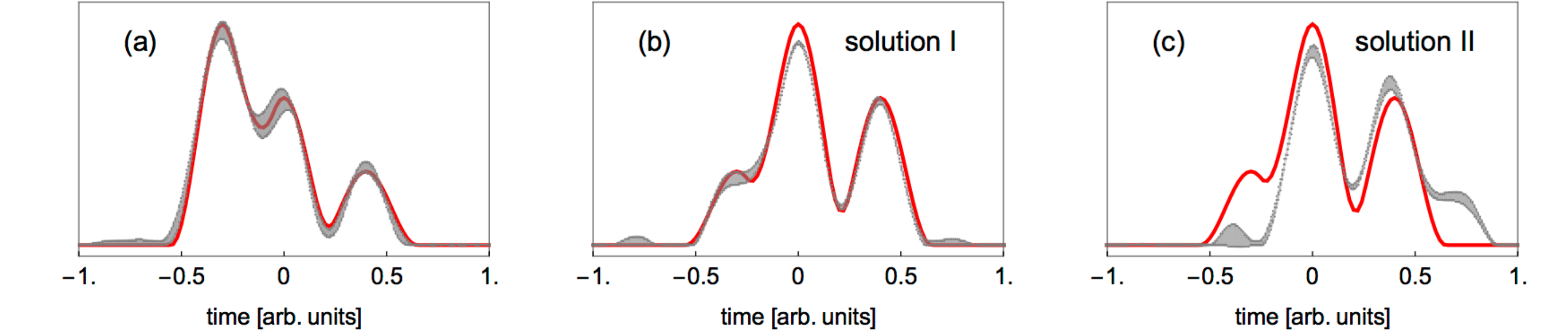} 	
\caption{\small{Iterative reconstruction of a bunch consisting of three   cosine-squared pulses  of equal width and with non-uniform spacing.  {\bf (a)} The largest peak is at the front.	The iterative method yields a perfect reconstruction. {\bf (b,\,c)} The largest peak is in the center. In 60 out of 100 iteration cycles a faithful reconstruction is achieved (solution I), in 40  out of 100 cycles 	the reconstruction is wrong (solution II).  }}
\label{3Cos-noneq-IT}
	\end{figure}

\noindent {\bf General remarks on iterative phase retrieval with random initial phases}\\
Without  constraints on the time-domain profile, the form factor modulus can be combined with an arbitrary set of phases, yielding  infinitely many different temporal profiles. In general, these profiles will not fulfill the mandatory constraint that the longitudinal particle density has to be non-negative for all times. Using this constraint  in the Gerchberg-Saxton-Loop is sufficient to reduce the  number of possible solutions considerably.
To investigate the uncertainties in the resulting time profile, which are caused by the randomness of the start phases, we follow a procedure proposed in \cite{Pelliccia}. The iterative loop is started 100 times, each time  with a new set of random phases, and the resulting profiles are averaged. 
 This averaging has to be done with care since the reconstructed  time profiles $\rho_j(t)$ ($j=1 \ldots 100$) will have arbitrary time shifts with respect to each other and sign-reversals of the time direction will happen (see Fig.\,\ref{time-revers-shift}). These ambiguities must be removed before averaging. To this end one optimizes the correlation coefficient between any two $\rho_i(t)$, $\rho_j(t+\delta t)$  by varying the time offset $\delta t$ and by trying if time reversal $\rho_j(t) \rightarrow \rho_j(-t)$ improves the agreement. The 2\,$\sigma$ band of the properly adjusted profiles is shown as a grey band in  
 Figs.\,\ref{Aku-Gauss-Cos-iter}, \ref{Two-Cos-IT}, \ref{3Cos-eq-IT} and \ref{3Cos-noneq-IT}.
 
 \vspace{2mm}
\begin{figure}[ht]
\centering
\includegraphics[width=17cm]{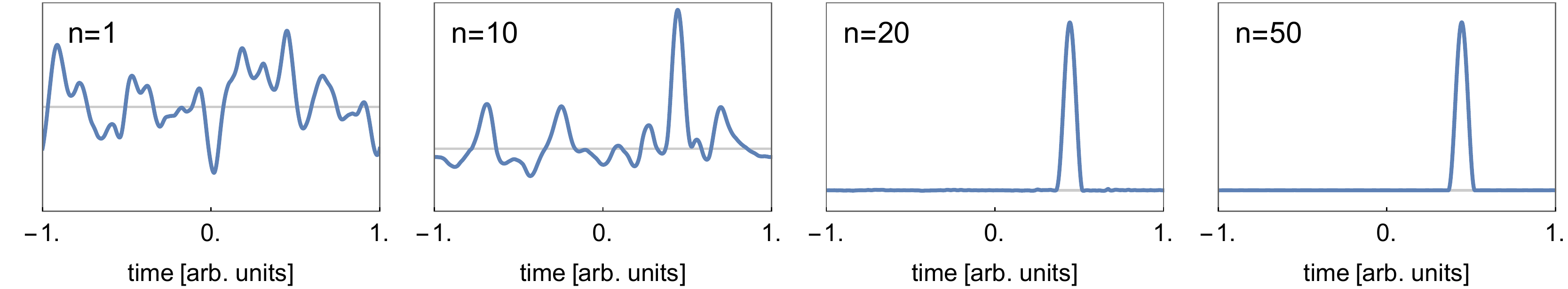} 	
\caption{\small{Iterative reconstruction of a cosine-squared charge distribution. Shown are the iteration steps $n=1$, $n=10$, $n=20$ and $n=50$.}}
\label{Gauss-iter1-50}
	\end{figure}
\noindent A nice demonstration of the progressing improvement in  iterative bunch shape reconstruction is presented in  Fig.\,\ref{Gauss-iter1-50}. The original bunch shape is  a  cosine-squared  pulse.  Starting with  random phases the time signal $g_n(t)$ is initially very spiky and has large undershoots. The negative values are quickly eliminated by the time-domain constraints, and  the input distribution $\rho(t)$ is well reproduced after about 20 iterations.
The speed of convergence is found to depend on 	the initial conditions and on the complexity of the profile, typically some 100  iterations are needed. 
	
\vspace{2mm}

\noindent  It turns out that three classes of solutions can be observed. \vspace{1mm}

\noindent Class A: A unique solution  with basically no variation of the resulting time  profile. \\
Class B: One distinct shape of the profile but with a more or less pronounced uncertainty band.\\
 Class C: Several distinct profiles with their respective  uncertainty bands.\vspace{2mm}

\noindent We study now the same bunch shapes as in the previous section. \\

\noindent (1) A single Gaussian or a truncated cosine-squared profile lead to class A solutions. Here  the iterative phase reconstruction yields a unique result in very good agreement with the actual profile, as shown in Fig.\,\ref{Aku-Gauss-Cos-iter}b and Fig.\,\ref{Aku-Gauss-Cos-iter}c. However, quite a different result is obtained for the step-exponential function $f_1(t)$ which yields a class B solution. The steep initial rise is badly reproduced and a series of randomly fluctuating time profiles is observed (Fig.\,\ref{Aku-Gauss-Cos-iter}a). The average profile is superimposed with  artificial structures which appear even in front of the step. The analytic KK  reconstruction is far superior in this case. \vspace{1mm}

\noindent (2) Now we consider bunches with two superimposed cosine-squared  peaks of different width. When the narrow peak is at the front,  solutions of class C are obtained, as demonstrated by figures \ref{Two-Cos-IT}a and  \ref{Two-Cos-IT}b. Two distinct sets of phases with very narrow variability are found with equal probability. One of them corresponds to a time profile resembling the input profile, but with  artificial wiggles,  while the second set leads to a completely different profile featuring three peaks. Notice that both solutions have the same form factor modulus and fulfill the constraint of positive charge density. 
Again, in this case the analytic KK reconstruction is far superior since it reproduces exactly the original time profile (see Fig.\,\ref{2cos-diffwidth}a). 
When the narrow peak and the broad peak are centered with respect to each other  (Fig.\,\ref{Two-Cos-IT}c), the solution belongs to class A, but unfortunately it is wrong, just like the KK reconstruction depicted in Fig.\,\ref{2cos-diffwidth}b. So neither the KK method nor the iterative method is capable of reconstructing this shape. \vspace{2mm}

\noindent (3) For ``triple-peak'' structures we find different behaviors depending on the details of the structure. We have seen in Fig.\,\ref{3Cos2-KK} that a bunch consisting of three  cosine-squared pulses of equal width and with uniform spacing is faithfully reconstructed by the KK method. The iterative reconstruction leads to curious results. If the large peak is in the center one gets a class B solution with a good reproduction of the original shape, see Fig.\,\ref{3Cos-eq-IT}c.  However, when the large peak is at the front, one finds a class C solution: in about 2/3 of the 100 iteration cycles a faithful reconstruction is achieved (Fig.\,\ref{3Cos-eq-IT}a) while in the remaining cycles  the reconstruction is wrong~(Fig.\,\ref{3Cos-eq-IT}b).\\
Next we study a bunch consisting of three  cosine-squared pulses of equal width and with non-uniform spacing, compare Fig.\,\ref{3Cos2-KK-nonequi}. When the largest peak is at the front,	the iterative method yields a perfect reconstruction (Fig.\,\ref{3Cos-noneq-IT}a). When the 
largest peak is in the center one finds a class C solution.  In 60 out of 100 iteration cycles a faithful reconstruction is achieved (Fig.\,\ref{3Cos-noneq-IT}b) but in the remaining cycles	the reconstruction is wrong~(Fig.\,\ref{3Cos-noneq-IT}c).

\vspace{2mm}

\noindent The above examples demonstrate	explicitly that the iterative method suffers from the same ambiguities as the dispersion-relation method. Moreover, it becomes evident that the intrinsic ambiguity of  phase reconstruction from the magnitude of the form factor cannot be resolved  by  combining different phase retrieval methods.

\clearpage

\section{Infrared and THz  Spectrometer}\label{spectrometer}

In the description of the multichannel infrared and THz  spectrometer CRISP (Coherent transition Radiation Intensity SPectrometer) 
 we follow  a previous publication \cite{Wesch-NIM} but address also more recent developments. An important  step was the calibration \cite{Toke}  of the completely assembled multichannel-spectrometer which was carried out in 2016 at the infrared free-electron laser FELIX in The Netherlands.

\subsection{Blazed reflection gratings}
Coherent radiation from short electron bunches extends over a wide range in wavelength, from a few micrometers up to about 1\,mm.  Gratings are useful to disperse the polychromatic radiation into its spectral components.  The free spectral range of a grating is defined by the requirement that different diffraction orders do not overlap. Since light of wavelength $\lambda$,  diffracted in first order,  will coincide with light of wavelength $\lambda/2$, diffracted in second order, the ratio of the longest and the shortest wavelength in the free spectral range is close to two. Hence  many different gratings  are needed to cover the full spectral range of coherent transition radiation. Overlap of different orders can be avoided by passing  the radiation through a bandwidth-limiting device before it impinges on a grating.    It will be shown below that this bandwidth limitation can be accomplished by a preceding grating. 

\begin{figure}[ht]
	 \centering
		\includegraphics[width=15cm]{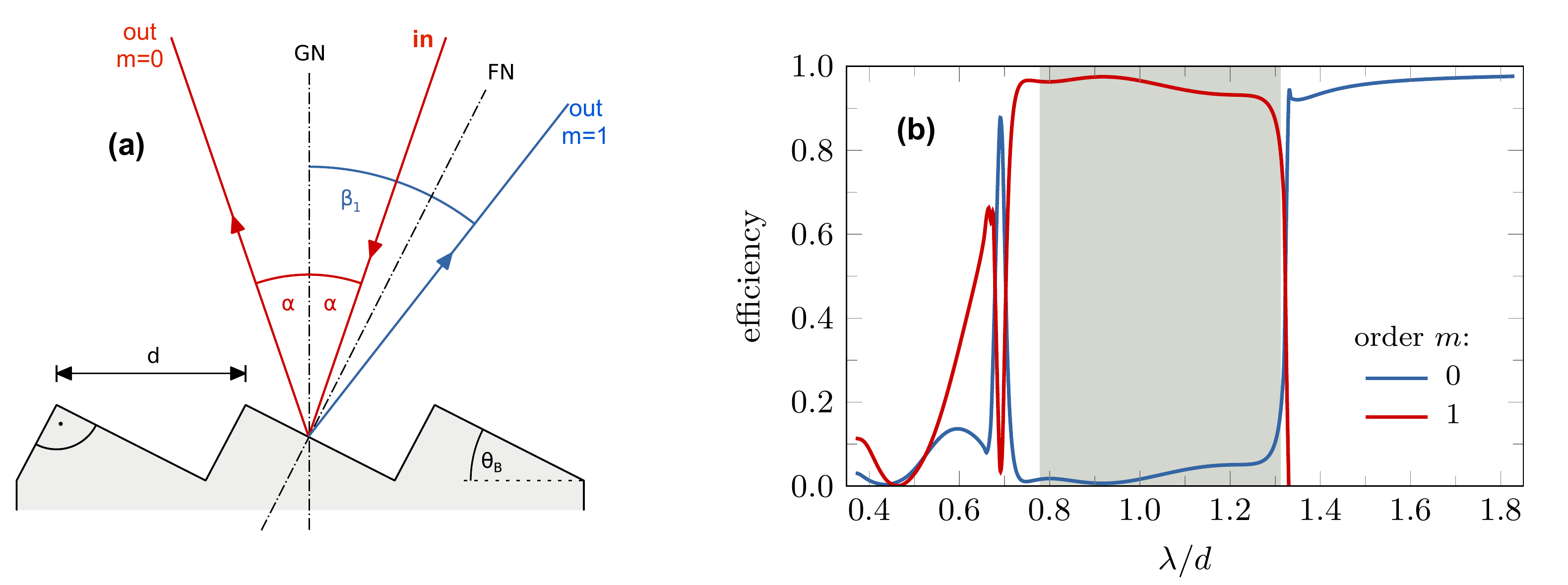}
	 \caption{\small{{\bf (a)} Principle of a blazed reflection grating.  For optimum efficiency, the incident ray and the first-order diffracted ray
	  (for $\lambda=1.045\,d$) have to obey the law of reflection at each facet.  
\newline	 {\bf (b)} Efficiency curve of a gold-plated reflection grating for radiation polarized perpendicular to the grooves, computed with the code {\it PCGrate} (solid red curve) for first-order diffraction ($m=1$). The wavelength range of first-order diffraction  is  $0.78\,d<\lambda<1.31\,d$, it  is  marked by the shaded area. The computed efficiency is about $90\%$ and almost   flat.
The blue curve shows the computed efficiency for zero-order diffraction. For wavelengths $\lambda > 1.33\,d$ the grating acts as a plane mirror with a reflectivity of about $95\%$.}}
	\label{blazed-grating}
	\end{figure}

A transmission grating with a large number of narrow slits distributes the radiation power almost evenly among  many diffraction orders.  Much superior are blazed reflection grating with triangular grooves as shown in Fig.~\ref{blazed-grating}a. They obey the grating equation
\begin{equation}\label{grating-equation}
	d \, (\sin{\alpha} + \sin{\beta_m}) = m \,\lambda 
	\end{equation}
	where $d$ is the distance between adjacent grooves, $m$ is the diffraction order and $\alpha$ the angle between the incident ray and the grating normal.
 To optimize the intensity for first-order diffraction ($m=1$), the  angle $\alpha$ is chosen such that the incident ray and the first-order diffracted ray (with a  wavelength of $\lambda=1.045\,d$,  the center  wavelength of the shaded  area shown in  Fig.~\ref{blazed-grating}b)   enclose equal angles with respect to the facet normal FN   \cite{grating-handbook}. This implies 
$$\theta_B-\alpha=\beta_1-\theta_B~~~\Rightarrow ~~\alpha=2\,\theta_B-\beta_1  $$ 
where $\theta_B$ is the blaze angle ($\theta_B=27^\circ$ in our case).\vspace{1mm}

\noindent Diffraction effects vanish if the wavelength becomes too large. The incidence angle is $\alpha=19^\circ$  in our spectrometer setup, hence the largest possible value of $\sin{\alpha} + \sin{\beta_m} $ is 1.33. This implies that for  wavelengths   $\lambda > 1.33 \,d$ the grating equation~(\ref{grating-equation}) can only be satisfied with $m=0$ which means that no diffracted wave exists. The grating acts then  as a simple plane mirror:
$$\sin{\alpha} + \sin{\beta_0}=0~~~\Rightarrow~~\beta_0=-\alpha \,.$$
This
 ``specular reflection'' of long wavelengths is utilized in the multistage spectrometer described below.
 
 The efficiency of a grating in a given diffraction order $m$ is defined as the ratio of diffracted light energy to incident energy. It was computed with the commercial code {\it PCGrate-S6.1} by I.I.G. Inc. 
In Fig.~\ref{blazed-grating}b,  the efficiency as a function of wavelength is shown  for the  diffraction orders $m=1$ and $m=0$.
 Short-wavelength radiation with  $\lambda <0.78\,d$ must be removed by a preceding grating stage  to avoid overlap of different diffraction orders.

\subsection{Multiple grating configuration}
\begin{figure}[ht!]
	  \centering
		\includegraphics[width=17cm]{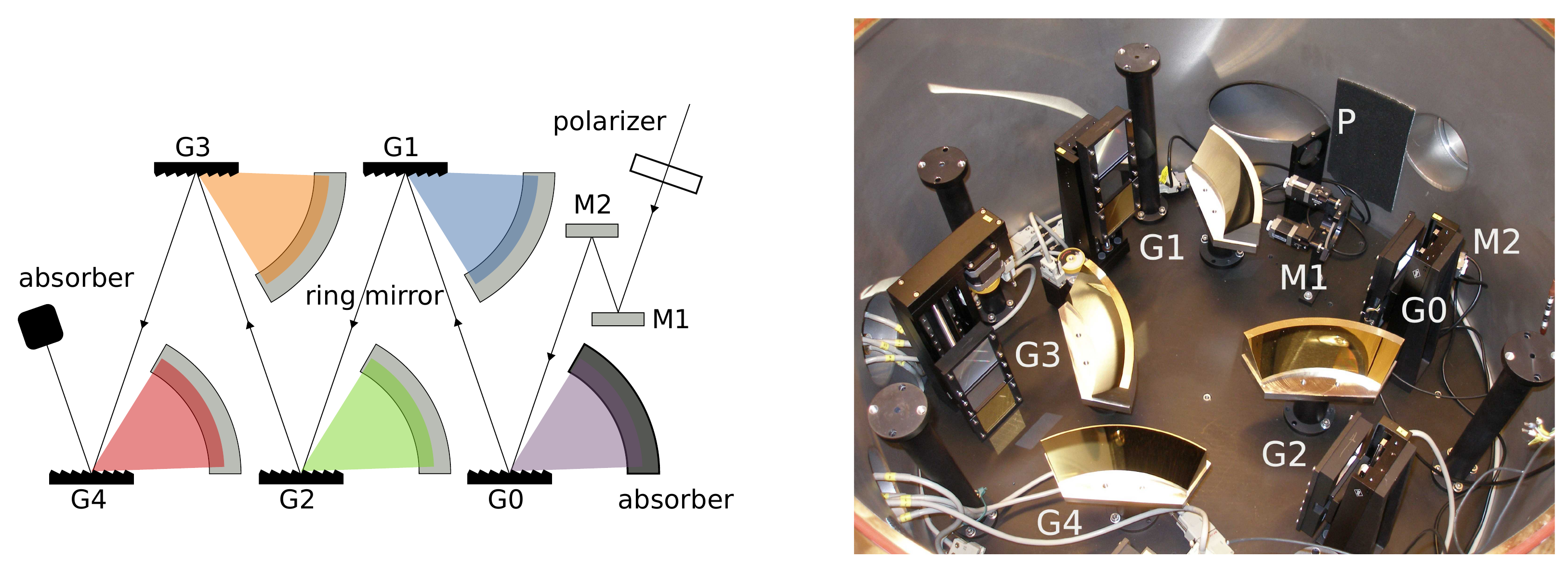}
	  \caption{\small{Schematic view and photo of the staged spectrometer equipped with five reflection gratings. The spectrometer is mounted in a vacuum vessel to avoid the absorption of THz waves in  air of normal humidity. The detector arrays are not yet mounted above the focusing mirrors. Grating G4 is just outside the photo but its mirror can be seen. P is the polarizer, and M1, M2 are the input alignment mirrors.}}
	\label{5grating-spectr}
	\end{figure}

\begin{figure}[ht!]
	 \centering
	\includegraphics[width=14cm]{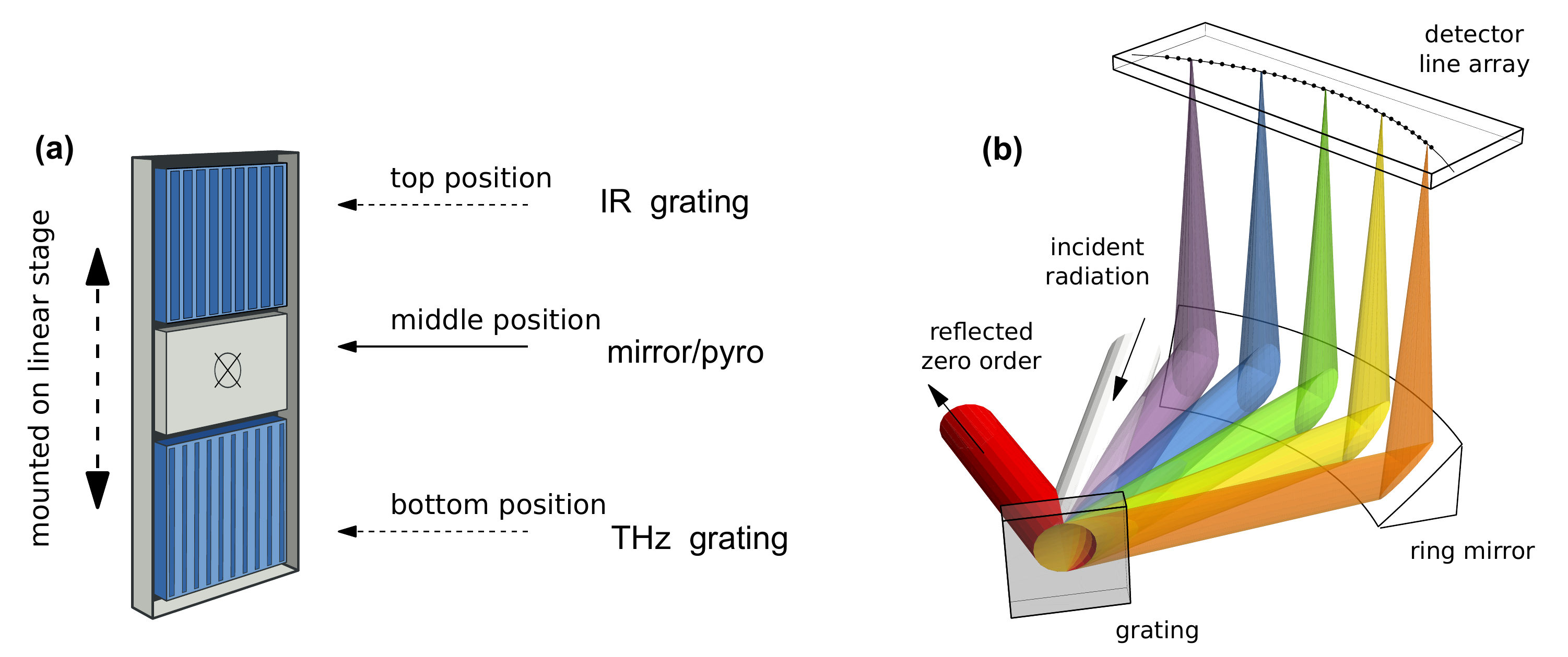}
	 \caption{ \small{{\bf (a)} Corresponding pairs of gratings are mounted on vertical translation stages with the aluminum-plated infrared grating  in the upper position and the gold-plated THz (far-infrared) grating    in the lower position. Between each  pair there is either a plane mirror (for G1, G2, G3) or a pyroelectric detector (for G0 and G4) which are used for alignment purposes. {\bf (b)} Arrangement of the grating, the ring mirror and the array of 30 pyroelectric  detectors. The light dispersion and focusing have been computed with a ray tracing code, for clarity this is shown for only 5 of the 30 wavelength channels. }
	 }
	\label{gratingpair-5rays}
	\end{figure}

The spectrometer is equipped with five  consecutive  reflection gratings, G0 to G4 (see Fig.~\ref{5grating-spectr}). Each grating exists in two variants, one for the infrared (IR) regime, the other for the THz regime. The parameters are summarized in Table~\ref{grating-param}. The IR and THz gratings are mounted on top of each other in  vertical translation stages (Fig.\,\ref{gratingpair-5rays}a). Between each grating pair there is either a mirror (for G1, G2 and G3) or a pyroelectric detector (for G0 and G4), these  are needed for alignment.

	\begin{table}[ht] 
	\caption[]{\small{Parameters of the gratings. The triangular grooves have a blaze angle of $\theta_B=27^\circ$. The   distance   between two grooves is called $d$.    The minimum and maximum wavelengths of the free spectral range for first-order diffraction are called $\lambda_{\mathrm{min}}$ and $\lambda_{\mathrm{max}}$.  The wavelength above which the grating acts as a plane mirror is called $\lambda_0$. All dimensions are quoted in $\mu$m. The coarse gratings with 
	$d \ge 58.82\,\mu$m are gold-plated and were custom-made by Kugler Precision, the fine gratings with $d \le 33.33\,\mu$m are aluminum-plated and were purchased from Newport Corporation.}}
	\label{grating-param}
	\begin{center} 
	\begin{tabular}{c c}
	IR mode $5.1-43.5\,\mu$m &~~~~~~~~~~~~~~~~~~~~~THz mode $45.3-434.5\,\mu$m
\end{tabular}

\begin{tabular}{|c| c| c| c| c|c|c| c| c| c| c|}
\hline
 grating	&	$d$ &   $\lambda_{\mathrm{min}}$&	
$\lambda_{\mathrm{max}}$& $\lambda_{\rm 0}$&~~~~~~~ &grating	&	$d$ &   $\lambda_{\mathrm{min}}$&	
$\lambda_{\mathrm{max}}$& $\lambda_{\rm 0}$\\
\hline
G0	&	4.17	& -				& -	&5.5		&	&	G0	&	33.33	& -				& -	&44				\\	
\hline 
G1&	6.67	& 5.13				& 8.77			&8.8	&&G1&	58.82	& 45.3				& 77.4			&77.6	\\
\hline 
G2	&	11.11	&8.56				& 14.6	& 14.7			&	&G2	&	100.0	&77.0				& 131.5	& 132				\\
\hline 
G3	&	20.0	& 15.4				& 26.3	& 26.4			&	& G3	&	181.8	&140.0				& 239.1	& 240				\\
\hline 
G4	&	33.33 &	27.5				& 43.5	& -		&  &G4	&	333.3 &	256.7				& 434.5	& 440			\\
\hline 
\end{tabular}
\end{center}
\end{table} 

In the following we describe the THz configuration, the infrared configuration works correspondingly.  The incident radiation is passed through a  polarization filter (HDPE thin film THz polarizer  by TYDEX) to select the polarization component perpendicular to the grooves of the gratings, and is then directed towards grating G0 which acts as a  bandwidth-limiting device:  short-wavelength radiation ($\lambda < \lambda_0=44\,\mu$m) is dispersed by G0 and guided to an absorber, long-wavelength radiation ($\lambda>\lambda_0$) is specularly reflected towards G1  which is the first grating stage of the spectrometer.    Radiation in the  range $\lambda_{\mathrm{min}}=45.3\,\mu$m$<\lambda<\lambda_{\mathrm{max}}=77.4\,\mu$m is dispersed by G1 in first-order and focused by a ring mirror onto a multi-channel detector array, while  radiation with $\lambda>\lambda_0=77.6\,\mu$m is specularly reflected and sent to G2. The subsequent gratings work similarly and disperse the wavelength intervals 
$[77.0,131.5]\,\mu$m (G2),  $[140.0,239.1]\,\mu$m (G3), and $[256.7, 434.5]\,\mu$m (G4).

For each of the gratings G1 to G4, the first-order diffracted radiation is recorded in an array of  30  pyroelectric detectors which are arranged on a  circular arc covering $57^\circ$.
 A ring-shaped parabolic mirror focuses the light onto this arc. The computed light dispersion and focusing  is shown schematically in Fig.\,\ref{gratingpair-5rays}b for  5 of the 30 wavelength channels.  
	
\vspace{20mm}	
	
\subsection{Pyroelectric detectors}	
\begin{figure}[htb!]
	 \centering
		\includegraphics[width=15cm]{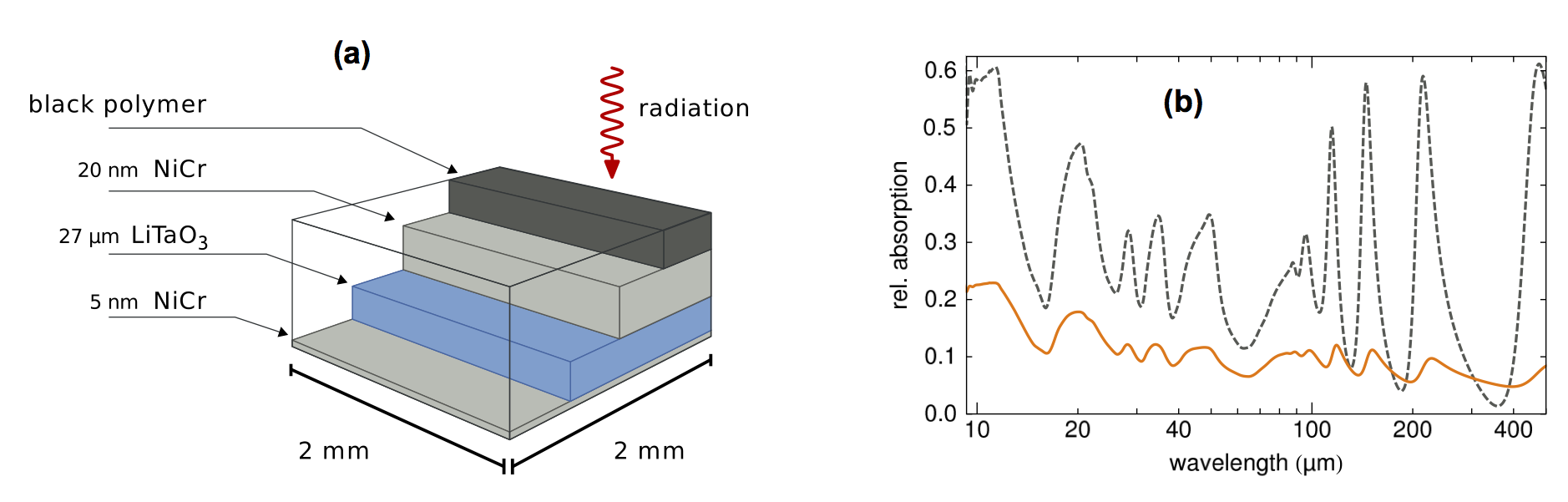}
	\caption{\small{{\bf (a)} Layout of the pyroelectric detector LIM-107-X005.
	 {\bf (b)} Computed infrared absorption as a function of wavelength. Solid yellow curve:  27\,$\mu$m  LiTaO$_3$ detector with optimized  coatings for minimum internal reflections (20\,nm NiCr at front surface and 5\,nm NiCr at back surface). Dotted black curve:  27\,$\mu$m  LiTaO$_3$ detector with standard coatings:  a 5\,nm NiCr layer at the front surface and a  thick gold layer at the back surface.}}
	\label{pyro-layout}
	\end{figure}

A critical component of the broadband single-shot spectrometer is a detector featuring high sensitivity over the entire  infrared  and far-infrared (THz) regime, from $\mu$m  to  mm wavelengths. Bolometric devices, responding to the deposited radiation energy through a temperature rise, are capable of covering such a wide wavelength range. 
A special pyroelectric detector has been developed to our specification by an industrial company (InfraTec). 
This sensor possesses sufficient sensitivity for the application in a coherent transition radiation spectrometer and has a fast thermal response. The layout of the detector is shown in Fig.~\ref{pyro-layout}a.
	It consists of a 27\,$\mu$m thick lithium tantalate (LiTaO$_3$) crystal with an active area of 2 $\times$ 2\,mm$^2$. The front surface is covered with a NiCr electrode  of 20\,nm thickness instead of  the more conventional 5\,nm. The backside electrode is a 5\,nm NiCr layer instead of the conventional thick gold electrode.  
The combination of a comparatively  thick front electrode and a thin backside metallization  suppresses internal reflections which are the origin of the strong wavelength-dependent efficiency oscillations observed in conventional pyroelectric detectors. The beneficial effect of the novel surface layer structure is illustrated in Fig.~\ref{pyro-layout}b. To enhance  absorption below 100\,$\mu$m the front electrode is covered with a black polymer layer which is  transparent above 100\,$\mu$m. \vspace{2mm}
	
\noindent  The thermal expansion of the  pyroelectric crystal, due to the absorption of radiation, creates a surface charge which is converted into a voltage signal by the charge-sensitive preamplifier (Cremat CR110). The preamplifier and the twisted-pair line driver amplifier are mounted on the electronics board inside the vacuum vessel. Line receiver, Gaussian shaping amplifier and ADC (analog-to-digital converter) are located outside  the vacuum vessel.
	 The commercial  preamplifier Cremat CR110 generates pulses with a rise time of 10\,ns and a decay time of 140\,$\mu$s. This is adequate for  repetition rates of 10\,Hz or less. A Gaussian shaping amplifier (Cremat CR200 with $4\,\mu$s shaping time) is used to optimize the signal-to-noise ratio. The shaped signals are digitized with 120 parallel ADCs with 9 MHz clock rate, 14 bit resolution and 50 MHz analog bandwidth. 
	
\subsection{Transition radiation beamline and spectrometer response function}\label{202m-beamline}
\noindent {\bf CTR beamline}\\
Transition radiation is produced on a screen inside the ultrahigh vacuum beam pipe of the FLASH linac at the 202\,m position. The screen is tilted by $45^\circ$ hence backward TR is emitted perpendicular  to the electron beam axis.
The  screen has a rectangular shape  ($16\times 25\,$mm$^2$) and   consists of a $380\,\mu$m  thick polished silicon wafer which is coated with a 150\,nm aluminum layer at the front surface. It is a so-called ``off-axis'' screen, positioned outside the nominal electron beam axis.  Selected electron bunches can be steered onto the TR screen using a fast kicker magnet. 	 This  permits high-resolution diagnostics on a single bunch out of a long  train without impeding the FEL gain process for the unkicked bunches.\vspace{2mm}
\begin{figure}[htb!]	 
	 \centering
	\includegraphics[width=16cm]{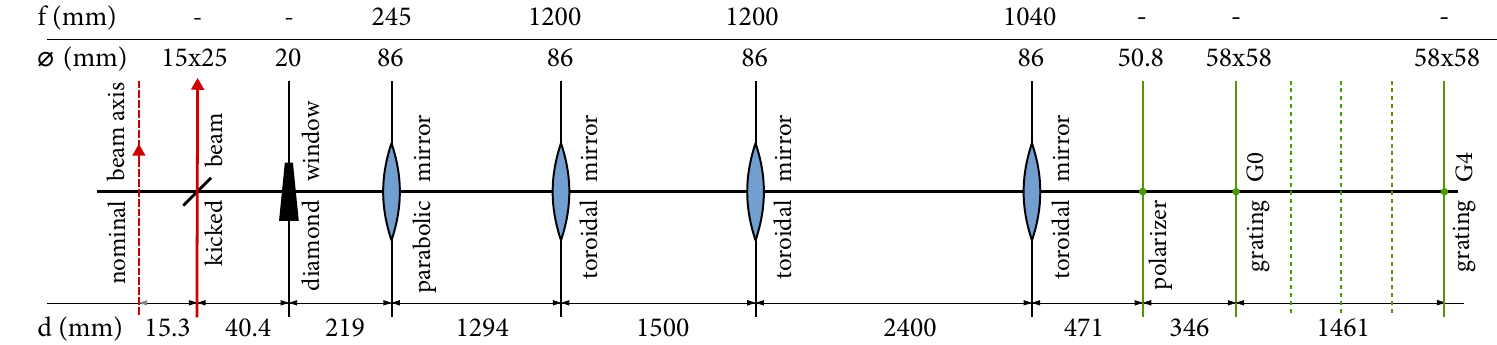}
	 \caption{\small{ Optical design of the  CTR beamline. The focusing parabolic or toroidal mirrors M1 to M4 are shown as lenses with their respective positions,  focal lengths, and diameters. Before entering the staged grating arrangement,  the radially polarized transition radiation is passed through a linear polarizer to select the field component perpendicular to the grooves of the blazed gratings.	} }
	\label{CTR-beamline-optics}
	\end{figure}
	\begin{figure}[htb!]	 
	 \centering
	\includegraphics[width=7cm]{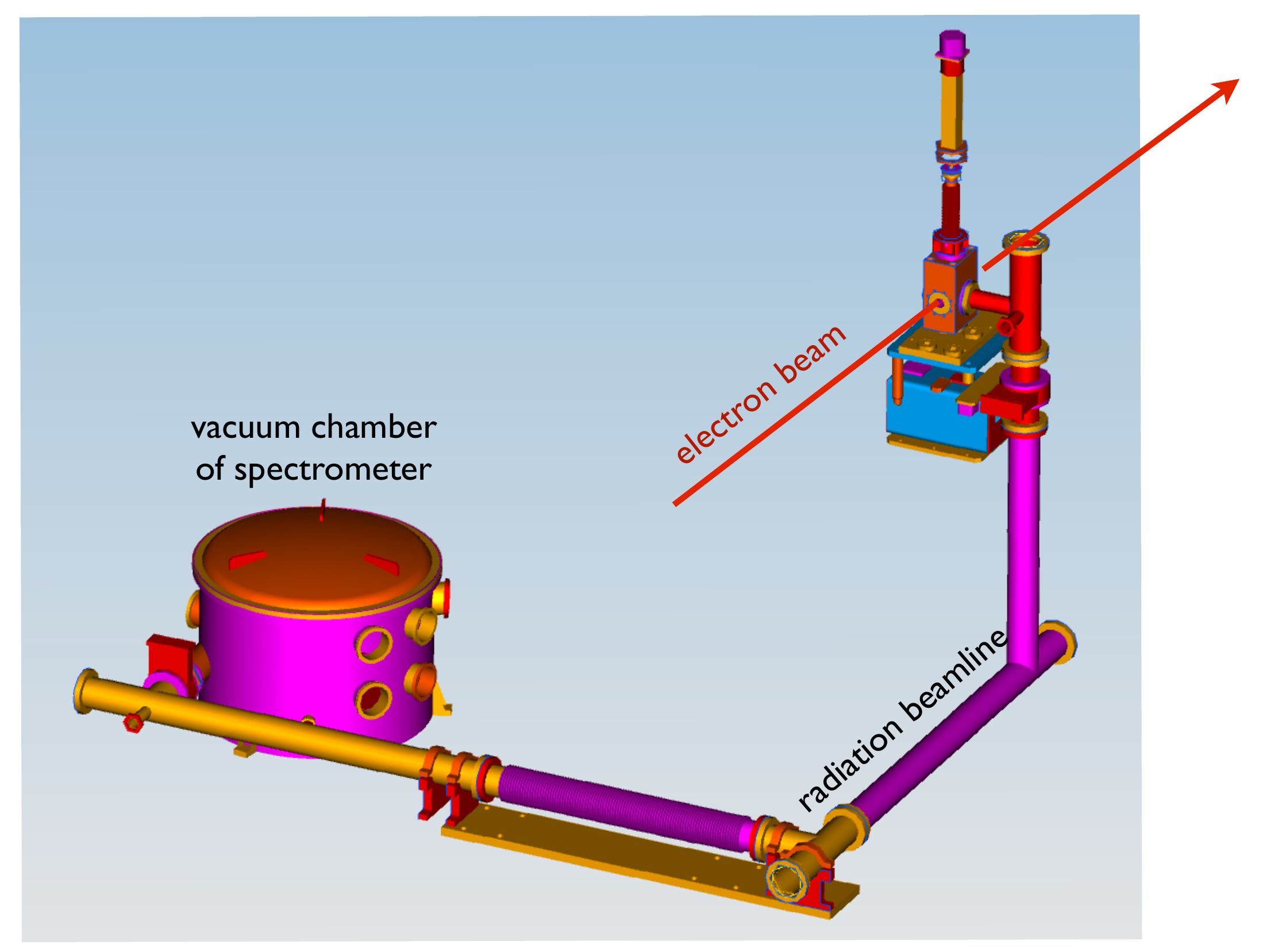}
	 \caption{\small{Technical layout of the CTR beamline at the 202\,m position. }
		 }
	\label{CTR-beamline-drawing}
	\end{figure}

The radiation  is coupled out through a  0.5\,mm thick window made from chemical vapor deposition (CVD) diamond. In contrast to standard window materials such as glass, quartz or polyethylene, CVD diamond   has almost negligible absorption in the entire spectral range of transition radiation, from visible light up to millimeter waves, except for a narrow absorption band around $5\,\mu$m where lattice vibrations are excited.   The radiation  is transported to the spectrometer  by an optical system consisting    of four focusing and four plane mirrors.   The optical layout is shown  in Fig.\,\ref{CTR-beamline-optics}, a three-dimensional technical drawing  in Fig.\,\ref{CTR-beamline-drawing}.   The design criteria are identical to those of the CTR beamline  at the 141\,m position of the FLASH linac and have been documented in Ref.\,\cite{Casalbuoni-III}, they need not be repeated here. Beamline and spectrometer are mounted in vacuum vessels to avoid the strong absorption of THz waves in air of normal humidity. 

 \vspace{20mm}

\noindent {\bf Response function}\\
The calibration of a broadband   spectrometer, covering the wavelength range from $4\,\mu$m to $400\,\mu$m,  is a demanding task. There exists no ``table-top'' radiation source which can be tuned over such  wide a range. A broadband  accelerator-based source is the  free-electron laser FELIX in The Netherlands, it provides monochromatic infrared radiation between a few $\mu$m and 
$135\,\mu$m. Our spectrometer CRISP was shipped to FELIX and calibrated with FEL radiation in 2016, see below. \\
The other essential component of the spectrometer setup  at the 202\,m position of the FLASH linac is the CTR beamline which guides the transition radiation from the TR screen to the spectrometer. This beamline is rigidly mounted in the linac tunnel and cannot be moved to an outside FEL laboratory to determine its wavelength-dependent transmission properties. Moreover, such a test would be pretty meaningless since FEL radiation and transition radiation have very different angular characteristics: the FEL beam is well-collimated and has little divergence while the TR beam has a fairly wide  angular divergence and requires large-aperture  mirrors in the beamline. 
A performance test of the entire system - CTR beamline plus CRISP spectrometer - has to be done {\it in situ}. \vspace{1mm}

\noindent An ideal test scenario would be to have a beam of pointlike electron bunches with precisely known charge $Q$, which generate  transition radiation of well-known emission characteristics. 
 A  realistic test scenario has to take into account that the beam optics in the FLASH linac is optimized for  high-gain FEL operation and  that an extremely small beam radius cannot be realized.   The 202\,m position of the linac is a good location for the spectrometer because here the  electron beam is round and   the horizontal and vertical beta functions are  reasonably small ($\beta_x \approx \beta_y \approx 7\,$ m). The rms beam radius is $\sigma \approx  100\,\mu$m.    \vspace{2mm}

\noindent To investigate the performance of the multichannel spectrometer we consider therefore a {\it reference bunch} of length zero with a charge of 
$Q_{\rm ref}=100\,$pC and a cylindrically symmetric  Gaussian transverse density distribution with $\sigma_{\rm ref}=100\,\mu$m.  The transition radiation produced by the bunch upon crossing  the TR screen can be accurately computed. The radiation passes through the beamline, where  losses occur due to diffraction and aperture limitations, and enters the spectrometer. Here it is decomposed into 120 spectral components (either in the IR regime or in the THz regime) that impinge on the corresponding pyroelectric detectors.   The amplifiers     produce  120 voltage signals $v_m$  which  are digitized by ADCs. The voltages $v_m$  are taken as the components of a  {\it voltage-signal vector}  $\boldsymbol{v}=(v_1, v_2 \ldots v_{120})$. There are two such voltage-signal vectors, one for the IR  regime, the other for the THz regime.  These two vectors are the system response  to the reference bunch. The same set of 120 pyro-detectors is used in the two operation modes, only the gratings are changed when switching from IR mode to THz mode or vice versa.\vspace{2mm}  

\noindent In each of the operation modes (IR or THz) the spectrometer defines 120 wavelength bins 
$\lambda_m \pm \delta \lambda_m $.
The primary  transition radiation energies within these bins    are called $U_m$.    Only a certain   fraction $u_m$ of the primary TR energy $U_m$ passes through the CTR beamline and arrives at the pyro-detector $m$:
$$u_m=P_{\rm trans}(\lambda_m)U_m$$
where $P_{\rm trans}(\lambda_m)$ is the wavelength-dependent transfer function (transmission) of the CTR beamline.\\
Thus the  overall response function of the multichannel spectrometer, as mounted at the rear end of the CTR beamline, consists of two parts: (1) the generation  of transition radiation,  the transmission  through  the CTR beamline and the grating stages as a function of wavelength, and (2) the wavelength-dependent response function of the pyroelectric detectors.\vspace{2mm}

\noindent {\underline{\it Part 1\,:}}  The generation  of transition radiation and its transmission through the  CTR beamline (Figs.\,\ref{CTR-beamline-optics} and \ref{CTR-beamline-drawing}) have been determined by an elaborate ``start-to-end'' simulation using the code {\it THzTransport}. 
The mathematical formalism is explained in Appendix D.
In the program, an infinitesimally short reference  bunch with a charge of 
$Q_{\rm ref}=100\,$pC and a cylindrically symmetric  Gaussian transverse density distribution with $\sigma_{\rm ref}=100\,\mu$m is used. 
The code {\it THzTransport} computes the transition radiation produced by the reference bunch by applying the  Weizs\"acker-Williams method of virtual  photons (see Section \ref{TransRad}). The electromagnetic field of a radially extended charged disc is used, hence  the suppression of short wavelengths  by the transverse form factor (see  Fig.\,\ref{Fig-F2trans}) is automatically taken into consideration. The 
spectral  radiation components are propagated from the TR source screen through the CTR beamline to the corresponding pyro-detector.  This is done for all wavelengths $\lambda_m$. The propagation   proceeds in a stepwise fashion: \vspace{1mm}

\noindent TR screen $\rightarrow$ diamond window $\rightarrow$  M1 $\rightarrow$  M2 $\rightarrow$  M3  $\rightarrow$  M4 $\rightarrow$ spectrometer.\vspace{2mm}

\noindent  All apertures and near-field diffraction effects are taken into consideration. Within the staged-grating spectrometer, the simulation distinguishes which grating guides the radiation contained in the wavelength bin $\lambda_m \pm \delta \lambda_m $ to the corresponding pyro-detector $m$,  taking the focusing by the ring mirror into account. The computed spectral energy $u_m$ impinging onto  the pyro-detector is converted into a voltage signal $v_m$ using the calibration described in Part 2.
\begin{figure}[htb!]	 
	 \centering
		\includegraphics[width=10cm]{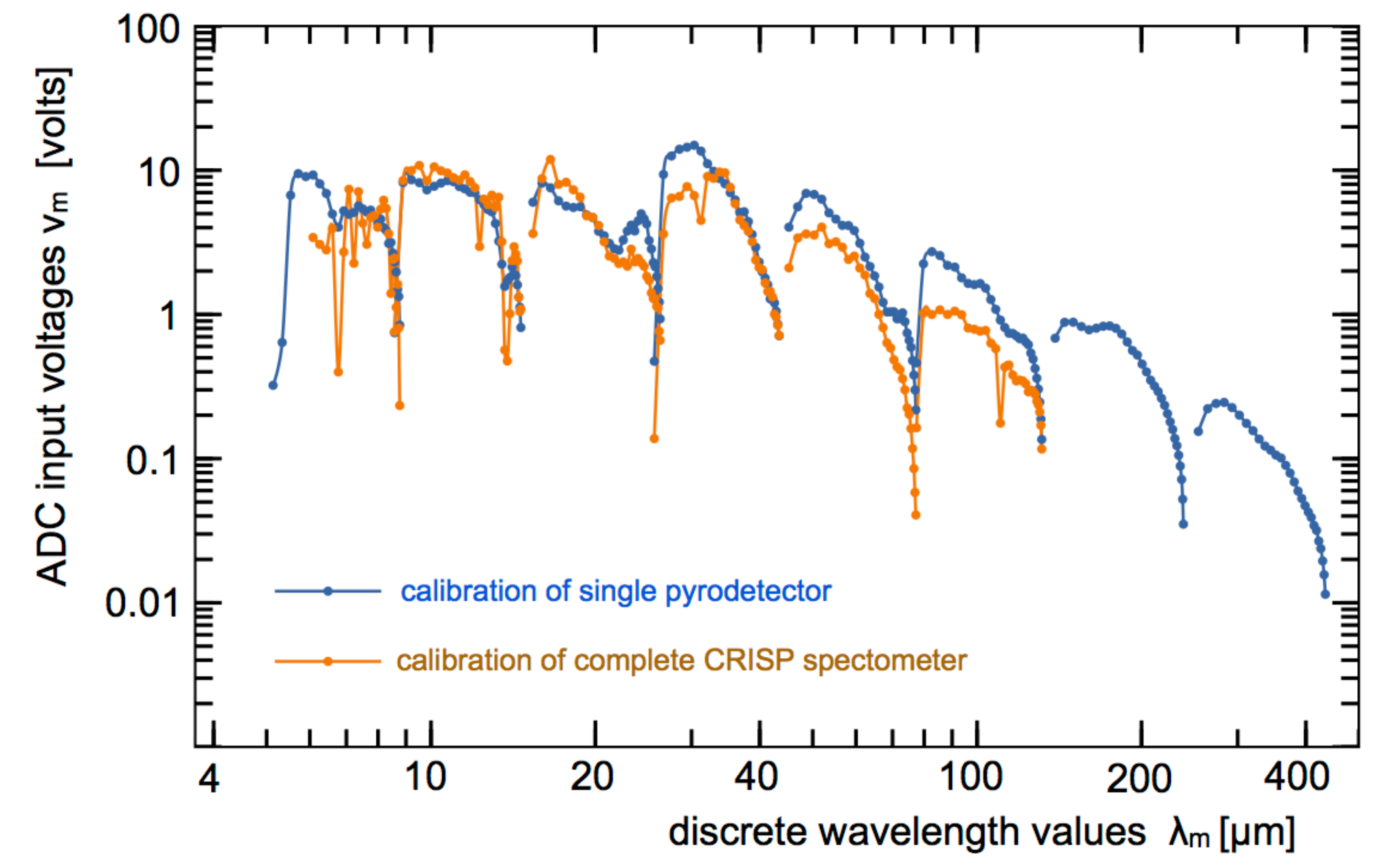}
	 \caption{\small{Blue dots: First determination of the spectrometer response function, derived from the comprehensive {\it THzTransport} simulation described in {\it Part 1}  and the measured response  of a single prototype pyro-detector \cite{Behrens}. 
	The  grating efficiency  was  taken from theory, see Fig.\,\ref{blazed-grating}b.
	 Yellow dots: Improved response function after calibration of the completely assembled CRISP spectrometer  at the infrared free-electron laser FELIX  (for details see Ref.\,\cite{Toke}). This overall calibration determines the responses  of all 120 pyroelectric detectors as well as  the grating efficiencies. The sawtooth-like structure is caused by the variation of  acceptance within each detector array. }
 }
	\label{spectrometer-response-prelim}
	\end{figure}	

\vspace{1mm}

\noindent {\underline{\it Part 2\,:}}
The calibration of a prototype pyroelectric detector was carried out in 2008 \cite{Behrens}. Combining this calibration with the numerical simulation described in {\it Part1} yields the preliminary  spectrometer response function plotted   in 
Fig.\,\ref{spectrometer-response-prelim} (blue dots).
The  spectral response  of the completely assembled CRISP spectrometer was determined in 2016    \cite{Toke} with tunable monochromatic infrared radiation from the free-electron laser FELIX in the range from $6\,\mu$m to $135\,\mu$m. Starting from the measured FEL power at the entrance aperture of the spectrometer and the known transverse intensity distribution of  the FEL radiation, the monochromatic FEL beam  was propagated by a {\it THzTransport} simulation through the polarizer,  the grating stages and the focusing ring mirror up to the pyro-detector corresponding to the selected FEL wavelength. In this way 
the response  of each of the 120 pyro-detectors was calibrated by establishing the relation between the known incident radiation energy $u_m$ and the measured output voltage $v_m$, both in the IR configuration  and in the THz configuration. 
The efficiency of the gratings was part of this calibration.

 \vspace{2mm}

 The overall spectrometer calibration with  FEL radiation cannot be simply  taken over to the TR diagnostic station at FLASH.    The  input of the well-collimated FEL radiation  into the spectrometer is very different from the transition radiation input. The quite divergent   transition radiation wave has to propagate through the CTR beamline before  entering the spectrometer. For this reason, the pyro-detector calibration at FELIX must be combined with the above-mentioned  ``start-to-end''  simulation  (for details see Refs.\,\cite{Toke, Wesch-thesis}). Having computed the TR energies $u_m$ impinging onto the pyroelectric detectors, the  detector  calibration at  FELIX is utilized to convert these energies  $u_m$ into voltages $v_m$. The improved spectrometer response function based on this overall calibration  is also shown in Fig.\,\ref{spectrometer-response-prelim} (yellow dots). The rather similar wavelength-dependencies of the blue and yellow curves allows to extrapolate the  pyro-detector calibration  into the wavelength range from $135\,\mu$m to $420\,\mu$m where no FEL radiation was available at the FELIX facility. \vspace{2mm}

\begin{figure}[htb!]	 
	 \centering
		\includegraphics[width=11cm]{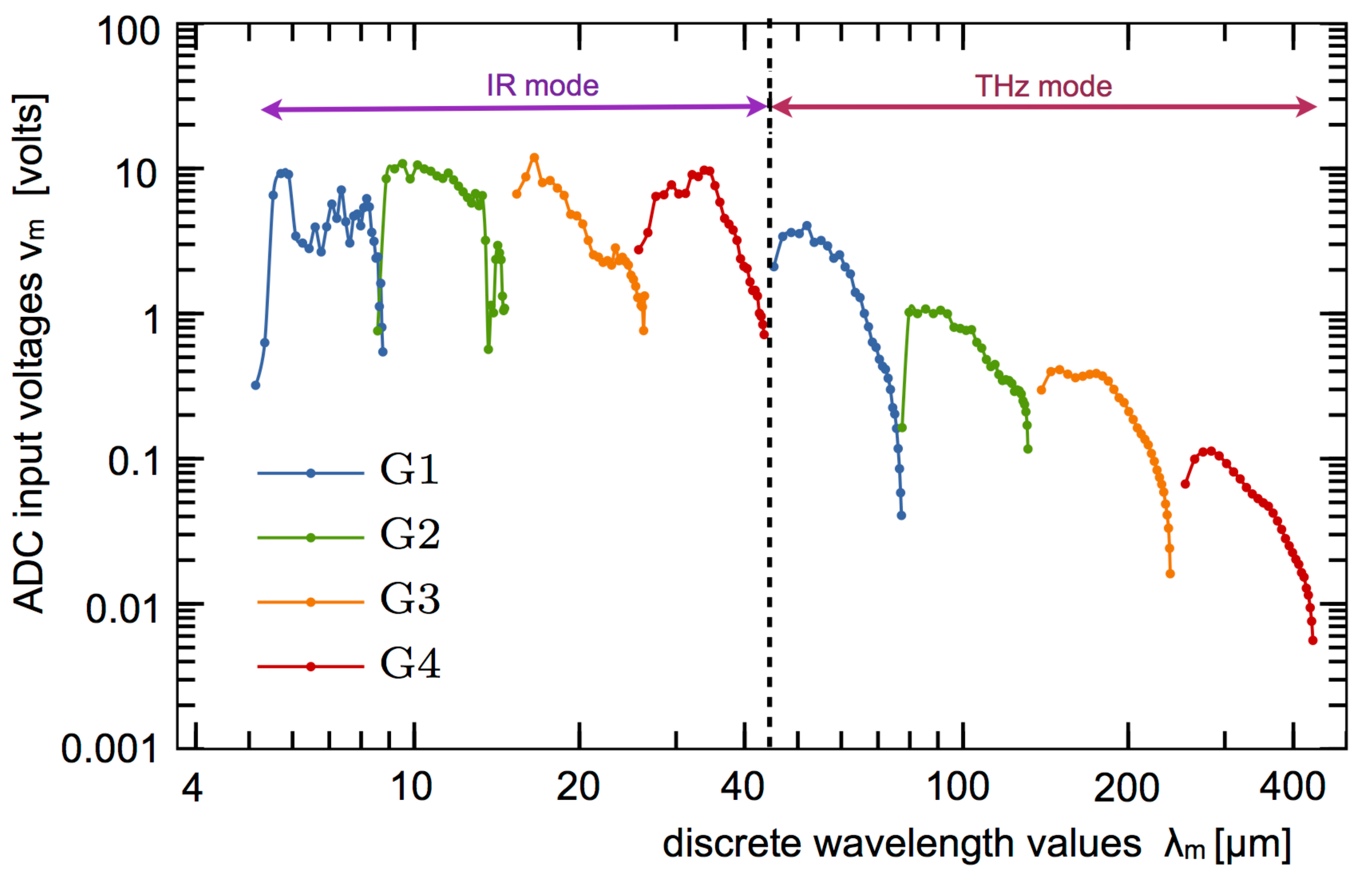}
	 \caption{\small{The overall response function $R_{\rm ref}(\lambda)$  of the multichannel spectrometer as installed at the rear end of the CTR beamline  (all corrections applied). The discrete wavelengths $\lambda_1, \lambda_2,\ldots \lambda_{120}$ (in either the IR mode or the THz mode) are the center values of the wavelength bins $\lambda_m \pm \delta \lambda_m$ associated with the 120 pyro-detectors (see text).  The response  function has been computed using the parameters of the reference bunch:
	 length zero, rms radius $\sigma_{\rm ref}=100\,\mu$m, charge $Q_{\rm ref}=100\,$pC.}}
	\label{spectrometer-response}
	\end{figure}	
\noindent Meanwhile the 2016 data have been critically re-evaluated and a number of additional subtle corrections have been applied.  The resulting overall response function $R_{\rm ref}(\lambda)$  is shown in Fig.~\ref{spectrometer-response}. This  response function  of the entire {\it System} (defined as  the combination of TR screen, diamond window, CTR beamline and CRISP spectrometer)  has been computed using the parameters of the reference bunch. It relates indirectly   the  spectral transition radiation energy $U_m$ inside  the wavelength bin $\lambda_m \pm \delta\lambda_m$  and the ADC input voltage $v_m$ of the corresponding pyroelectric detector. The defining equation 
\begin{equation}
v_m=R_{\rm ref}(\lambda_m)~~~\mathrm{with}~~m=1,2\ldots 120 ~~~~\mathrm{(either~ IR~ or~ THz~ mode})
\label{resp-funct-ref}
\end{equation}
looks deceivingly simple, but one has to keep in mind that all complications - the elaborate mathematical computations and the difficult calibration procedure -  are  hidden in the response function.\vspace{2mm}

\noindent The {\it System} response to an  arbitrary bunch with the same rms radius  $\sigma=\sigma_{\rm ref}$ but with a different charge $Q$ and a finite length  (longitudinal form factor $F(\omega)=|{\cal F}(\omega)|$) can be easily written down.  The
radiation energy $U'_m$ in the wavelength bin $\lambda_m \pm \delta\lambda_m$, generated by this bunch, and the ADC input voltage $v'_m$ of pyro-detector number $m$ are related to the same quantities of the reference bunch by
$$U'_m=\frac{Q^2}{Q_{\rm ref}^2}\,|F(\omega_m)|^2 U_m \,,~~~v'_m=\frac{Q^2}{Q_{\rm ref}^2}\,|F(\omega_m)|^2 v_m
~~~\mathrm{with}~~\omega_m=\frac{2\pi c}{\lambda_m}\,.$$
Inserting for $v_m$ the above relation (\ref{resp-funct-ref}) we get the important equation
\begin{equation}\boxed{~
 |F(\omega_m)|^2\equiv \left|F\left(\frac{2\pi c}{\lambda_m}\right)\right|^2 =\frac{Q_{\rm ref}^2}{Q^2}\frac{v'_m}{R_{\rm ref}(\lambda_m)}~~~~(m=1,2 \ldots 120) 
 ~~~~\mathrm{(either~ IR~ or~ THz~ mode})~}
 \label{Eq-response-function}
\end{equation}
which allows us to derive the longitudinal form factor  from the measured  voltage-signal vector $\boldsymbol{v'}=(v'_1, v'_2 \ldots v'_{120})$ with the help of the response function 
$R_{\rm ref}(\lambda)$.  Note that  this equation must be used twice, in the IR mode and in the THz mode, in order to cover the full spectral range.  
 
  \vspace{3mm}
  
\noindent  {\bf Impact of electron beam radius} \\
\begin{figure}[htb!]
\centering
\includegraphics[angle=0,width=9cm]{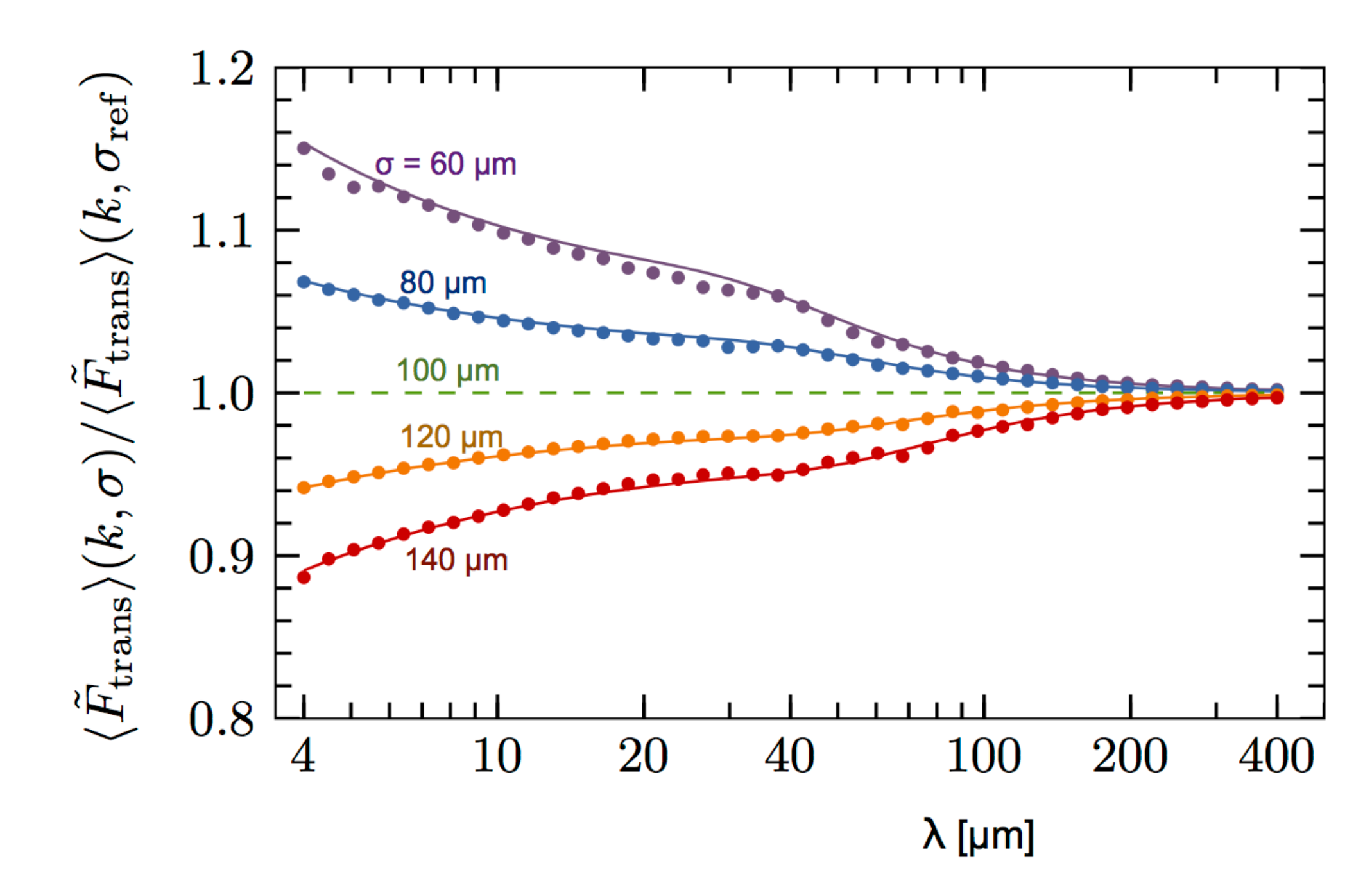}
  \caption[]{\small{ Influence  of the electron beam radius on the rms transverse form factor\,(\ref{FFav}). 
  The ratio $\langle \tilde{F}_{\rm trans}\rangle(k, \sigma)/
  \langle \tilde{F}_{\rm trans}\rangle(k, \sigma_{\rm ref})$ is plotted as a function of wavelength ($\lambda=2\pi /k$).  The reference  radius is 
  $\sigma_{\rm ref}=100\,\mu$m. The continuous curves result from   computations based on the near-field formula (\ref{GF-nearfield}) and an  aperture angle of 100\,mrad (compare Fig.\,\ref{Fig-F2trans}b).  The dots come from a numerical modeling of the radiation transport through the CTR beamline  to the spectrometer.}
   }
\label{FFtrans-spect}
\end{figure}
\noindent As said above, the response function  $R_{\rm ref}(\lambda)$ has been computed for the reference  rms electron beam radius  of $\sigma_{\rm ref}=100\,\mu$m. When the electron energy  or the optics of the accelerator   is changed, the beam radius at the spectrometer position will change as well. It his hence of interest to know how critically the rms transverse form factor (\ref{FFav})  depends on $\sigma$. To get an impression, we plot in  Fig.\,\ref{FFtrans-spect} 
the ratio
$$
\frac{\langle \tilde{F}_{\rm trans}\rangle(k,\sigma)}{\langle \tilde{F}_{\rm trans}\rangle(k,\sigma_{\rm ref})}~~~~~~
\mathrm{with}~~~k=\frac{2\pi}{\lambda}$$
 as a function of wavelength
for rms radii between $60\,\mu$m and $140\,\mu$m. 
This figure shows that the impact of the electron beam radius is weak: a $20\%$~uncertainty in the knowledge of the beam radius changes  the rms  transverse form factor
 by just  $5\%$ at the shortest wavelength of $4\,\mu$m and by much less at larger wavelengths.

\clearpage

\section{Results on bunch shape reconstruction at FLASH by CTR spectroscopy }

The parallel readout of the CRISP spectrometer permits the measurement of the CTR spectrum generated by a single bunch,  either in the infrared mode or  in the THz mode. To cover the full spectral range from $4\,\mu$m to $420\,\mu$m,  measurements with the two grating sets are needed (see Section \ref{spectrometer}). The stability of the accelerator is generally quite high and the fluctuations in the CTR spectrum within the required data taking time of a few minutes are smaller than the uncertainties caused by the preamplifier noise.
For the results shown here, the form factors have been derived as averages of 200 single bunches, kicked from consecutive bunch trains at 10\,Hz repetition rate. The measurement series  with the two grating sets are carried out consecutively and they are individually averaged.
The averaging procedure reduces detector and amplifier noise as well as statistical fluctuations,   and it extends  the applicability of spectroscopic bunch shape analysis
 to low bunch charges and large bunch lengths.

\subsection{Computation of form factor}

Consider now  bunches with an rms radius of $\sigma=\sigma_{\rm ref}=100\,\mu$m but with unknown charge $Q$ and unknown longitudinal charge density profile. The spectrometer measures the  deposited transition radiation energies $u'_m$  in  its  240 discrete wavelengths bins 
$\lambda_m \pm \delta \lambda_m$ (120 in the IR regime and 120 in the THz regime).  The pyroelectric LaTiO$_3$ crystals and the amplifiers convert  these energies into voltages $v'_m$ which are digitized and recorded. 
Knowing the  bunch charge $Q$, the absolute square $|F(\omega)|^2$ of the longitudinal form factor  is computed at the discrete frequencies 
$\omega_m=2\pi c/\lambda_m$ with the help of Eq.\,(\ref{Eq-response-function}). 
\vspace{2mm}
 \begin{figure}[ht!]
		\centering 
\includegraphics[width=15cm]{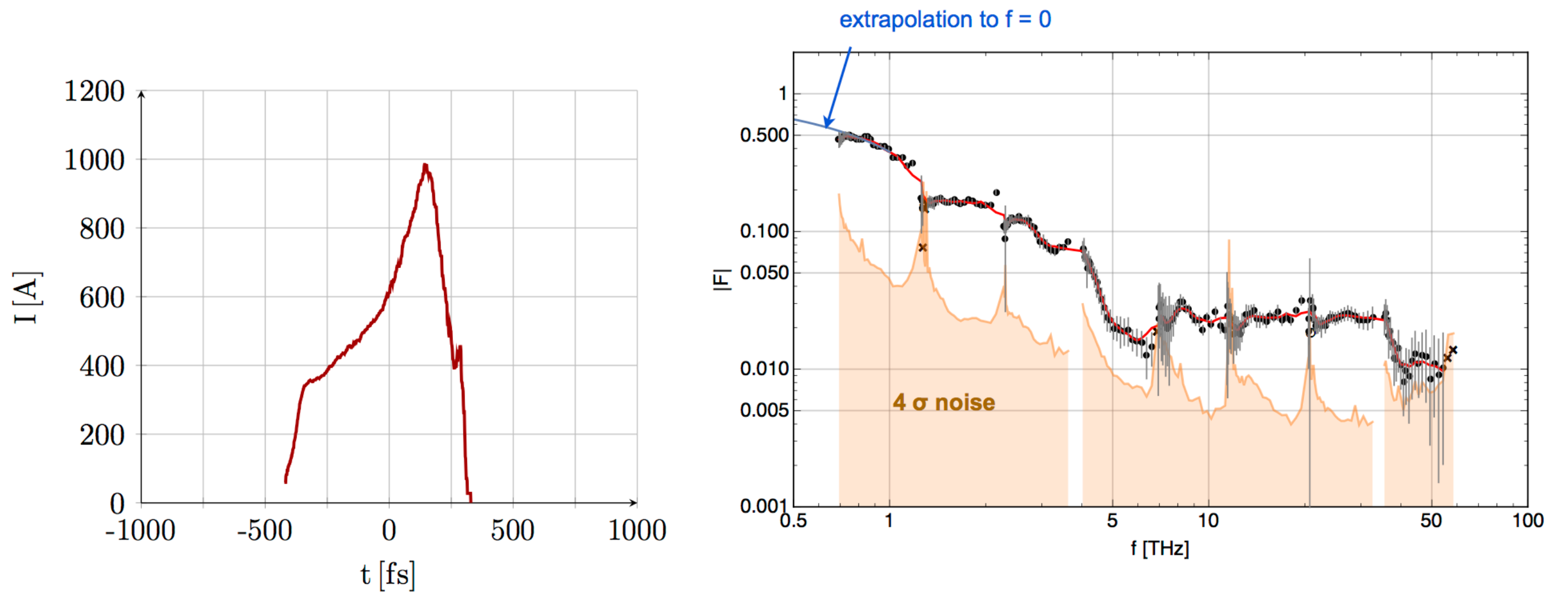} 
		\caption{ \small{A  measured TDS time profile and the corresponding  form factor as determined from the spectroscopic data.  $F(\omega)=|{\cal F}(\omega)|$  is plotted  as a function of $f=\omega/2\pi$.  The smooth red curve is a spline-interpolation which suppresses strong  fluctuations in the 5-10 THz region and above 40 THz. The grey vertical lines indicate the  shot-to-shot fluctuations in the sequence of 200 bunches. The shaded area shows the 4\,$\sigma$ limit of the noise signal. 
The blue curve is the extrapolation to $f=0$. Note that the frequency axis is logarithmic, hence the condition $F(0)=1$ cannot be seen in this graph, but it  is indeed fulfilled.}
		}				
\label{TDS-FFSignal-1}	
		\end{figure}

\noindent Several data processing steps have to be applied which are illustrated in Fig.\,\ref{TDS-FFSignal-1} where an experimentally determined  form factor  is shown as a function of frequency. \\
 a) Only ``significant'' data points are retained  with a voltage signal
 at least 4 standard deviations above noise. The noise level is determined by measurements without electron beam.  \\
 b) Points with a strong excursion are removed. \\
 c) The data are extrapolated  into the  frequency  range not covered by the spectrometer. The low-frequency extrapolation is made with a Chebyshev polynomial of second order (parabola) and the basic condition $F(\omega) \rightarrow 1$ for $f=\omega/2\pi \rightarrow 0$ is imposed as a constraint. 
It is very reassuring that the form factors can be smoothly extrapolated to $F(0)=1$ without any  scaling factors. This means that the absolute calibration of the {\it System} (TR screen - CTR beamline - spectrometer) is known to a level of about $20\%$.  \vspace{1mm}

\noindent  The extrapolation to high frequencies is uncritical since the measured form factor at the highest frequencies is normally so small that it has very little influence on the  bunch shape reconstruction. It can be done with an inverse power law.

\subsection{Bunch shape reconstruction methods}\label{reco-iter}
The Kramers-Kronig phase is computed by numerical integration of   formula (\ref{KK-phase-sect2}) with a cutoff frequency of 
$\omega_{\rm cut}=2 \pi\cdot 300\,$THz. For this purpose, the measured form factor values are interpolated by a spline function and extrapolated to high and low frequencies as mentioned above.
The inverse Fourier integral is also evaluated by numerical integration. In both cases, we use a Gauss-Kronrod integration scheme with locally adaptive subintervals. \vspace{2mm}

\noindent In the iterative phase reconstruction algorithm, the experimentally determined magnitude $F(\omega)$ of the   form factor,  including the extrapolations to low and high frequency,  is evaluated at the discrete points of a frequency grid with 4000 grid points.  The limiting parameter sets  used are\vspace{1mm}

Set 1: $-100\,$THz$\,\le f \le +100\,$THz,   step width  50\,GHz,   time range 10\,ps with 5\,fs resolution\\
\indent Set 2: $-600\,$THz$\,\le f \le +600\,$THz,  step width  300\,GHz,   time range  1.6\,ps  with 0.8\,fs resolution\vspace{1mm}

\noindent or a suitable intermediate set. 
The frequency range and step width of the discretization are adapted to the expected width of the time profile. The condition is imposed that there be at least 200 points within the FWHM region, otherwise  the maximum frequency and bin width of the grid are adjusted appropriately. The Fourier transformations of the iterative loop are done by an FFT algorithm, the inverse  Fourier transformation by IFFT.\vspace{1mm}

The following steps depend on the choice of the initial phases. If one starts with either a constant phase or the KK phase, the procedure is straightforward. The phase factors $\exp(-i\,\Phi(\omega_j)t)$ are evaluated  at the discrete grid frequencies $\omega_j=2\pi f_j$ and then the IFFT is carried out. For the case of ``random'' start phases some more effort is needed. Following a proposal by Fienup \cite{Fienup}, the computational steps  are as follows:\\
1) In step 1 all phases are put to zero and the IFFT is applied to the $F_j=F(\omega_j)$. This yields a symmetric time profile whose maximum  is then normalized to 1.   \\
2) In step 2 a threshold of 0.2 is imposed. All points of the time distribution whose values are below this threshold are put to zero. The points above threshold are replaced by random  numbers between zero and one.  \\
3) In step 3 an FFT is applied to the discrete time distribution generated in step 2.  The resulting phases $\Phi_j$ are ``quasi-random'' (since their origin is a randomized time distribution), and these phases are used for starting the Gerchberg-Saxton loop.  This method of generating  randomized start-phases avoids the unphysical procedure of assigning a completely random start phase to each frequency point,  and moreover it improves the speed of convergence.

The progress of convergence of the iteration loop is monitored by comparing the modulus of the reconstructed form factor with the measured values. The iteration is assumed to have converged if  the Pearson correlation coefficient deviates from 1 by less than $10^{-4}$ (or if the change  between subsequent iterations is below $10^{-4}$). \vspace{2mm}

\noindent The iterative phase retrieval procedure has several  variants which can be used to improve the convergence, speed and stability of the result \cite{Fienup}. We have studied these variants in great detail but this  is outside the scope of this paper and will be presented in a dedicated publication.
We just mention that  procedures like  ``shrink wrapping''  or  ``bubble wrapping'' , promoted in the past \cite{Marchesini}, did not  improve the results.\vspace{2mm}

\noindent To mitigate  fluctuations in the time profile, resulting from the randomness of the start phases, we follow a procedure proposed in \cite{Pelliccia}. The iterative loop is started about 50 times with new quasi-random phases and the resulting profiles are averaged. This averaging has to be done with care since the reconstructed  time profiles $\rho_j(t)$ ($j=1 \ldots 50$) will have arbitrary time shifts with respect to each other and sign-reversals of the time direction will happen (see Fig.\,\ref{time-revers-shift}). These ambiguities must be removed before averaging. To this end one optimizes the correlation coefficient between any two $\rho_i(t)$, $\rho_j(t+\delta t)$  by varying the time offset $\delta t$ and by trying if time reversal $\rho_j(t) \rightarrow \rho_j(-t)$ improves the agreement.

 Averaging is a  means to identify significant structures. Structures with a high likelihood will appear repeatedly  in many  iteration cycles and  survive the averaging while those with a low likelihood  will fluctuate  from one iteration cycle to the next   and average to zero.

\subsection{Experimental results}
\begin{figure}[ht!]
		\centering 
\includegraphics[width=10cm]{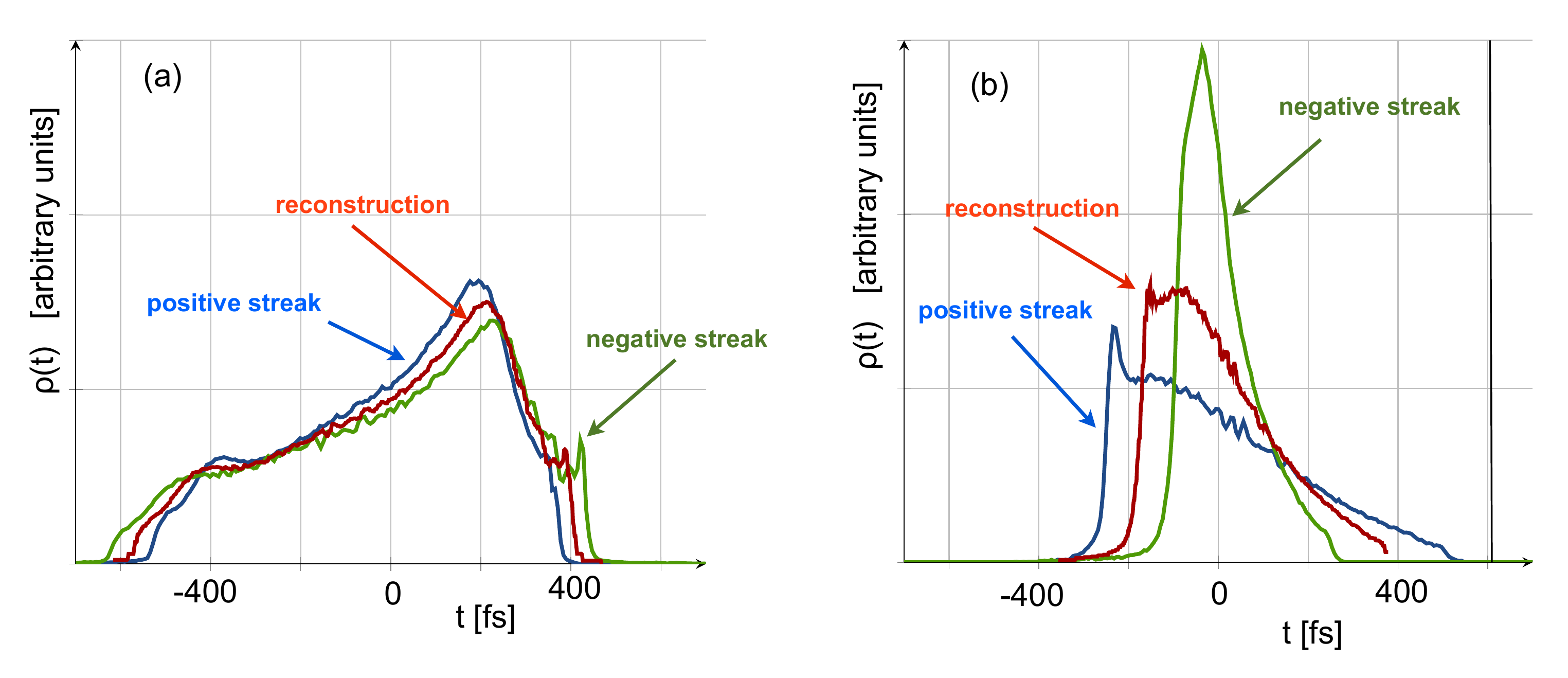} 
		\caption{ \small{Time profiles measured with the TDS. The positive streak direction  is shown by the blue curve, the  negative streak direction by the green curve,  and the  reconstructed time profile by the red curve.
	 {\bf (a)} Bunch with moderate compression.  {\bf (b)} Bunch with strong compression and a steep initial rise. }
		}				
\label{TDSstreak}	
		\end{figure}

Many different bunch shapes can be realized  at  FLASH by  varying the off-crest phase of the 1.3\,GHz  RF field in the accelerating cavities preceding the bunch compressor, and by choosing appropriate values for the amplitude and phase of the 3.9\,GHz  RF field in the third-harmonic cavity. Here we present  four examples of longitudinal bunch shape reconstruction from spectroscopic measurements. The computed time profiles are compared with the time profiles recorded with the transversely deflecting microwave structure TDS \cite{TDS2}, whenever available.   The TDS is basically an ultrafast oscilloscope tube with a resolution in the 10\,fs regime.  In principle it is a single-shot device that should be  suited to faithfully record the longitudinal particle density $\rho(t)$ of a single electron bunch. In practice this is not the case for the bunches having passed the magnetic bunch compressor chicanes since these  might have acquired a nonvanishing average transverse momentum and other internal correlations that vary along the bunch axis. As a consequence, the streak image  on the TDS view screen depends on the streak direction, see Fig.\,\ref{TDSstreak}.  

This effect can be taken care of by streaking a first bunch in positive  direction and a second bunch in negative direction.   The most likely bunch profile is obtained   from these data by a tomographic reconstruction algorithm developed at SLAC  (unpublished). The TDS time profiles shown in the following figures result from the  tomographic reconstruction. The difference between the two streak directions is small for  wide bunches with little structure but becomes pronounced for strongly compressed bunches featuring sharp structures \cite{Minjie}. Presently systematic studies are being carried out which will be reported elsewhere.

 \noindent $\underline{\it Example~1}$\\ 
 \begin{figure}[htb!]
		\centering 
\includegraphics[width=14cm]{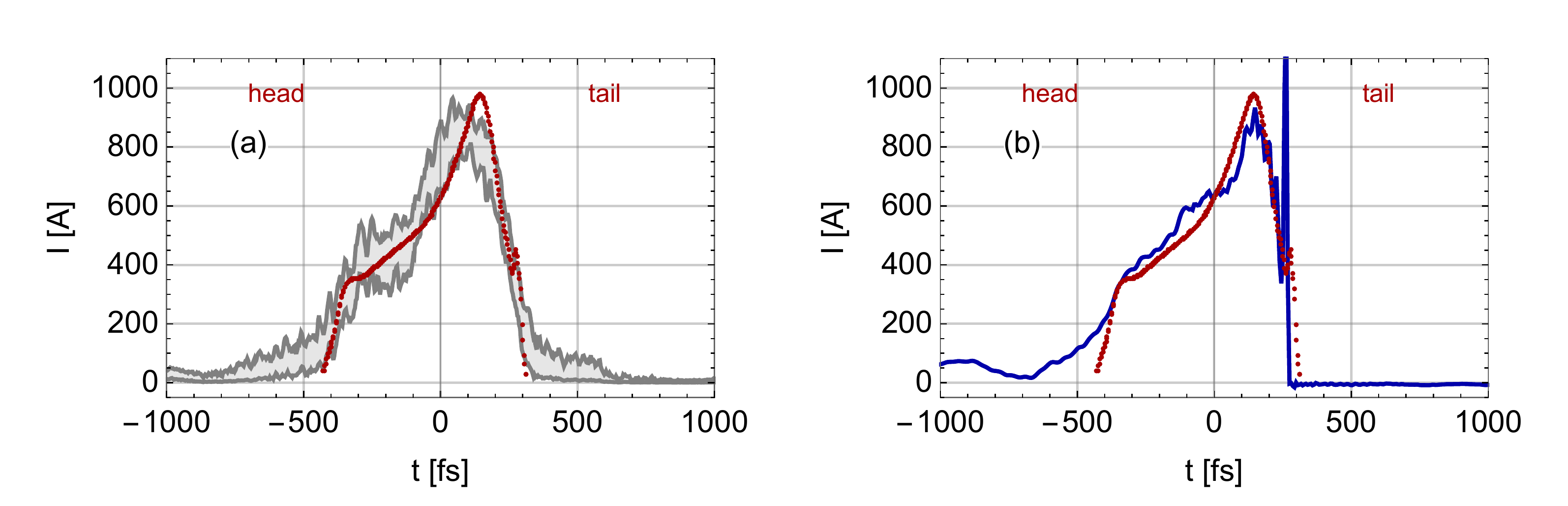} 
		\caption{ \small{Reconstructed shapes of the type-1 bunches. Parameters: electron energy $E_e=700\,$MeV, bunch charge $Q=390\,$pC, TDS resolution $\sigma_{\rm \small{TDS}}=29\,$fs. Red curves:  TDS measurement.  {\bf (a)} Iterative phase retrieval with quasi-random initial phases.  {\bf (b)} Analytic phase retrieval using the Kramers-Kronig phase. }
		}				
\label{Signal1}	
		\end{figure}   
In this example bunches were generated with a  steep decay in the tail region. 	 The  TDS profile and the longitudinal form factor have already been shown in Fig.\,\ref{TDS-FFSignal-1}. 
The iterative reconstruction  with quasi-random start phases yields time profiles which are in good agreement with the TDS profile, see Fig.\,\ref{Signal1}.
However, the analytic KK reconstruction  generates a profile with an extremely sharp spike in the tail region.  The TDS resolution is insufficient to decide whether this spike is real or an artefact.  \vspace{3mm}

\begin{figure}[htb!]
		\centering
		\includegraphics[width=15cm]{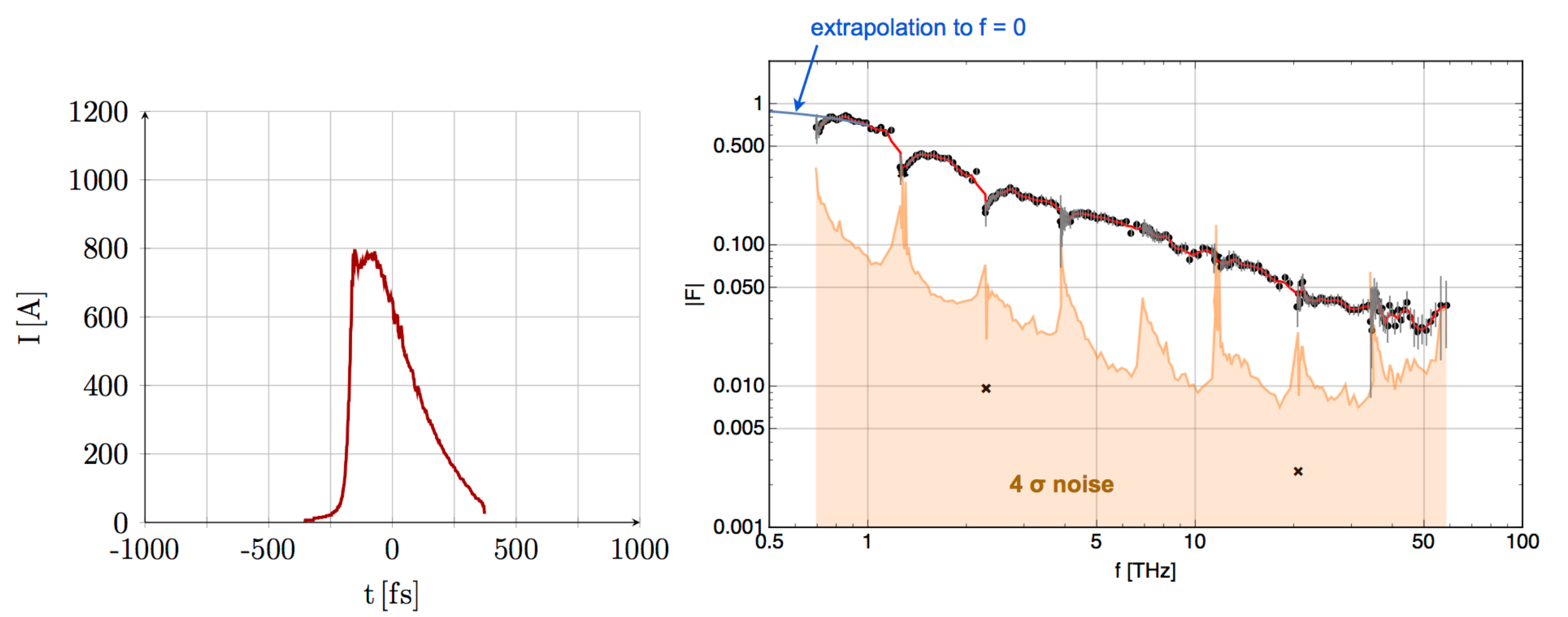} 
\caption{ \small{TDS time profile and  measured form factor $F(\omega)$ of type-2 bunches  (gray dots), plotted  as a function of $f=\omega/2\pi$, and the smooth spline-interpolation (red curve) which suppresses strong local fluctuations around 50 THz. The shaded area shows the 4\,$\sigma$ limit of the noise signal. 
Also shown is the extrapolation to $f=0$ (blue line). }}
\label{TDS-FFSignal2}	
		\end{figure}

\noindent $\underline{\it Example~2}$\\   
 In  another measurement series   bunches were produced with a  steep rise and a roughly exponential decay, resembling Akutowicz's function $f_1(t)$. 
 The  TDS time profile and the form factor are shown in Fig.\,\ref{TDS-FFSignal2}. On a logarithmic scale, $F(\omega)$ exhibits a smooth drop over 1.5 orders of magnitude. The form factor is well above the $4\,\sigma$ limit of the noise and is measurable up to about 40\,THz.  The data in the 40 - 50  THz region have large point-to-point fluctuations and are of little use in signal reconstruction.
 \begin{figure}[htb!]
		\centering 
\includegraphics[width=14cm]{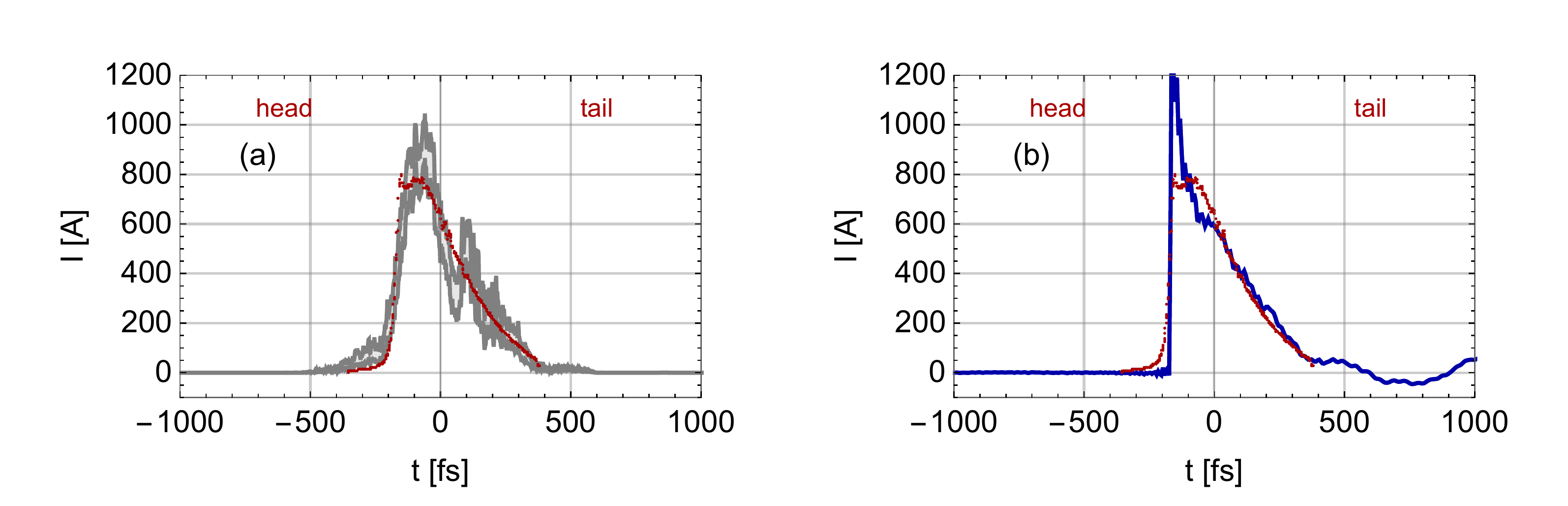} 
		\caption{ \small{Reconstructed shapes  of bunches  with a  steep rise and a roughly exponential decay. Parameters: electron energy $E_e=540\,$MeV, bunch charge $Q=240\,$pC, TDS resolution $\sigma_{\rm \small{TDS}}=49\,$fs. The bunch current $I(t)$ is plotted as a function of time. The bunch head is at the left side (early times). Red curves:  TDS measurement (compare Fig.\,\ref{TDSstreak}b).  {\bf (a)} Iterative phase retrieval with quasi-random initial phases.  {\bf (b)} Analytic phase retrieval using the Kramers-Kronig phase.}
		}				
\label{Signal2}	
		\end{figure}
 The reconstructed time profiles $I(t)$ of the bunch current are presented in Fig.\,\ref{Signal2}. The TDS profile (red curve)	reveals that the steep rise occurs at the bunch head (early times) while the exponential drop is in the tail region. 
The iterative reconstruction  with quasi-random start phases yields time profiles with a fairly large uncertainty band, and the steep rise is washed out. This is  a general observation:  the iterative method with random start-phases tends to wash out sharp structures, especially after averaging over many repetitions. 
 The analytic reconstruction with the KK phase works very well, which may not be a  surprise in view of the excellent KK reconstruction of the 
 Akutowic  function $f_1(t)$ (Fig.\,\ref{Gauss-Cosine}a).  The initial rise is even steeper than in the TDS time profile.

\vspace{10mm}

\begin{figure}[ht!]
		\centering 
\includegraphics[width=15cm]{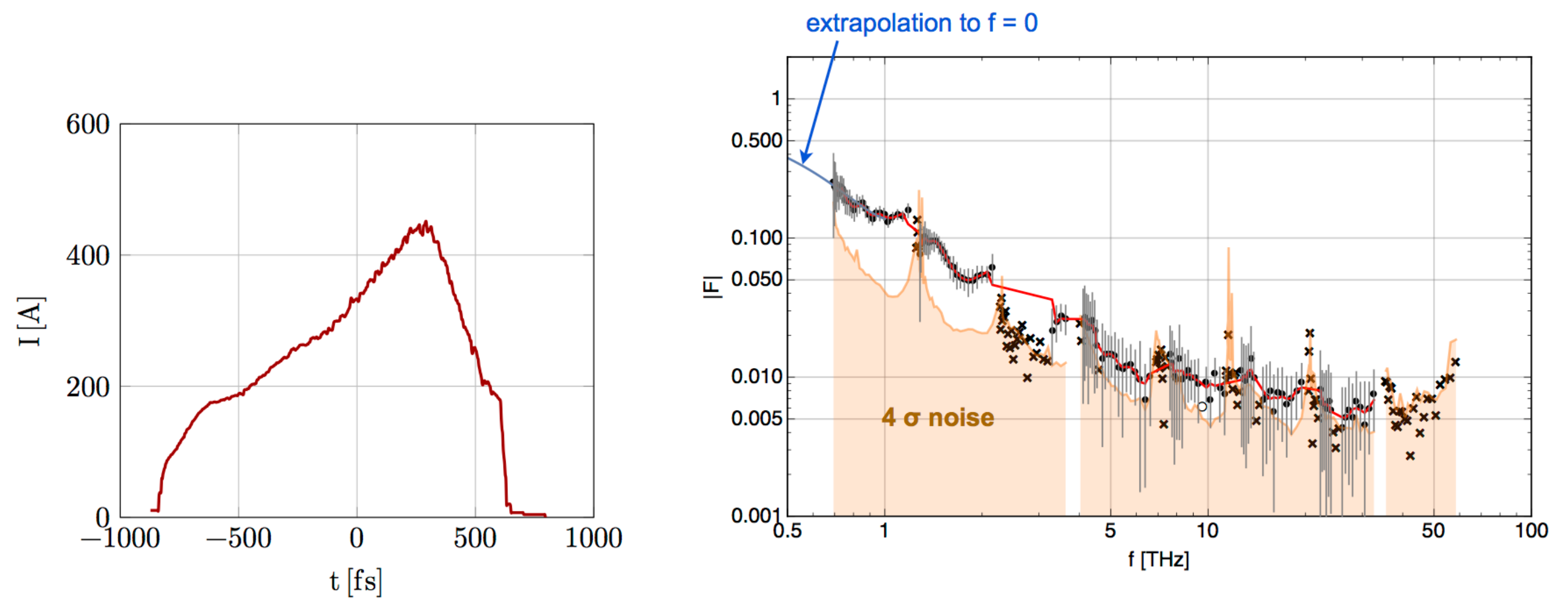} 
		\caption{ \small{ TDS time profile  and form factor of bunch type 3. Remark on the extrapolation to $f=0$ (blue curve):   the frequency axis is logarithmic and the minimum frequency in this figure is $f_{\rm min}=0.5\,$THz. The grey vertical lines,  which become very  pronounced above 5 THz, indicate the  shot-to-shot fluctuations of the measurement. }
		}				
\label{TDS-FFSignal3}	
		\end{figure}

\noindent $\underline{\it Example~3}$\\ 
Example 3 illustrates the response to a rather long bunch. 
 The measured form factor (Fig.\,\ref{TDS-FFSignal3}) drops rapidly with frequency, falling below 0.1 slightly above 1 THz. In the IR regime (above 5 THz) it oscillates rather strongly from shot to shot and hardly exceeds the sensitivity limit. Nevertheless, the reconstruction of the bunch profile is possible with both the Kramers-Kronig and the iterative method. The overall shape and bunch length are again in very good agreement  with the TDS measurement. As expected, the Kramers-Kronig phase compiles the high frequency content to sharp structures at the end of the bunch tail while the iterative reconstruction results in a  smoother average profile. 
The total bunch length of about 1 ps reaches the upper limit for this bunch charge. 
		\begin{figure}[ht!]
		\centering 
\includegraphics[width=14cm]{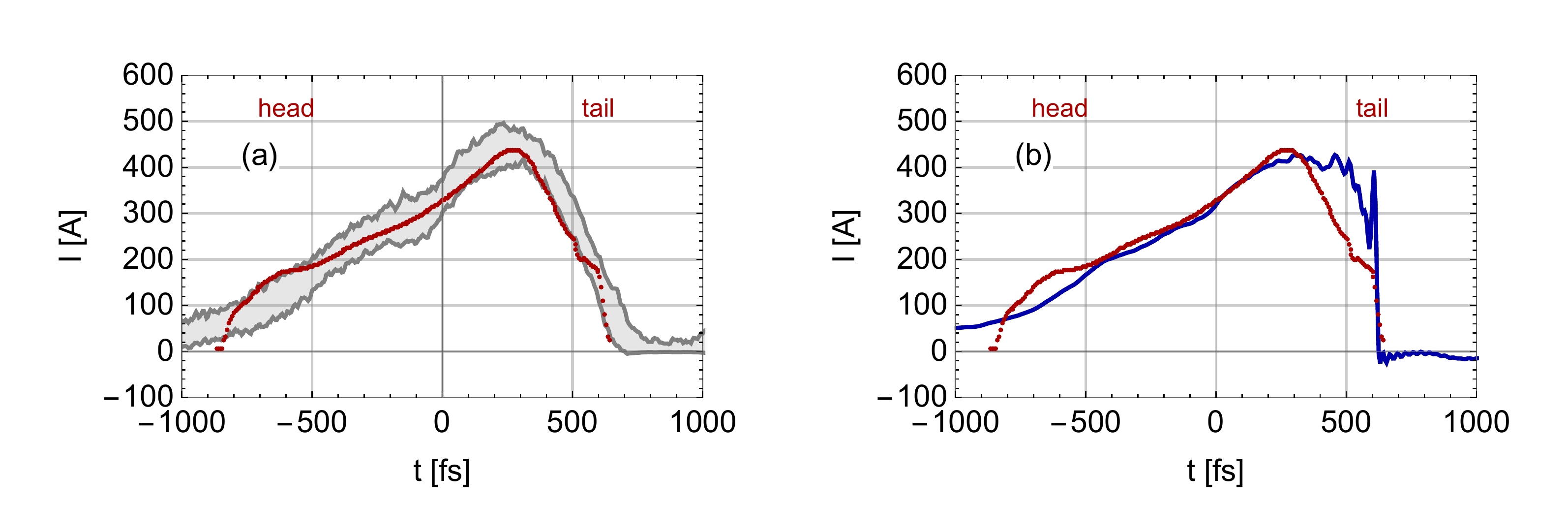} 
		\caption{ \small{Reconstructed shapes  of  bunch type 3. Parameters: electron energy $E_e=710\,$MeV, bunch charge $Q=400\,$pC, TDS resolution $\sigma_{\rm \small{TDS}}=45\,$fs. Red curves:  TDS measurement.  {\bf (a)} Iterative phase retrieval with quasi-random initial phases.  {\bf (b)} Analytic phase retrieval using the Kramers-Kronig phase. }
		}				
\label{Signal3}	
		\end{figure}

\begin{figure}[ht!]
		\centering 
\includegraphics[width=13cm]{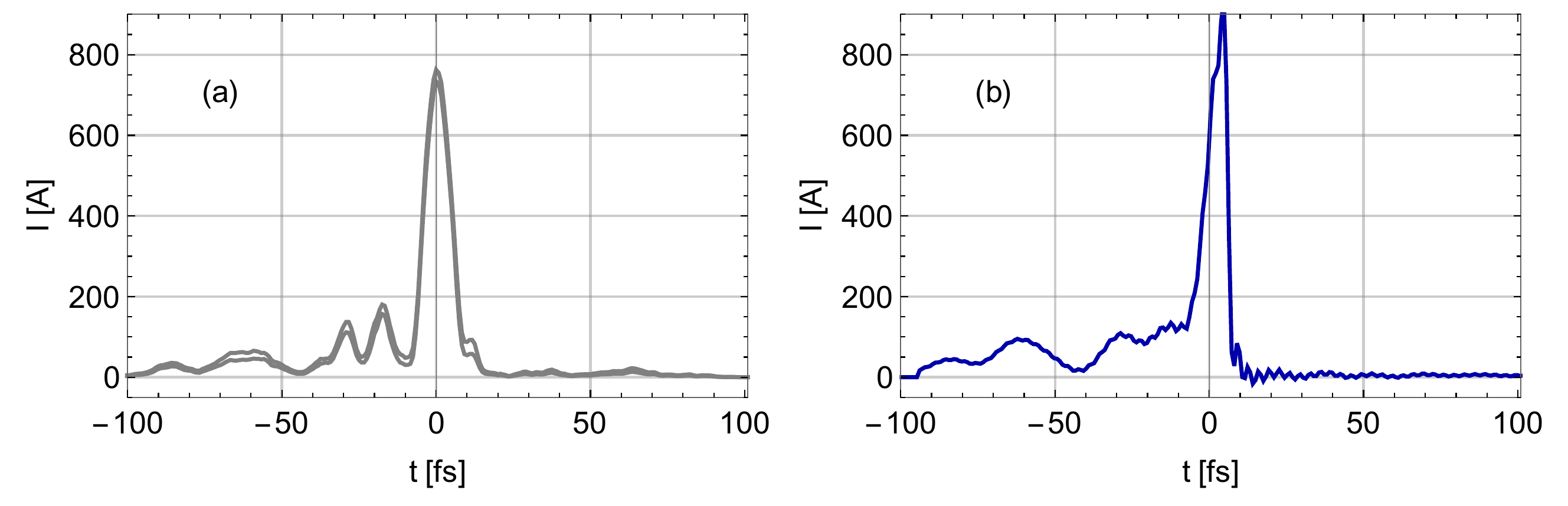}
		\caption{ \small{Reconstructed shapes of an ultrashort bunch (type 4). The full width at half maximum is 8 fs. This narrow peak cannot be resolved by the 2.86 GHz  TDS that is presently installed at FLASH.  {\bf (a)} Iterative phase retrieval with quasi-random initial phases.  {\bf (b)} Analytic phase retrieval using the Kramers-Kronig phase.}
		}				
\label{Signal4}	
		\end{figure}

\begin{figure}[ht!]
		\centering 
\includegraphics[width=8cm]{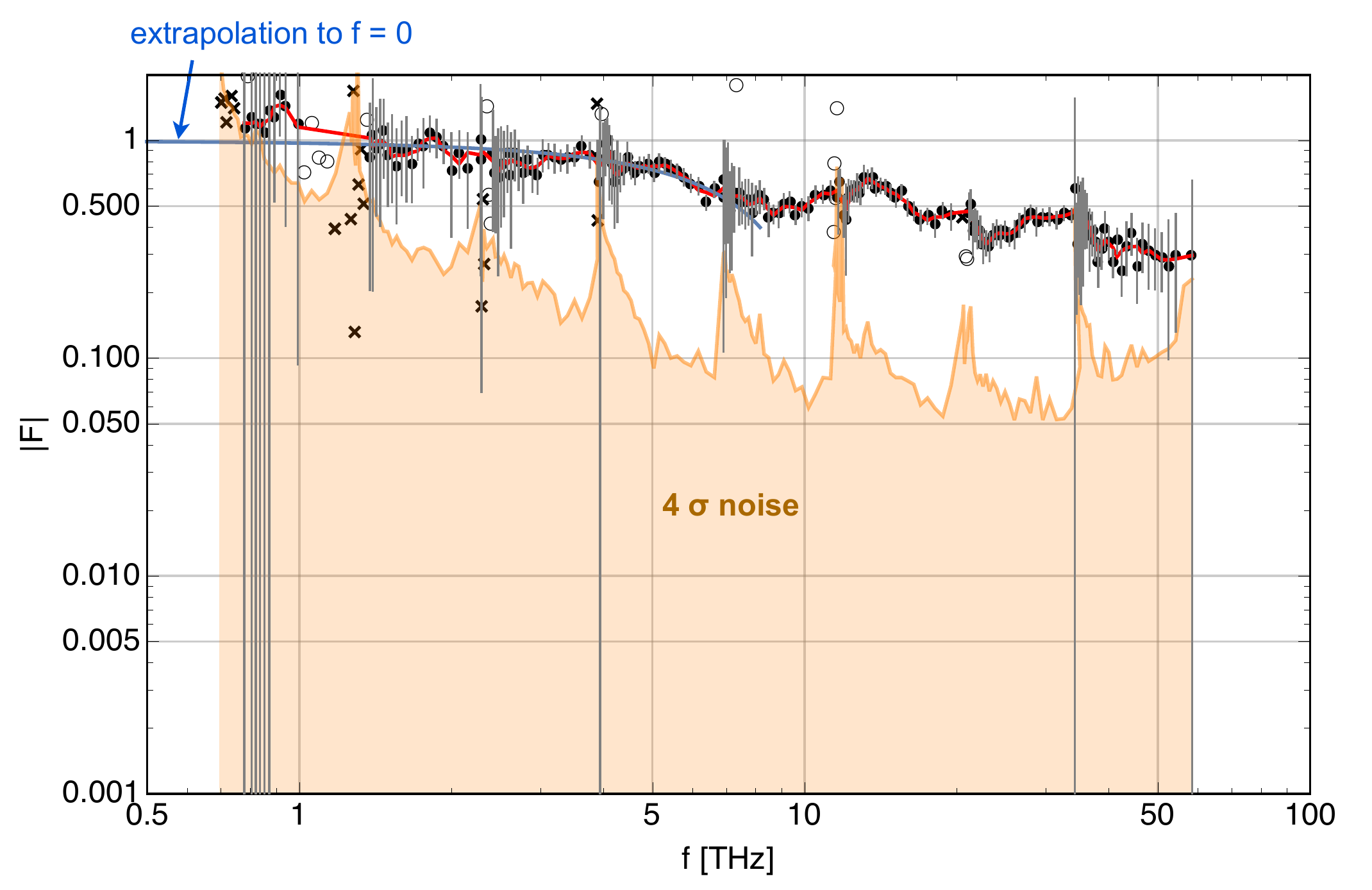} 
		\caption{ \small{The form factor of the ultrashort bunch drops very slowly with frequency.}
		}				
\label{FFSignal4}	
		\end{figure}

\noindent $\underline{\it Example~4}$\\ 
Example 4 finally demonstrates the possibility to measure extremely short bunches of very low charge. The total charge was only 14 pC  but the bunch was highly compressed. The electron energy was $E_e=706\,$MeV.
The form factor (Fig.\,\ref{FFSignal4}) falls off very slowly, by just a factor of two between 1 THz and 40 THz. It is still well above 0.1 for the highest measured frequency of 60 THz. The reconstructed bunch profiles (Fig.\,\ref{Signal4}) show a narrow peak with a FWHM of about 8 fs. The two reconstruction algorithms are in good agreement. A comparison with the TDS is meaningless in this case since the  bunch length is less than the resolution limit of the TDS (about 10 fs) even with optimized accelerator  optics. 
\vspace{10mm}		

\begin{figure}[htb!]
		\centering 
\includegraphics[width=16cm]{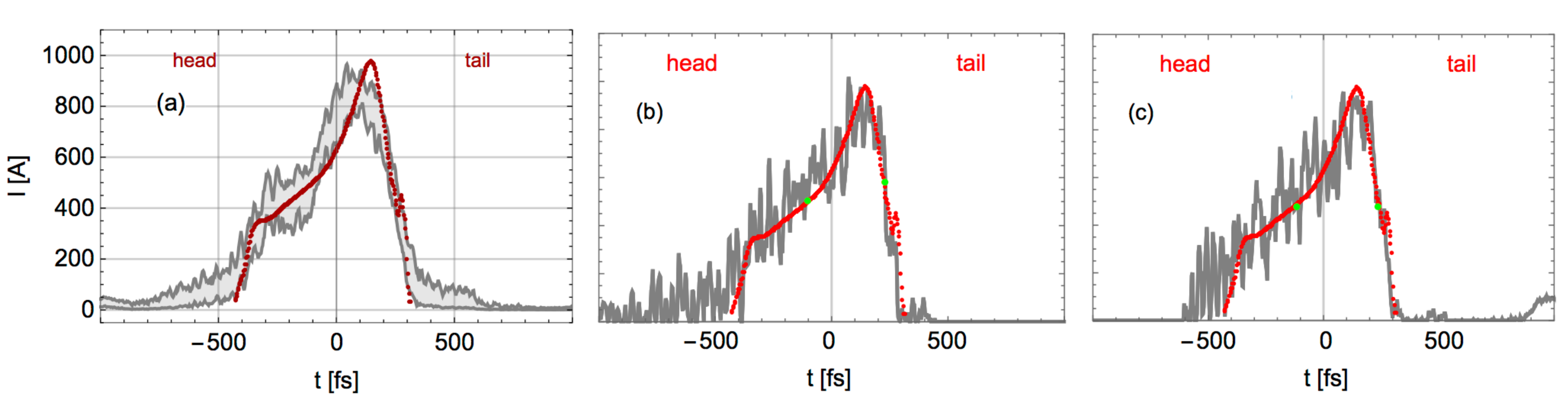} 
		\caption{ \small{Reconstructed shapes of the type-1 bunches, starting with quasi-random phases.  Red curves:  TDS measurement.  {\bf (a)} Average profile of 100 iteration cycles.  {\bf (b,\,c)} Two single-cycle profiles. The rear slope is definitely steeper in the single-cycle cases and agrees well with the slope of the TDS profile.}
		}				
\label{Signal1-av-single}	
		\end{figure}

\noindent {\bf Impact of averaging}\\
The averaging method, applied by us in the iterative reconstruction starting with quasi-random phases, has a tendency to smear out steep slopes. This is illustrated in Fig.\,\ref{Signal1-av-single}	where we compare the averaged profile of type-1 bunches with two single-cycle iterations.

\subsection{Summary and conclusions}
Our four examples demonstrate the great potential of broadband CTR spectroscopy for longitudinal bunch diagnostics. The bunch shapes that have been reconstructed from the measured form factors are in good agreement with the shapes determined  by the time-domain instrument TDS. The inverse Fourier transformation using the Kramers-Kronig phase yields  good results but the same is true for the iterative method starting   either from the KK phase or from randomized phases. The ambiguities which occur frequently  in the model calculations are not observed here, probably because the investigated bunches feature essentially only a single peak. A remarkable result is that the spectroscopic method enables the resolution of very short time structures (below 10\,fs) where the present  2.86\,GHz TDS meets its resolution limit. This is evident in Fig.\,\ref{Signal4}. 	

\noindent We have shown that the frequency-domain technique of coherent transition radiation spectroscopy is highly competitive with 
 high-resolution time-domain techniques, provided a broadband multi-channel spectrometer with single-shot capability is used. Both methods complement each other and their combination is vital for obtaining faithful results. Dedicated studies are underway to exploit these possibilities.\vspace{30mm}

\noindent {\bf Acknowledgments}\\
We are indebted to  A.F.G. van der Meer for drawing our attention to the staged grating concept and to Hossein Delsim-Hashemi for his invaluable contributions to the design and commissioning of a prototype multistage spectrometer. We thank   Kai Ludwig and Bernd Beyer  for their important contributions to the spectrometer and the CTR beamline.  The support of the FLASH team  during our experimental studies is gratefully acknowledged.   Thanks are due to P. Smirnov, P. G\"ottlicher, M. Hoffmann, A. Schleiermacher, P. Pototzki and V. Rybnikov for help and advice.

\clearpage

\section{Appendix A: Dispersion Relations}\label{AppendixA}

A dispersion relation is an integral formula relating a dispersive process to an absorptive process. 
An example is the relation between  the refractive index $n(\omega)$ of an optical medium and its extinction coefficient $k(\omega)$, or the relation between real part and imaginary part of the dielectric function  of a solid. 
Dispersion relations follow rigorously from causality. In this appendix we follow closely the book {\it Optical Properties of Solids} by F. Wooten \cite{Wooten} but we go into more detail and present several mathematical proofs that are missing in \cite{Wooten}. A comprehensive treatment can be found in \cite{Toll}. 

\subsection{Basics of complex analysis}\label{complex-analysis}
{\bf Cauchy-Riemann  equations}\\
The set of complex numbers $z=x+iy$ is called ${\bf  C}$. These numbers can be depicted as points in a plane $(x, iy)$. A function $f(z)$ is called analytic (or holomorphic)  in an open subset $U$ of ${\bf  C}$
if the differential quotient exists
\begin{equation}
\frac{df}{dz}=\lim_{\Delta z\rightarrow 0} \frac{f(z+\Delta z)-f(z)}{\Delta z}~~~\mathrm{for~all}~~z\in U\,.
\end{equation} 
The differential quotient  is defined in the same way as in real analysis, but the limit can be  approached from many different directions. This has far-reaching consequences.
Separating the complex function into its real and imaginary parts
\begin{equation}
f(z)=f(x+iy)=u(x,y)+iv(x,y)
\end{equation} 
one can prove that $f(z)$ is analytic if and only if the 
  {\bf Cauchy-Riemann differential equations} are fulfilled
 \begin{equation}
 \frac{\partial u}{\partial x}=\frac{\partial v}{\partial y} \,,~~~~~~\frac{\partial u}{\partial y}=-\frac{\partial v}{\partial x} \,.
 \end{equation}
 Analytic functions have many remarkable properties that are not valid for real functions. For example the derivative of an analytic function is again analytic  which means that analytic functions are infinitely often differentiable.
\vspace{2mm}

\noindent {\bf Cauchy Integral Theorem and Residue Theorem}\\
The Cauchy integral theorem  is an important statement about line integrals in the complex plane ${\bf  C}$. If two different paths connect the same start and end points, and if the function is analytic in an open set containing the two paths, then these two path integrals of the function yield the same value. The theorem is usually formulated for closed paths: \\ Let $U$ be an open subset of ${\bf  C}$ which is simply connected, let $f : U \rightarrow  {\bf  C}$ be an analytic function, and let 
$\Gamma$ be a closed loop. Then the line integral over the closed loop vanishes		 		
\begin{equation}\label{Cauchy}
\oint_\Gamma f(z)dz=0 ~~~~~~~~~~~~Cauchy~Integral~Theorem\,.
\end{equation}

\noindent Consider now the  function $g(z)=f(z)/(z-z_0)$  where $z_0$ is an arbitrary point  inside the closed loop $\Gamma$. The function $g(z)$ is analytic except for a small vicinity around the pole at $z_0$.  
The Residue Theorem states
\begin{equation}\label{residue}
\oint_\Gamma \frac{f(z)}{z-z_0}dz=2\pi\,i f(z_0) \equiv 2\pi\,i \,\mathrm{Res}(f,z_0)~~~Residue~Theorem\,.~
\end{equation}
This is easy to verify for a small circle centered at $z_0$ whose radius $a$ tends to zero. On the circle we have
\begin{eqnarray}
z&=&z_0+a e^{i \theta}, ~~dz=i\,a e^{i  \theta} d \theta \nonumber \\
\lim_{a \rightarrow 0}\oint_{\rm circ} \frac{f(z)}{z-z_0}dz&= & f(z_0) \lim_{a \rightarrow 0} \int_0^{2\pi} \frac{1}{a e^{i  \theta}}\,i\,a e^{i  \theta} d \theta =2\pi\,i f(z_0)\,.  \nonumber
\end{eqnarray}\vspace{2mm}

\begin{figure}[t]
		\centering
		\includegraphics[width=8cm]{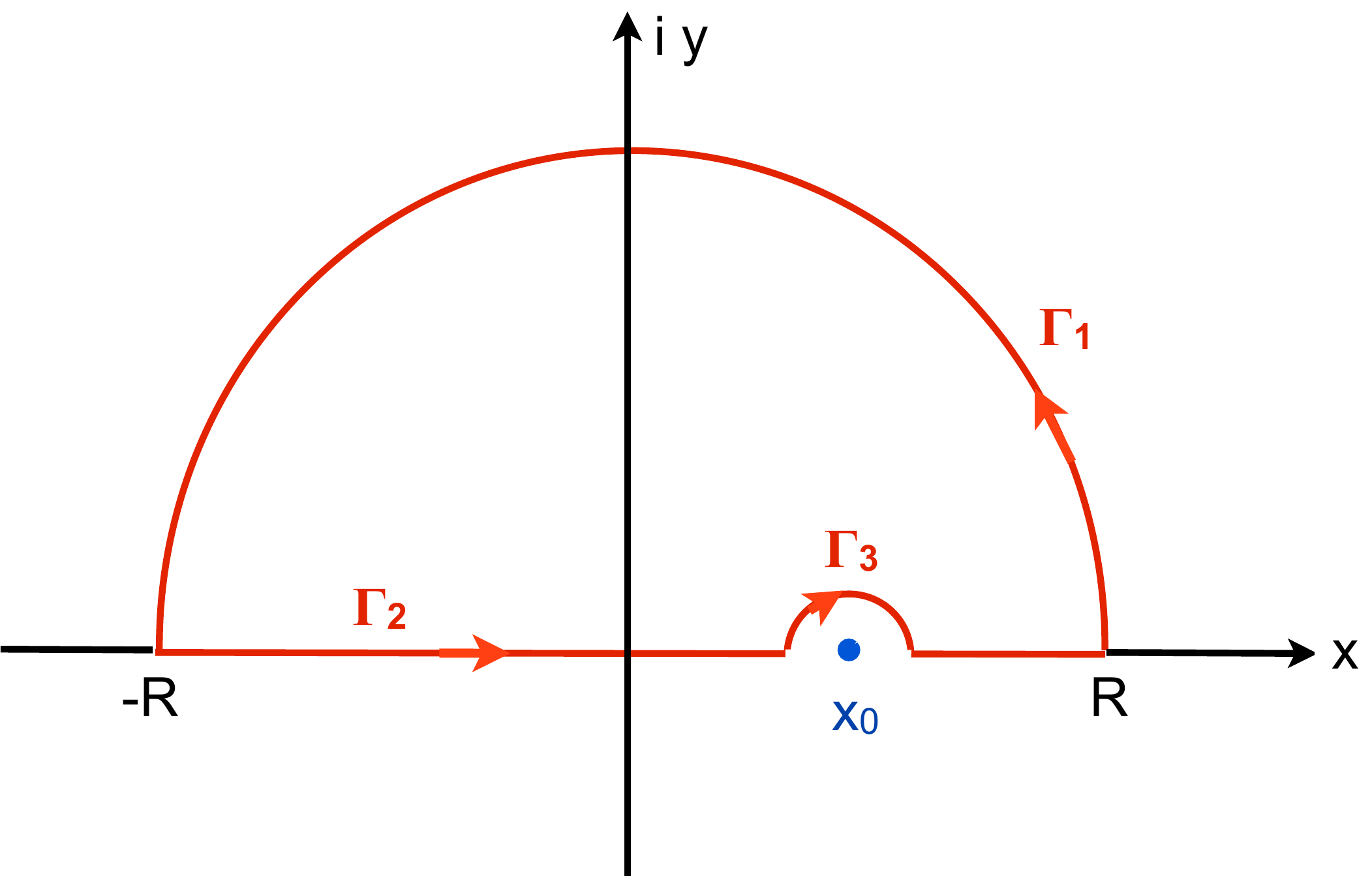} 
		\caption{\small{The closed integration path $\Gamma$. }	}				
		\label{Gamma-path}
		\end{figure}

\noindent {\bf Computation of  an important  integral}\\
In the next section \ref{response-function} an integral of the type
$$\int_{-\infty}^\infty \frac{e^{ix}}{x-x_0}dx$$
appears. The integrand has a singularity at $x=x_0$. We want to show how such  integrals along the real axis can be evaluated by going into the complex plane and using the Cauchy and Residue Theorems.
For this purpose we consider a closed integration path $\Gamma$ consisting of three parts  (see Fig.\,\ref{Gamma-path}): \\(1) A large semicircle $\Gamma_1$ of radius $R$ which is centered at the origin $z=0$,\\
 (2) a straight line $\Gamma_2$ along the real axis from $x=-R$ to $x=x_0-\varepsilon$ and from $x=x_0+\varepsilon$ to $x=+R$,
 \\ (3) a small semicircle $\Gamma_3$ of radius $\varepsilon$ which is centered at $x_0$.
 \vspace{2mm}

\noindent {\bf Step 1}
Consider a function $f(z)$ which  is analytic in the upper half plane $y \ge 0$, except for a finite number of poles, and which vanishes asymptotically
\begin{equation}\label{asymptotic-zero}
\lim_{|z| \rightarrow   \infty} f(z)=0~~~~\mathrm{for}~~~y \ge 0\,.
\end{equation}

\noindent {\it Statement\,}: The line integral of the function $f(z)e^{iz}$ along the semicircle $\Gamma_1$ tends to zero in the limit $R \rightarrow \infty$:
\begin{equation}\label{semicircle1}
\lim_{R\rightarrow \infty} \int_{\Gamma_1} f(z) e^{iz} dz=0 \,.
\end{equation}
This statement is by no means obvious. Because of (\ref{asymptotic-zero}), the integrand  tends to zero with increasing radius of the semicircle,
 $|f(z)e^{iz}| \rightarrow 0$ for $|z| \rightarrow \infty$ and $y>0$,  but at the same time the path length of the semicircle tends to infinity. \vspace{2mm}
 
 \noindent {\it Proof\,}: The proof of (\ref{semicircle1}) goes as follows (see H. Cartan \cite{Cartan}).
 On the semicircle we have
 $$z=R e^{i \theta}\,,~~dz=i R e^{i \theta} d\theta\,,~~~|e^{iz}|=e^{-R \sin {\theta}} \,.$$
 An upper limit for the integral is 
 $$\left| \int_{\Gamma_1} f(z) e^{iz} dz \right| \le  \int_0^\pi  |f(R e^{i \theta} )| e^{-R \sin {\theta}} R d\theta \le M(R) \int_0^\pi e^{-R \sin {\theta}} R d\theta
 = 2 M(R) \int_0^{\pi/2} e^{-R \sin {\theta}} R d\theta $$
with  
$$M(R)=\max_{0 \le \theta \le \pi}(|f(R e^{i \theta})|)\,. $$
In the interval $0\le \theta \le \pi/2$ one has $\sin{\theta} \ge 2 \theta /\pi$, hence
$$\int_0^{\pi/2} e^{-R \sin {\theta}} R d\theta \le \int_0^{\pi/2} e^{-2\theta R/\pi} R d\theta=\pi/2 \,.$$ 
The result is
$$\left| \int_{\Gamma_1} f(z) e^{iz} dz \right|  \le \pi\, M(R)
~~~~~\mathrm{and}~~
\lim_{R\rightarrow \infty} M(R)=0\,.$$
Therefore
$$\lim_{R\rightarrow \infty} \int_{\Gamma_1} f(z) e^{iz} dz=0 ~~~~~~~~~\mathrm{qed}\,.$$
Specifically, the function $f(z)=1/(z-x_0)$ fulfills the condition (\ref{asymptotic-zero}). Hence
\begin{equation}\label{semicircle1b}
\lim_{R\rightarrow \infty} \int_{\Gamma_1} \frac{ e^{iz}}{z-x_0} dz=0 \,.
\end{equation}
 \vspace{3mm}

\noindent {\bf Step 2}
Next we want to evaluate the closed-loop  integral
$$\oint_\Gamma \frac{e^{iz} }{z-x_0} dz\,.$$
The function $1/(z-x_0)$ has a pole at $z=x_0$ but is analytic elsewhere. Next we show that 
 $e^{i  z}$ is analytic in the entire complex plane. We prove this for the more general case
$e^{i  z \tau }$ where $\tau$ is a real number. For this purpose we write
$$e^{i z \tau}=e^{i (x+i\,y) \tau}=u(x,y)+i\,v(x,y)\,.$$
The real functions $u(x,y)$ and $v(x,y)$ are
$$u(x,y)=\cos(x \tau)\,e^{-y \tau},~~~~v(x,y)=\sin(x \tau)\,e^{-y \tau} \,.$$
It is easy to verify that  they fulfill the Cauchy-Riemann differential equations
$$
\frac{\partial u}{\partial x}=-\tau\,\sin(x \tau)\,e^{-y \tau}=\frac{\partial v}{\partial y} \,, ~~~~
\frac{\partial u}{\partial y}=-\tau\,\cos(x \tau)\,e^{-y \tau}=-\frac{\partial v}{\partial x} \,.$$
This proves that 
 $f(z)=e^{i z \tau}=e^{i (x+i\,y) \tau}=u(x,y)+i\,v(x,y)$
 is an analytic function of the complex variable $z=x+i\,y$.
\vspace{3mm}

\noindent
The pole at $x_0$ is avoided by choosing the closed integration path $\Gamma$ shown in Fig.\,\ref{Gamma-path}.
There is no singularity inside the loop, hence from Cauchy's formula (\ref{Cauchy})
$$
\oint_\Gamma \frac{e^{iz} }{z-x_0}  dz=\int_{\Gamma_1} \frac{e^{iz} }{z-x_0}  dz+
\int_{\Gamma_2} \frac{e^{iz} }{z-x_0} dz+\int_{\Gamma_3} \frac{e^{iz} }{z-x_0}  dz =0 \,.
$$
The three path integrals are evaluated in the limit $\varepsilon \rightarrow 0$ and $R \rightarrow \infty$. 

\noindent Because of (\ref{semicircle1b})  we get
$$\int_{\Gamma_2} \frac{e^{iz} }{z-x_0}  dz=-\int_{\Gamma_3} \frac{e^{iz} }{z-x_0} dz \,.$$
The integral over the small semicircle is readily evaluated
\begin{eqnarray}
z&=&x_0+\varepsilon \, e^{i \theta}, ~~dz=i\,\varepsilon\, e^{i \theta} d\theta \nonumber \\
\lim_{\varepsilon \rightarrow 0}\int_{\Gamma_3} \frac{e^{iz}}{z-x_0}dz&=& e^{i x_0} \lim_{\varepsilon \rightarrow 0} \int_{2\pi}^{\pi} \frac{1}{\varepsilon\, e^{i \theta}}\,i\,\varepsilon e^{i \theta} d\theta =-\pi\,i e^{i x_0}\,.  
\label{Gamma3-Integral}
\end{eqnarray}
The integral over the path $\Gamma_2$ is therefore in the limit $R\rightarrow \infty$ and $\varepsilon \rightarrow 0$
$$
\int_{\Gamma_2}  \frac{e^{iz}}{z-x_0}dz=\lim_{\varepsilon \rightarrow 0} \left[ \int_{-\infty}^{x_0-\varepsilon}\frac{e^{iz}}{z-x_0}dz +\int_{x_0+\varepsilon}^{\infty}\frac{e^{iz}}{z-x_0}dz\right]
 =i\,\pi\, e^{i x_0}\,.  
$$
The path $\Gamma_2$ is along the real axis, hence $z=x$ on $\Gamma_2$. It is customary to define the principal value of an integral, denoted by the letter ${\cal P}$,
 by approaching  the singularity at $x_0$  symmetrically from both sides:
 \begin{equation}
 {\cal P} \int_{-\infty}^{\infty}\frac{e^{ix}}{x-x_0}dx =\lim_{\varepsilon \rightarrow 0} \left[ \int_{-\infty}^{x_0-\varepsilon}\frac{e^{ix}}{x-x_0}dx +\int_{x_0+\varepsilon}^{\infty}\frac{e^{ix}}{x-x_0}dx\right]
 \label{princ-val-int}
 \end{equation}
 The important result is 
 \begin{equation}     \label{principal-value}
 {\cal P} \int_{-\infty}^{\infty}\frac{e^{ix}}{x-x_0}dx 
 =i\,\pi\, e^{i x_0}\,.  
\end{equation}

\subsection{Dispersion relations as a consequence of causality}\label{response-function}

\noindent {\bf Linear response functions}\\
The response  $X$ of a {\it linear}  system  depends linearly on the stimulus $S$. The response can be calculated by a convolution integral
\begin{equation}
X(t)=\int_{-\infty}^{+\infty} G(t-t')S(t')dt'\,.
\label{response-integral}
\end{equation}
$G(t-t')$ is called the response function. 
An example of a linear system is  a dielectric medium in which the induced polarization (the response) depends linearly on the applied electric field (the stimulus): 
$$\boldsymbol{P}=\chi_e \varepsilon_0 \boldsymbol{E}\,.$$
The electric susceptibility $\chi_e$ is the response function in this case.
(Nonlinear systems, for example nonlinear crystals  for frequency doubling of laser light, are not treated here).
\vspace{2mm}

\noindent Causality puts an important constraint on the response function. Consider for example an electromagnetic wave pulse hitting the surface of a   dielectric slab where it is reflected. The stimulus is the incident pulse, the response is the reflected pulse, and both are related by (\ref{response-integral}). 
There cannot be a reflected pulse before the incident pulse arrives, hence we must request
\begin{equation}
G(t-t') =G(\tau)=0~~~~~~\mathrm{for} ~~~\tau=t-t'<0\,. 
\end{equation}

\noindent The Fourier transform of the response function 
\begin{equation}\label{Fourier-Response}
\tilde{G}(\omega)=\int_0^\infty G(\tau) e^{i \omega \tau } d\tau\,.
\end{equation}
 is in general a complex-valued function. 
\vspace{3mm}

\noindent {\bf Complex frequency plane}\\
Using the theory of analytic functions we want to  derive a {\it dispersion relation} between real part and imaginary part of $\tilde{G}(\omega)$.
For that purpose we define complex frequencies
 $$\hat{\omega}=\omega_{\rm r} + i\,\omega_{\rm i}$$
The Fourier integral  (\ref{Fourier-Response}) defines $\tilde{G}(\omega)$  as a function of the real variable $\omega$. We generalize this definition to comprise  complex frequencies as well:
\begin{equation}
\tilde{G}(\hat{\omega})=\tilde{G}(\omega_{\rm r} + i\,\omega_{\rm i})=\int_0^\infty  G(\tau) e^{i \omega_{\rm r} \tau} e^{- \omega_{\rm i} \tau} d\tau\,.
\end{equation}
We know from the previous section that $\exp(i \hat{\omega} \tau ) $ is an analytic function of the complex 
variable $\hat{\omega}=\omega_{\rm r} + i\,\omega_{\rm i}\,$,  and hence the response function $\tilde{G}(\hat{\omega})$ is  analytic.
Moreover, because of $\tau \ge 0$,  $|\tilde{G}(\hat{\omega})|$ is bounded in the upper half of the complex $\hat{\omega}$ plane and tends to zero for $\omega_{\rm i} \rightarrow \infty$. Therefore  the line integral along the semicircle $\Gamma_1$ tends to zero in the limit $R \rightarrow \infty$, and
we are allowed to use
Eq.\,(\ref{principal-value}):
\begin{equation}
{\cal{P}}  \int_{-\infty}^\infty \frac{\tilde{G}(\omega)}{\omega-\omega_0}d\omega= i \pi \tilde{G}(\omega_0) \,.
\end{equation}
Separating real and imaginary part we obtain the dispersion relations
\begin{equation}
\Re(\tilde{G}(\omega_0))=\frac{1}{\pi}\,{\cal{P}} \int_{-\infty}^\infty \frac{\Im(\tilde{G}(\omega))}{\omega-\omega_0}d\omega \,,~~~\Im(\tilde{G}(\omega_0))=-\frac{1}{\pi}\,{\cal{P}}  \int_{-\infty}^\infty \frac{\Re(\tilde{G}(\omega))}{\omega-\omega_0}d\omega \,
\label{disp-relation}
\end{equation}
which  show that real part and imaginary part of the response function are intimately connected. The real part  can be computed by a principal value integral over the imaginary part, and  the imaginary part  can be computed by a principal value integral over the real part. \\
An important application is the relation between real part and imaginary part of the electric susceptibility 
$\chi_e(\omega)=\varepsilon(\omega)-1=\varepsilon_1(\omega)-1+ i\,\varepsilon_2(\omega)$ of a solid (see \cite{Wooten})
\begin{equation}
\varepsilon_1(\omega)-1=\frac{1}{\pi}\,{\cal{P}} \int_{-\infty}^\infty \frac{\varepsilon_2(\omega')}
{\omega'-\omega}d\omega' \,,~~~\varepsilon_2(\omega)=-\frac{1}{\pi}\,{\cal{P}}  \int_{-\infty}^\infty 
\frac{\varepsilon_1(\omega')-1}{\omega'-\omega}d\omega' \,.
\label{disp-relation-epsilon}
\end{equation}
These are the famous Kramers-Kronig dispersion relations for the dielectric function of a solid body.
In the above formulas we have replaced $\omega_0$ by $\omega$ and renamed the integration variable from $\omega$ to $\omega'$.\vspace{3mm}

\noindent{\bf Truncated bunch shapes}\\
In order to satisfy the causality condition  in our model calculations on bunch shape reconstruction, we restrict ourselves to truncated functions by requiring that  the particle density   in the bunch vanishes identically   below a time threshold:  $\rho(t)\equiv 0$ for all times $t<t_{\rm min}$.  An example is the cosine-squared wave (\ref{rho-trunc-cos-squared}) shown in Fig.\,\ref{Gauss-Cosine}.
Simple other choices  are truncated rectangular or triangular functions. 

\noindent  Note that a Gaussian function violates causality because it extends over the full range $-\infty <t < + \infty$. Hence it may not be a surprise that we encounter problems with truly Gaussian-shaped time profiles. A truncated Gaussian 
 preserves causality, this important example will be studied in sect.\,\ref{Gauss-KK-Gamma1} and Appendix C.

\subsection{Alternative proof of the dispersion relations between real and imaginary part }
We discuss now an alternative  proof of the dispersion relations between real and imaginary part which   emphasizes the fundamental role of causality.
Any real function $f(t)$ can be expressed as the sum of an even and an odd function:
$$f(t)=f_{\rm even}(t)+f_{\rm odd}(t)~~\mathrm{with}~~f_{\rm even}(t)=(f(t)+f(-t))/2\,,~~~f_{\rm odd}(t)=(f(t)-f(-t))/2\,.$$

\noindent Now we impose the requirement of causality and request that $f(t)$ vanishes  for all $t<0$. 
Then the following relations hold
\begin{equation}
f_{\rm even}(t)=S(t)\, f_{\rm odd}(t)\,,~~~f_{\rm odd}(t)=S(t)\, f_{\rm even}(t)
\label{even-odd}
\end{equation}
where $S(t)$ is  the signum function
$$S(t)=-1~~~\mathrm{for}~~~t<0 \,,~~~S(t)=+1~~~\mathrm{for}~~~t>0 \,.$$
The signum function is the limiting case of a quadratically integrable function
$$S(t)=\lim_{a \rightarrow 0}S_a(t)~~~\mathrm{with}~~S_a(t)=-e^{at} ~\mathrm{for}~t<0 \,,~~S_a(t)=e^{-at}~\mathrm{for}~t>0\,.$$
The Fourier transform of $S_a(t)$ can be easily computed
$$
\tilde{S}_a(\omega)=\int_{-\infty}^\infty S_a(t)e^{i \omega t} dt=    -\int_{-\infty}^0 e^{at}e^{i \omega t} dt +\int_0^{+\infty} e^{-at}e^{i \omega t}  dt
=-\frac{1}{a+i\omega}-\frac{1}{-a+i\omega} \,,$$
so the Fourier transform of the signum function is
\begin{equation}
\tilde{S}(\omega)=\lim_{a\rightarrow 0} \tilde{S}_a(\omega)= \frac{2\,i}{\omega} \,.
\label{tildeS}
\end{equation}
From $e^{i\omega t}=\cos{\omega t} + i\,\sin{\omega t}$ follows that an even real  function $f_{\rm even}(t)$ has a real Fourier transform,  and an odd real  function $f_{\rm odd}(t)$ has a purely imaginary Fourier transform $\tilde{f}_{\rm odd}(\omega)$.
Therefore we obtain, using (\ref{even-odd})
\begin{equation}
\Re(\tilde{f}(\omega))=\tilde{f}_{\rm even}(\omega)={\cal FT}\{S(t)\, f_{\rm odd}(t)\}
\,,~~~\Im(\tilde{f}(\omega))=\frac{1}{i}\,\tilde{f}_{\rm odd}(\omega)=\frac{1}{i}\,{\cal FT}\{S(t)\, f_{\rm even}(t)\}
\label{Re-Im}
\end{equation}
where ${\cal FT}\{S(t)\, fç(t)\}$ denotes the Fourier transform of the product function $S(t)\, f_{\rm odd}(t)$,
and \newline ${\cal FT}\{S(t)\, f_{\rm even}(t)\}$ the Fourier transform of $S(t)\, f_{\rm even}(t)$.
\vspace{3mm}

\noindent The Fourier transform of the product of two time-domain functions is evaluated using the  following mathematical theorem:

\noindent {\it Theorem\,:} Let $f(t)$ and $g(t)$ be quadratically integrable functions with Fourier transforms $\tilde{f}(\omega)$ and  $\tilde{g}(\omega)$.
 The Fourier transform of the product function $h(t)=f(t)\,g(t)$ is given by the convolution integral of $\tilde{f}(\omega)$ and  $\tilde{g}(\omega)$:
 \begin{equation}
 \tilde{h}(\omega_0)=\frac{1}{2\pi}\int \tilde{f}(\omega_0-\omega)\tilde{g}(\omega)d\omega\,.
 \label{Eq-convolution}
 \end{equation}
 {\it  Proof\,}:
The Fourier transform of  $h(t)=f(t)\,g(t)$ is defined by
$$ \tilde{h}(\omega_0)=\int f(t)g(t) \,e^{i\omega_0 t}dt\,.$$
Inserting for $g(t)$ its inverse Fourier transform
$$g(t)=\frac{1}{2\pi}\int \tilde{g}(\omega)\,e^{-i\omega t}d\omega$$
one finds
$$ \tilde{h}(\omega_0)=\frac{1}{2\pi}\,\int \left\{\,\int f(t)e^{i(\omega_0-\omega)t}dt \cdot \tilde{g}(\omega)\,\right\}\,d\omega
=\frac{1}{2\pi}\int \tilde{f}(\omega_0-\omega)\tilde{g}(\omega)d\omega~~~\mathrm{qed.}$$
Applying this theorem in the equations (\ref{Re-Im})  we get
\begin{eqnarray}
\Re(\tilde{f}(\omega_0))&=&\tilde{f}_{\rm even}(\omega_0)=\frac{1}{2\pi}\int \tilde{S}(\omega_0-\omega)
\tilde{f}_{\rm odd}(\omega)d\omega=\frac{i}{2\pi}\int \tilde{S}(\omega_0-\omega)
\Im(\tilde{f}(\omega))d\omega\,, \nonumber \\
\Im(\tilde{f}(\omega_0))&=&\frac{1}{i}\tilde{f}_{\rm odd}(\omega_0)=-\frac{i}{2\pi}\int \tilde{S}(\omega_0-\omega)
\tilde{f}_{\rm even}(\omega)d\omega=-\frac{i}{2\pi}\int \tilde{S}(\omega_0-\omega)
\Re(\tilde{f}(\omega))d\omega\,. \nonumber
\end{eqnarray}
Inserting the Fourier transform  of the signum function, 
$\tilde{S}(\omega_0-\omega)=2i/(\omega_0-\omega)$, we finally obtain the dispersion relations (\ref{disp-relation})
\begin{equation}
\Re(\tilde{f}(\omega_0))=\frac{1}{\pi}\int \frac{\Im(\tilde{f}(\omega))}{\omega-\omega_0}d\omega\,,  ~~~~
\Im(\tilde{f}(\omega_0))=-\frac{1}{\pi}\int \frac{\Re(\tilde{f}(\omega))}{\omega-\omega_0}d\omega\,.
\end{equation}

\subsection{Dispersion relation for absolute magnitude and phase} \label{Disp-ampl-phase}
The dispersion relation (\ref{disp-relation}) requires the knowledge of either  the real part or the imaginary part of the response function. However, in many cases only the absolute magnitude is known, so it is desirable  to derive a relation allowing to compute the phase. Wooten \cite{Wooten}  discusses the mathematical procedures  for the reflection of an electromagnetic wave from a solid.  
A similar  task arises for the complex form factor 
$${\cal F}(\omega)=F(\omega) \exp(i \Phi(\omega))\,.$$
Here $\omega$ is a real variable.
 The spectral  intensity emitted by an electron bunch via the processes of transition, diffraction or synchrotron radiation is recorded with a  spectrometer  or another device  measuring just the intensity.  From these data one derives the absolute square of the bunch form factor, $|{\cal F}(\omega)|^2$. To determine the phase $\Phi(\omega)$, we adopt  Wooten's treatment of the complex reflectivity and  derive a dispersion relation for
 \begin{equation}
 \ln({\cal F}(\omega))=\ln(F(\omega))+i \Phi(\omega)\,.
 \label{log-phase}
 \end{equation}
 
\begin{figure}[t]
		\centering
		\includegraphics[width=9cm]{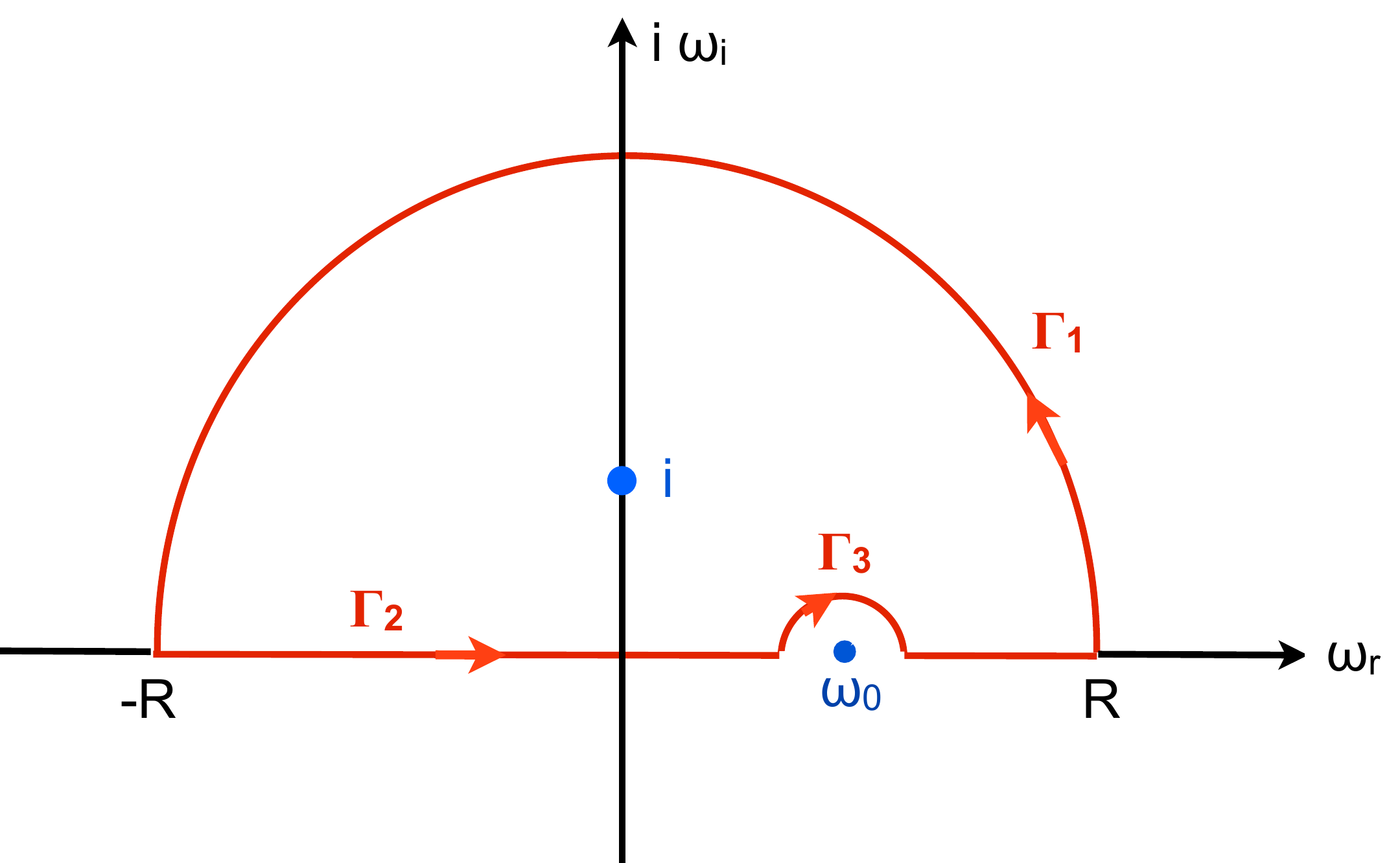} 
		\caption{\small{ The closed integration path $\Gamma$ for the auxiliary function (\ref{aux-function}). The difference to Fig.\,\ref{Gamma-path} is the pole on the imaginary axis at	$\hat{\omega}=i$.}	}			
		\label{Weg-log(omega)}
		\end{figure}

\noindent Now we continue this expression into the complex $\hat{\omega}$ plane.
 We want to apply the Residue Theorem for  the closed loop shown in Fig.\,\ref{Weg-log(omega)}. This needs a lot of work, and some problems arise. 

\vspace{2mm}

\noindent {\bf Prerequisites for the applicability of the  Residue Theorem}\\
(1) $\ln({\cal F}(\hat{\omega}))$ is not allowed to have a singularity in the upper half plane. Therefore  we must require 
${\cal F}(\hat{\omega}) \ne 0$ for any $\hat{\omega}$ with a positive imaginary part.  This condition, however, is violated in many cases. Complex zeros in the upper half plane may indeed exist, and they have a strong impact on the phase determination. We show in Appendix A, sect.\,\ref{Sect-Blaschke} how these zeros can be handled and demonstrate the mathematical procedure with many examples in 
Appendix B.\vspace{2mm}

\noindent (2) A severe problem  is that the form factor drops to zero at infinite frequency, hence $\ln({\cal F}(\hat{\omega}))$ diverges for $|\hat{\omega}| \rightarrow \infty$. 
This means that the line integral of $\ln({\cal F}(\hat{\omega}))$ over the large semicircle $\Gamma_1$ diverges as well. To circumvent this difficulty one defines the following auxiliary function
\begin{equation}\label{aux-function}
f_{\rm aux}(\hat{\omega})=\frac{(1+\omega_0 \hat{\omega})\ln({\cal F}(\hat{\omega}))}{(1+\hat{\omega}^2)(\omega_0 -\hat{\omega})}
=\frac{1}{\hat{\omega}-i}\,\frac{(1+\omega_0 \hat{\omega})\ln({\cal F}(\hat{\omega}))}{(\hat{\omega}+i)(\omega_0 -\hat{\omega})}
\end{equation}
with a real positive frequency $\omega_0$. Following  \cite{Wooten}
we furthermore make the  assumption that asymptotically  ${\cal F}(\hat{\omega})$ obeys an inverse power law:
\begin{equation} \label{inverse-powerlaw}
{\cal F}(\hat{\omega}) = b \,\hat{\omega}^{-s}~~\mathrm{with~a~positive~number}~s~~~(|\hat{\omega}| \rightarrow \infty).
\end{equation}
Then the function $f_{\rm aux}$ behaves asymptotically as
$$ f_{\rm aux}(\hat{\omega}) \approx  \frac{(1+\omega_0 \hat{\omega})(-s\ln(\hat{\omega})+\ln(b))}{(1+\hat{\omega}^2)(\omega_0 -\hat{\omega})}  \,.$$
Condition (\ref{inverse-powerlaw}) is met by the form factor  (\ref{FF-cosine}) of the cosine-squared wave cycle and the form factors of other truncated time-domain functions. However, it is violated for Gaussian form factors, as will be shown later. \\

If (\ref{inverse-powerlaw}) is satisfied one gets on  the large semicircle $\Gamma_1$ 
$$\hat{\omega}=R e^{i\theta}\,,~~|f_{\rm aux}(\hat{\omega})| \approx s \,\omega_0  \left| \frac{\ln(R)}{R^2} \right|\,.$$
The pathlength of $\Gamma_1$ is $\pi R$, hence the integral of $f_{\rm aux}(\hat{\omega})$ over this semicircle vanishes if R tends to infinity:
\begin{equation}\label{large-semicircle}
\int_{\Gamma_1} f_{\rm aux}(\hat{\omega})d\hat{\omega} \approx \pi \,s \,\omega_0  \left| \frac{\ln(R)}{R} \right| \rightarrow 0 \,.
\end{equation}
\vspace{4mm}

\noindent Next we compute the integral over the small semicircle $\Gamma_3$ which is centered at the real frequency $\omega_0$ and has an infinitesimal radius $\varepsilon$. On the semicircle we have 
$$\hat{\omega}=\omega_0+\varepsilon \, e^{i \theta}, ~~d\hat{\omega}=i\,\varepsilon\, e^{i \theta} d\theta\,,~~~~~ \lim_{\varepsilon \rightarrow 0}  f_{\rm aux}(\hat{\omega})= \frac{\ln({\cal F}(\omega_0))}{\omega_0 -\hat{\omega}}\,.$$
\noindent In analogy to Eq.\,(\ref{Gamma3-Integral})  we find therefore
\begin{eqnarray}
\lim_{\varepsilon \rightarrow 0}\int_{\Gamma_3} f_{\rm aux}(\hat{\omega})d\hat{\omega} &=& \ln({\cal F}(\omega_0)) \lim_{\varepsilon \rightarrow 0} \int_{\pi}^{0} \frac{1}{(-\varepsilon\, e^{i \theta})}\,i\,\varepsilon e^{i \theta} d\theta  
 =i\,\pi\,\ln({\cal F}(\omega_0)) \nonumber\\
&=& i\,\pi\, \ln( F(\omega_0)) - \pi\,\Phi(\omega_0)\,.  
\label{Gamma3-Integral-faux}
\end{eqnarray}

Performing the integration of $f_{\rm aux}(\hat{\omega})$ along  the closed loop shown in Fig. \ref{Weg-log(omega)} we have to keep in mind that the auxiliary function has a pole at $\hat{\omega}=i$ inside the loop. The Residue Theorem yields
$$\oint_{\Gamma} f_{\rm aux}(\hat{\omega})d\hat{\omega}=\int_{\Gamma_1} f_{\rm aux}(\hat{\omega})d\hat{\omega}+\int_{\Gamma_2} f_{\rm aux}(\hat{\omega})d\hat{\omega}+\int_{\Gamma_3} f_{\rm aux}(\hat{\omega})d\hat{\omega}
=2 \pi i \frac{1+i \omega_0}{2i (\omega_0-i)}\ln({\cal F}(i))=i \pi \ln({\cal F}(i))\,.$$
What is ${\cal F}(i)$? From  the definition\,(\ref{Fourier-formfactor}) of the form factor follows that ${\cal F}(i)$ is a real number which we call $\alpha$:
$$
{\cal F}(i)=\int_{-\infty}^{\infty} \rho(t) \exp(-\,t) dt= \alpha \,.$$
Putting things together, using $\ln({\cal F}(\omega))=\ln(F(\omega)) +i \Phi(\omega)$, and separating real and imaginary parts we obtain for the integral along the path $\Gamma_2$ (this is the real axis, hence we write $\omega$ instead of $\hat{\omega}$):
$$\int_{\Gamma_2} f_{\rm aux}(\omega)d\omega\equiv {\cal P} \int_{-\infty}^\infty \left[\frac{(1+\omega_0 \omega)\ln(F(\omega))}{(1+\omega^2)(\omega_0 -\omega)} +i\,\frac{(1+\omega_0 \omega) \Phi(\omega)}{(1+\omega^2)(\omega_0 -\omega)}\right]d\omega=- i \pi \ln(F(\omega_0))+\pi \Phi(\omega_0)+i \pi \ln(\alpha)\,.$$
The real part of this equation yields  a first form of the desired dispersion relation
\begin{equation}
\Phi(\omega_0)= \frac{1}{\pi}\, {\cal P} \int_{-\infty}^\infty \frac{(1+\omega_0 \omega)\ln(F(\omega))}{(1+\omega^2)(\omega_0 -\omega)} d\omega \,.
\end{equation}
\vspace{3mm}

\noindent {\bf Restriction to positive frequencies}\\
The particle density  $\rho(t)$ is a real function  while the  Fourier transform ${\cal F}(\omega)$ is a  complex-valued function obeying the rule\,(\ref{FF-FF*}):
$${\cal F}^*(\omega)={\cal F}(-\omega)~~~\Rightarrow~~F(-\omega)=F(\omega)\,,~~\Phi(-\omega)=-\Phi(\omega)\,.$$
The principal value integral is split up in two parts
$${\cal P} \int_{-\infty}^\infty f_{\rm aux}(\omega)d\omega={\cal P} \int_{-\infty}^0 f_{\rm aux}(\omega)d\omega+{\cal P} \int_{0}^\infty f_{\rm aux}(\omega)d\omega \,.$$
In the integration over negative frequencies we make the replacement $u=-\omega$ and  use $F(-u)=F(u)$
$${\cal P} \int_{-\infty}^0\frac{(1+\omega_0 \omega)\ln(F(\omega))}{(1+\omega^2)(\omega_0 -\omega)}d\omega=
{\cal P} \int_{0}^\infty \frac{(1-\omega_0 u)\ln(F(-u))}{(1+u^2)(\omega_0+u)}\,du
={\cal P} \int_{0}^\infty \frac{(1-\omega_0 u)\ln(F(u))}{(1+u^2)(\omega_0+u)}du \,.$$
Now we rename the integration variable $u$ back into $\omega$. 
Using
$$\frac{1+\omega_0 \omega}{\omega_0 -\omega}+\frac{1-\omega_0 \omega}{\omega_0 +\omega}=
\frac{2 \omega_0 (1+\omega^2)}{\omega_0^2-\omega^2}$$
 one can combine the two integrals and obtains
 $$\Phi(\omega_0)= \frac{2 \omega_0}{\pi}\, {\cal P} \int_{0}^\infty \frac{\ln(F(\omega))}{\omega_0^2 -\omega^2} d\omega \,.$$
 The integrand has a singularity at $\omega=\omega_0$. To cancel it one subtracts the expression
 $$\frac{2 \omega_0 }{\pi}\, {\cal P} \int_{0}^\infty \frac{\ln(F(\omega_0))}{\omega_0^2 -\omega^2} d\omega=
 \frac{2 \omega_0 \ln(F(\omega_0))}{\pi}\, {\cal P} \int_{0}^\infty \frac{1}{\omega_0^2 -\omega^2} d\omega=0 \,.$$
 It is easy to verify that this integral vanishes. With $x=\omega/\omega_0$ the integral is of the type
$$
 {\cal P} \int_{0}^\infty \frac{1}{1-x^2}dx=\lim_{\varepsilon \rightarrow 0} \left[~~\int_0^{1-\varepsilon}+\int_{1+\varepsilon}^\infty\right]\left(\frac{1}{1+x} +\frac{1}{1-x} \right) dx
=\lim_{\varepsilon \rightarrow 0} \left[~  \ln(2-\varepsilon)-\ln(2+\varepsilon)~\right]=0. 
$$
So finally we arrive at the important dispersion relation
\begin{eqnarray} \label{phase-reflectivity}
\Phi(\omega_0)= \frac{2 \omega_0}{\pi}\, \mathcal {P} \int_{0}^\infty \frac{\ln(F(\omega))-\ln(F(\omega_0))}{\omega_0^2 -\omega^2} d\omega \,.
\end{eqnarray}

\subsection{Computation of  form factor phase via dispersion relation}
\subsubsection{Kramers-Kronig phase}
As said before, the longitudinal form factor is the Fourier transform of the normalized longitudinal charge density distribution
\begin{equation}\label{formfactor-definition}
{\cal F}(\omega)=\int_{-\infty}^\infty \rho(t) e^{i \omega t} dt~~~\mathrm{with}~~\int_{-\infty}^\infty \rho(t) dt=1 \,.
\end{equation}
Spectroscopic experiments at particle accelerators yield only the absolute magnitude of the  form factor, $F(\omega)=|{\cal F}(\omega)|$, but neither its real  part nor its imaginary part.  
Writing
\begin{equation}
{\cal F}(\omega)=F(\omega) \exp(i \Phi(\omega))
\end{equation}
formula (\ref{phase-reflectivity}) can be applied to compute the phase $\Phi(\omega)$. 
The {\it Kramers-Kronig phase} is given by
\begin{equation}
\Phi_{\rm KK}(\omega_0)=\frac{2 \omega_0}{\pi} \, \mathcal {P} \int_{0}^{\infty} 
\frac{\ln(| {\cal F}(\omega)|)-\ln( |{\cal F}(\omega_0)|) } {\omega_0^2-\omega^2}\,d\omega 
\label{KK-phase}
\end{equation} 
 provided the prerequisites made in the derivation  of this formula are fulfilled\footnote{Physics requires  that we make a slight modification in this formula. Backward transition radiation does not extend to infinite frequencies. High-energy X or gamma rays will never be observed in backward direction.  Therefore the upper integration limit  is far from infinity but rather a suitable cutoff frequency $\omega_{\rm cut}$  in the ultraviolet regime  which corresponds to the radius $R$ of the large semicircle if we choose a large but finite value for $R$.}.
  The phase (\ref{KK-phase}) is also called the  canonical phase or the minimal phase, the latter expression being somewhat misleading since $\Phi_{\rm KK}(\omega)$ does not correspond to a minimum in the mathematical sense. One has to be aware that there are commonly used  test functions which do not obey all  prerequisites made in the derivation of Eq.\,(\ref{KK-phase}) and require a special treatment.  We will study two such problematic cases: nonvanishing integral of the auxiliary function  over the large semicircle,  and form factors having  zeros in upper half of the complex frequency  plane.
 
\subsubsection{Problem 1: Integral over large semicircle maybe nonzero} \label{Gauss-KK-Gamma1}
An essential prerequisite for the validity of formula (\ref{KK-phase}) is that the integral of the auxiliary function along the large semicircle in Fig.\,\ref{Weg-log(omega)} must vanish. This requirement is not met by Gaussian form factors. However, since it is very convenient to represent the time profile of a short electron bunch by a Gaussian function  or a linear combination of Gaussians, it is worthwhile to investigate if  it is possible to preserve the validity of the KK phase formula (\ref{KK-phase}) by taking into account the line integral along the large semicircle as an extra contribution. We will show now that this is indeed possible.  \vspace{2mm}

\noindent {\bf (a) Single Gaussian}\\
Consider a single Gaussian which for simplicity is centered at $t=0$.
The form factor is also a Gaussian. 
$$\rho(t) =\frac{1}{\sqrt{2\pi}\sigma}\, \exp\left(-t^2/(2\sigma^2)\right)\,,~~~
{\cal F}(\omega)=\exp\left(-\sigma^2\omega^2/2\right)\,.$$
With this form factor  we are facing a mathematical problem. The logarithm is proportional to the square of frequency, 
$\ln({\cal F}(\omega)) =-\sigma^2\omega^2/2$, which implies that the integral of the auxiliary function (\ref{aux-function}) over the large semicircle $\Gamma_1$ in Fig.\,\ref{Weg-log(omega)} does not vanish.  Is it still  permitted to use the Kramers-Kronig formula for a Gaussian bunch?  To decide this question, it is obviously  necessary to compute explicitly the line integral along the half circle $\Gamma_1$. To this end we go into the complex frequency plane ($\hat{\omega}=\omega_{\rm r}+i\,\omega_{\rm i}$). 
The auxiliary function has a simple form 
\begin{equation}
f_{\rm aux}(\hat{\omega})=\frac{(1+\omega_0 \hat{\omega})\ln({\cal F}(\hat{\omega}))}{(1+\hat{\omega}^2)(\omega_0 -\hat{\omega})}
=\frac{(1+\omega_0 \hat{\omega})(-\sigma^2\hat{\omega}^2/2)}{(1+\hat{\omega}^2)(\omega_0 -\hat{\omega})}\,.
\label{faux-Gauss}
\end{equation}
Now we compute the line integral over the semicircle $\Gamma_1$. The radius $R$ is chosen to be very large, $R \gg \omega_0$, 
but we do not let $R$ tend to infinity. On the semicircle we can simplify $f_{\rm aux}\,$:
$$\hat{\omega}=R e^{i\theta} \,, ~~R \gg \omega_0~~\Rightarrow~~f_{\rm aux}(\hat{\omega}) \approx \omega_0 \sigma^2/2\,. $$
Thus we get in very good approximation
\begin{equation}
\int_{\Gamma_1}f_{\rm aux}(\hat{\omega}) d\hat{\omega}= \omega_0 \sigma^2/2\,\int_0^\pi \,i \,R\,e^{i\theta} d\theta
=+\omega_0 R \sigma^2
\label{IntGamma1}
\end{equation}
so the line integral over the semicircle $\Gamma_1$ is  indeed different from zero.\vspace{2mm}

\noindent Next we evaluate the line integral over the straight path $\Gamma_2$, the real axis.  Making use of the computational steps in  section \ref{Disp-ampl-phase} we obtain
$$\int_{\Gamma_2}f_{\rm aux}(\omega) d\omega=
2 \omega_0 \, \mathcal {P} \int_{0}^{R} 
\frac{\ln(| {\cal F}(\omega)|)-\ln( |{\cal F}(\omega_0)|) } {\omega_0^2-\omega^2}\,d\omega \,.$$
 Inserting the Gaussian form factor this expression  yields:
\begin{equation}
\int_{\Gamma_2}f_{\rm aux}(\omega) d\omega = -\omega_0 R \sigma^2\,.
\end{equation}
The most remarkable result is: in  case of a Gaussian form factor the line integral along the real axis is cancelled by the line integral along the large semicircle. The phase of the Gaussian form factor is identical to zero, as it should be.  \vspace{2mm}

\noindent {\bf (b) Truncated Gaussian}\\
As stated before, a truncated Gaussian preserves causality.  Hence one might expect that the integral of the auxiliary function over large semicircle   vanishes in this case.  We prove now that this is indeed the case.  
Consider the function
\begin{eqnarray}
\rho(t)&=&0~~~~\mathrm{for}~~~t<0~ \,,\nonumber \\
\rho(t)&=&\frac{1}{\sqrt{2\pi}\sigma}\exp\left(\frac{-(t-\tau)^2}{2 \sigma^2} \right)  ~~~~\mathrm{for}~~~t \ge 0\,.
\label{trunc-Gauss2}
\end{eqnarray}
If we choose for example $\tau=\alpha\,\sigma$ with $\alpha=5$, the truncated Gaussian and a normal Gaussian (centered at $t=\tau$) are hardly distinguishable. 
\begin{figure}[hb!]
\begin{center}
\includegraphics[width=8cm]{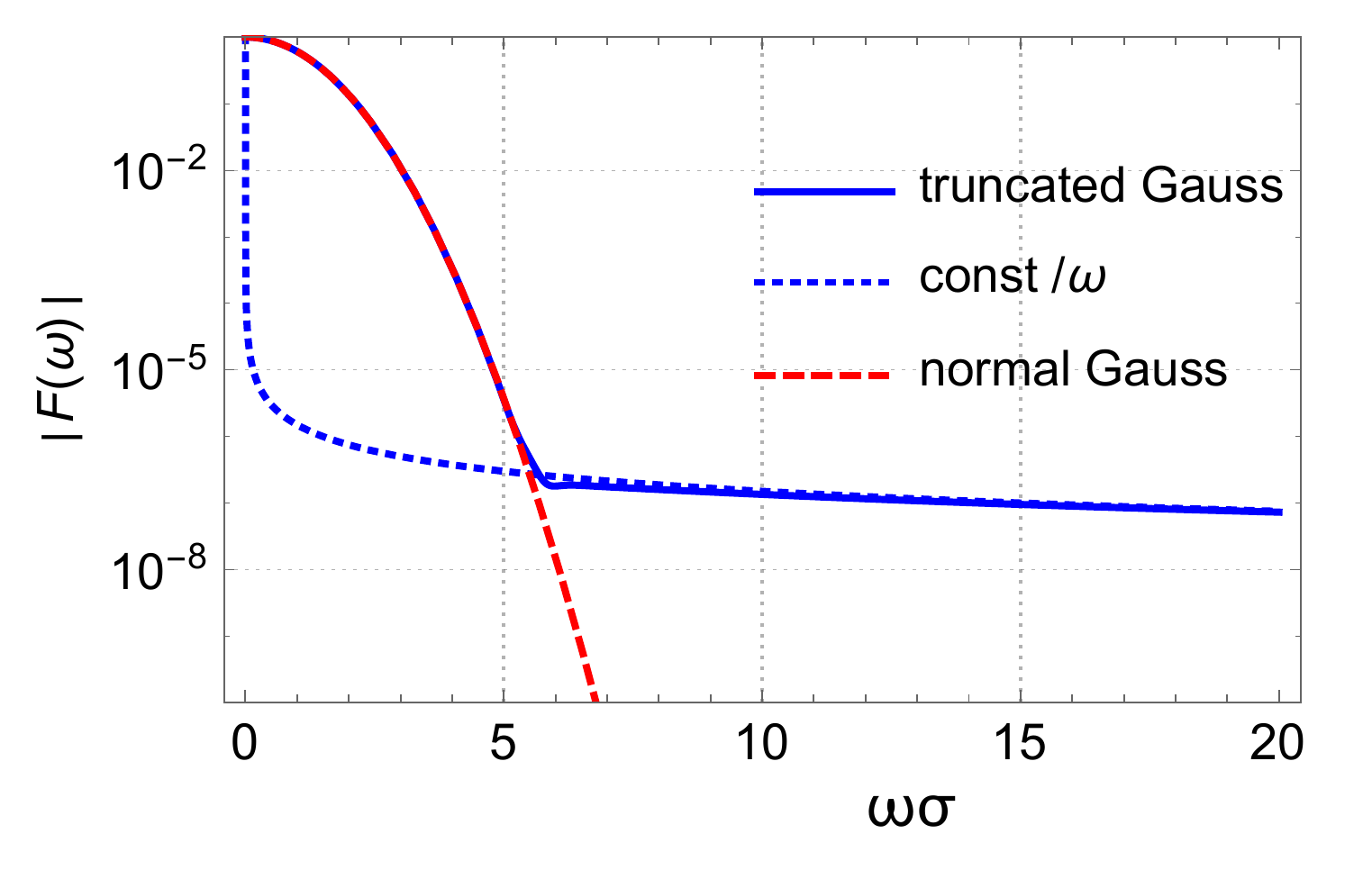}
\end{center}
\caption{\small{The magnitude $F_{\rm trunc}(\omega)$ of the  form factor of a truncated Gaussian (solid blue curve), plotted as a function of the dimensionless variable $\omega \sigma$, in comparison
with a normal Gaussian form factor (dashed red curve). At large values of  $\omega \sigma$,  the  form factor obeys an  inverse power law
$F_{\rm trunc}(\omega) \approx \,$const$/\omega$ (dotted blue curve). }
}
\label{truncatedGauss}
\end{figure} 

\noindent Their
 form factors are for real $\omega$
 \begin{eqnarray}\label{Ftrunc-ana}
 {\cal F}_{\rm trunc}(\omega)&=&\frac{1}{2}\exp\left(-\frac{\sigma^2\omega^2}{2}+i\,\omega \tau \right)
\left(1+\mathrm{erf}\left[\frac{\alpha+i\,\sigma\omega}{\sqrt{2}}\right]\right) ~~\mathrm{(erf\,=\,error~function)} \,, \\
{\cal F}_{\rm norm}(\omega)&=&\exp\left(-\frac{\sigma^2\omega^2}{2}+i\,\omega \tau\right)~~~\mathrm{with}~~\tau=\alpha \,\sigma\,.\nonumber
 \end{eqnarray}

The magnitudes  are plotted  in Fig.\,\ref{truncatedGauss}.  Both form factors  agree in the range
$0\le\omega \sigma\le \alpha $,  however at large frequencies
the form factor of the normal Gaussian  drops to tiny values while that  of the truncated Gaussian levels off. By series expansion of Eq.\,(\ref{Ftrunc-ana}) one finds that asymptotically $F_{\rm trunc}(\omega)$ obeys an  inverse power law:
\begin{equation}
F_{\rm trunc}(\omega) \approx \frac{\exp(-\alpha^2/2)}{\sqrt{2\pi}\sigma}\cdot\frac{1}{\omega} ~~~\mathrm{for}~~\omega \sigma \gg \alpha \,.
\label{FF-inverse-power}
\end{equation}
 Thus the  condition (\ref{inverse-powerlaw}) is fulfilled, here with $s=1$, and this  means that the integral  of the auxiliary function over large semicircle  $\Gamma_1$ vanishes. The important conclusion is:  the Kramers-Kronig phase formula (\ref{KK-phase}) is perfectly valid for truncated Gaussian time-domain functions.

\subsubsection{Problem 2: Complex zeros of the form factor}  \label{Sect-Blaschke}
The most important prerequisite for Eq.\,(\ref{KK-phase}) to hold is that the form factor must be different from zero in the entire  upper half plane  ($\Im(\omega_{\rm r} + i \omega_{\rm i})=\omega_{\rm i}>0$). If the form factor vanishes at some point inside the closed integration loop $\Gamma$ its logarithm has an essential singularity here (not a  simple pole), and the Residue Theorem can no longer be applied. The Kramers-Kronig formula becomes obsolete. Is there a way out of this difficulty? \vspace{2mm}

\noindent  Suppose now that the form factor has a zero of first order at some complex frequency $\hat{\omega}_1=a+ib$ in the right upper quarter of the $\hat{\omega}$ plane ($a, b>0$). An interesting observation is  that zeros occur always in pairs, at $\hat{\omega}_1=a+ib$ in the right upper quarter  and at $\hat{\omega}_1'=-a+ib$ in the left upper quarter. This follows from Eq.\,(\ref{FF-FF*}).
This pair of zeros in the upper half plane can be removed by defining a modified form factor
\begin{equation}\label{mod-formfactor}
{\cal F}_{\rm mod}(\hat{\omega})={\cal F}(\hat{\omega}) {\cal B}(\hat{\omega})
\end{equation}
with the so-called  Blaschke 
factor
\begin{equation}
{\cal B}(\hat{\omega})= \frac{\hat{\omega}-\hat{\omega}_1^*}{\hat{\omega}-\hat{\omega}_1}
\cdot \frac{\hat{\omega}-\hat{\omega}_1'^{*}}{\hat{\omega}-\hat{\omega}_1'}=\frac{\hat{\omega}-(a-ib)}{\hat{\omega}-(a+ib)}\cdot \frac{\hat{\omega}-(-a-ib)}{\hat{\omega}-(-a+ib)} \,.
\end{equation} 

\noindent {\bf Behavior on real axis}\\
On the real $\omega$ axis, i.e.   for $\hat{\omega}=\hat{\omega}^*=\omega$, 
 the absolute magnitude of the Blaschke factor  is 1:
\begin{equation} \label{Fmod}
|{\cal B}(\omega)|=\left|\frac{\omega-(a-ib)}{\omega-(a+ib)}\right|\cdot \left|\frac{\omega-(-a-ib)}{\omega-(-a+ib)}\right|=1~~\Rightarrow~|{\cal F}_{\rm mod}(\omega)|=|{\cal F}(\omega)|~~~
\mathrm{for\, real}~~~\omega \,.
\end{equation}
This is a very important equation. It means that the form factor ${\cal F}(\omega)$ and  the modified form factor ${\cal F}_{\rm mod}(\omega)$ describe  exactly the same  radiation spectrum. Knowing just the absolute magnitude from experiment one cannot decide which of the two form factors (or even another expression) is the correct one.  

\begin{figure}[htb]
		\centering
		\includegraphics[width=12cm]{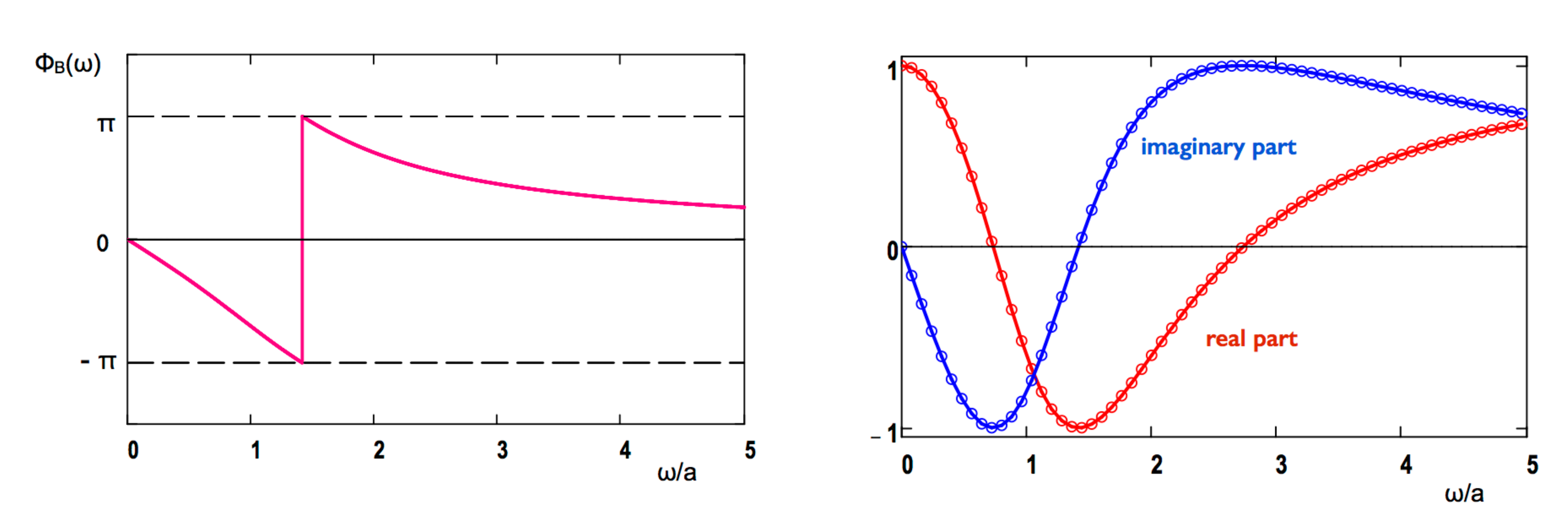} 
		\caption{\small{{\it Left\,}: The Blaschke phase $\Phi_{\rm B}(\omega)$ of a pair of zeros (at  $\hat{\omega}_1=a+ib$ and $\hat{\omega}'_1=-a+ib$)  for a ratio $b/a=1$. {\it Right\,}: Comparison of $\Re({\cal B}(\omega))$ (red curve) with $\cos{\Phi_{\rm B}(\omega)}$ (red circles) and comparison of $\Im({\cal B}(\omega))$ (blue curve) with 
		$\sin{\Phi_{\rm B}(\omega)}$ (blue circles).}	}				
		\label{Blaschke-Re-Im}
		\end{figure}

However, the phases of 
${\cal F}(\omega)$ and ${\cal F}_{\rm mod}(\omega)$ are different because the Blaschke factor is a complex, frequency-dependent number. Real and imaginary part of 
${\cal B}(\omega)$ are 
\begin{equation}
\Re({\cal B}(\omega))=\frac{\omega^4-2\omega^2(a^2+3b^2)+(a^2+b^2)^2}
{ [(\omega-a)^2+b^2][ (\omega+a)^2+b^2]}        \,,~~
\Im({\cal B}(\omega))=\frac{4\omega b(\omega^2-a^2-b^2)}
{ [(\omega-a)^2+b^2][ (\omega+a)^2+b^2]} \,.
\end{equation}
The phase of  ${\cal B}(\omega)$ is computed by the equation
\begin{equation}\label{Bla-phase}
\Phi_{\rm B}(\omega)=\arg({\cal B}(\omega))\,.
\end{equation}
This phase is not a continuous function of $\omega$ but exhibits a discontinuous jump by $2\pi$ at the frequency $\omega_{\rm jump}=\sqrt{a^2+b^2}$ where the imaginary part of the Blaschke factor vanishes. It is plotted in Fig.~\ref{Blaschke-Re-Im} as a function of the real scaled frequency $\omega/a$. 
The phase jump does not present any problem\footnote{The situation is rather more complicated if one uses the arctan function instead of the arg function to compute the phase of the Blaschke factor. Then two phase jumps, each by $\pi$,  are obtained near $\omega_{\rm jump}$ which  lead to errors when one computes the sine or cosine of the Blaschke phase. These deficiencies can be cured but the mathematical procedure is clumsy. }, the equalities 
 $\Re({\cal B}(\omega))=|{\cal B}(\omega)| \cos{\Phi_{\rm B}(\omega)}$ and 
 $\Im({\cal B}(\omega))=|{\cal B}(\omega)| \sin{\Phi_{\rm B}(\omega)}$ are perfectly well fulfilled.

\begin{figure}[htb]
		\centering
		\includegraphics[width=8cm]{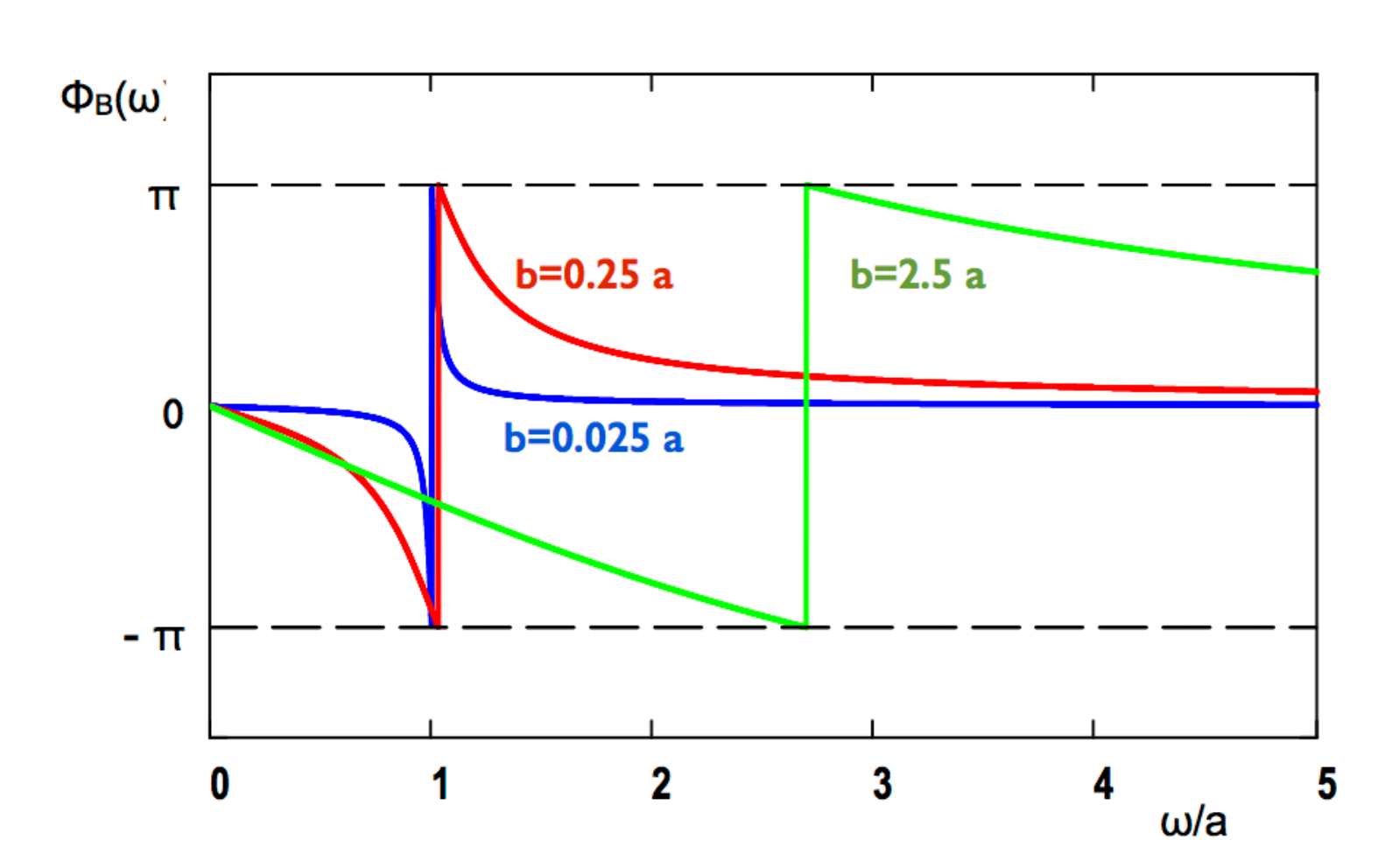} 
		\caption{ \small{The Blaschke phase $\Phi_{\rm B}(\omega)$ for various ratios $b/a$.}	}				
		\label{Blaschke-Phase-a-b}
		\end{figure}

\noindent In Fig.\,\ref{Blaschke-Phase-a-b} the function $\Phi_{\rm B}(\omega)$ is shown for various ratios $b/a$. 
In case of a very small imaginary part of the complex zero  $\hat{\omega}_1=a+ib$, i.e. for $b/a \ll 1$, the phase excursion $0 \rightarrow - \pi \rightarrow + \pi \rightarrow 0$ is very steep  and affects only a narrow frequency range. This means that complex zeros in the immediate vicinity of the real axis have a negligible influence on the 	bunch shape reconstruction. 

\noindent On the other hand, zeros with a very large imaginary part ($b \gg a$) lead to an almost linearly rising phase at $\omega \approx a$. They do not change the internal structure of the reconstructed bunch but merely shift its position on the time axis. \vspace{2mm}

\noindent Suppose now that ${\cal F}(\hat{\omega}) $ has a finite number of pairs of zeros in  the upper half  of the complex $\hat{\omega}$ plane, at the frequencies $\hat{\omega}_n=a_n+ib_n$ and $\hat{\omega}'_n=-a_n+ib_n$.  The {\it Blaschke product} is defined as  the product of the Blaschke factors  ${\cal B}_n(\omega)$ of these pairs.  The phases are defined for real $\omega$, they are additive. 
\begin{equation}
{\cal B}(\hat{\omega})=\prod_n {\cal B}_n(\hat{\omega})=\prod_n \frac{\hat{\omega}-(a_n-ib_n)}{\hat{\omega}-(a_n+ib_n)}\cdot \frac{\hat{\omega}-(-a_n-ib_n)}{\hat{\omega}-(-a_n+ib_n)} \,,~~~~~~\Phi_{\rm B}(\omega)=\sum_n \Phi_{{\rm B},n}(\omega)\,.
\label{Blaschke-phase-n}
\end{equation}
In analogy with Eq.\,(\ref{mod-formfactor}) the modified  form factor is given by
\begin{equation}\label{F-mod}
{\cal F}_{\rm mod}(\hat{\omega})={\cal F}(\hat{\omega}) {\cal B}(\hat{\omega})\,.
\end{equation}
This form factor has  no  zero in the upper half plane. It fulfills the requirements made in the derivation of formula (\ref{KK-phase}), and hence the phase of the modified  form factor is identical with the Kramers-Kronig phase, 
$\Phi_{\rm mod}(\omega)=\Phi_{\rm KK}(\omega)$. 
The phase of our original form factor, which we call {\it reconstruction phase} from now on, follows immediately from Eq.\,(\ref{F-mod})
\begin{equation}
\Phi_{\rm rec}(\omega)=\Phi_{\rm KK}(\omega)-\Phi_{\rm B}(\omega)\,.
\label{reconstr-phase}
\end{equation}

\clearpage

\section{Appendix B: Analytic Model Calculations with Causal Functions}

To assess the capabilities and the limitations of  the phase reconstruction of the bunch form factor by applying the theorems  of complex analysis, it is useful to employ the method for simple test functions $\rho(t)$ whose complex form factor is either known analytically or can be calculated by Fourier transformation. 

We have seen already that the Kramers-Kronig phase reconstruction works well for  bunches  with a single peak but leads to   puzzling results when two or more peaks are present: in some cases the original is perfectly reproduced, in other cases the reconstructed charge distribution is very different from the original. This will be illustrated with a number of examples, and we show the benefit of the Blaschke phase correction. In this appendix we restrict ourselves to test functions $\rho(t)$ which obey the basic requirement of causality. The same superpositions of cosine-squared pulses as in section \ref{ana-phase-retrieve} will be considered, and we will compute the Blaschke correction for these profiles. 
Another important topic is the relation between time reversal ($\rho(t) \rightarrow \rho(-t) $) and the location of the form factor zeros in the complex frequency plane.\vspace{1mm}

\noindent Gaussian time profiles, violating causality,  are the subject of Appendix C.

\subsection{Two cosine-squared pulses of different width}
\begin{figure}[htb!]
		\centering
		\includegraphics[width=17cm]{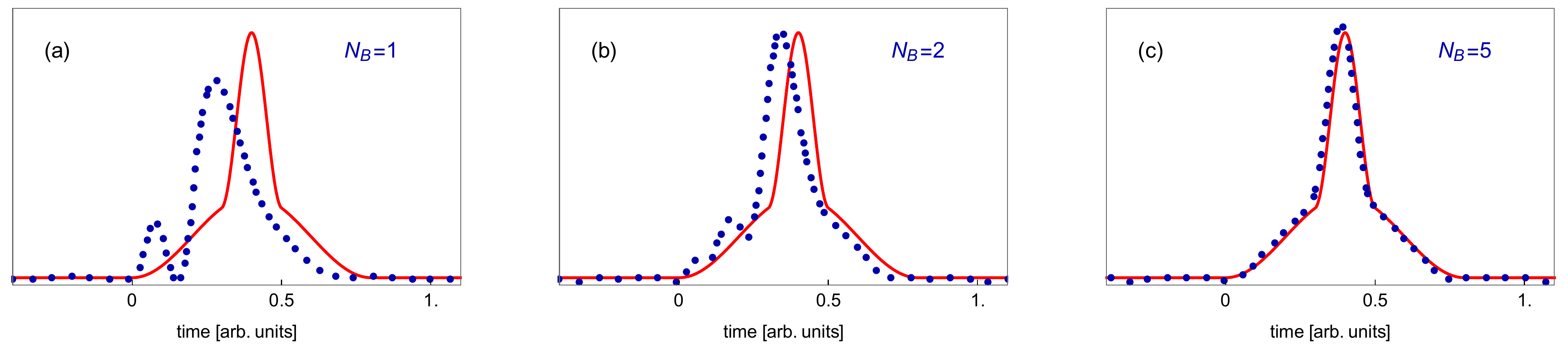} 	
\caption{\small{The Kramers-Kronig-Blaschke (KKB) reconstruction for $N_B=1, 2, 5$ zeros.}}
\label{two-cos2-nzeros}
	\end{figure}

Consider the superposition of two cosine-squared pulses of different amplitudes $A_1$, $A_2$, different widths $b_1<b_2$ and  center times $t_1$, $t_2$ which may be equal or different.
 The Kramers-Kronig reconstruction fails if the two pulses are centered with respect to each other, as shown in Fig.\,\ref{2cos-diffwidth}. To determine the Blaschke phase we have to find the complex zeros of the form factor which 
according to  Eq.\,(\ref{FF-cosine}) is given by
\begin{equation}
{\cal F}(\omega) =A_1\frac{\pi^2 \sin(\omega b_1)\exp( i \,\omega t_{1})}{\omega b_1 (\pi^2-\omega^2 b_1^2)} 
+A_2\frac{\pi^2 \sin(\omega b_2)\exp( i \,\omega t_{2})}{\omega b_2 (\pi^2-\omega^2 b_2^2)} \,.
\label{FF-two-cos2}
\end{equation}
We consider the case that the two pulses are centered with respect to each other and choose $t_1=t_2=b_2$. Unlike the Akutowicz case, treated in section \ref{ana-phase-retrieve}, the complex frequencies $\hat{\omega}_n$ where ${\cal F}(\hat{\omega})$ vanishes cannot be computed analytically but are found by a numerical search procedure.
\begin{figure}[ht]
		\centering
		\includegraphics[width=17cm]{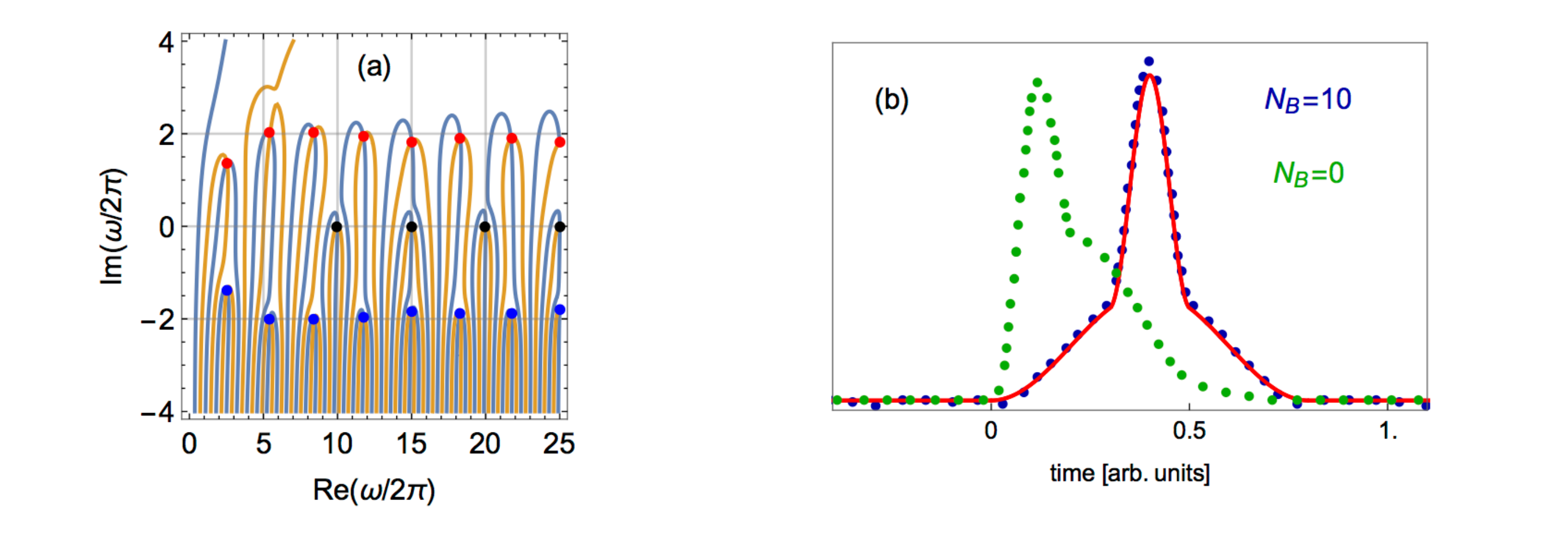} 	
\caption{\small{Reconstruction of a bunch consisting of two cosine-squared pulses of different width ($b_1<b_2$) which are both centered at $t=b_2\,$.
{\bf (a)} Plot of the contour lines $\Re({\cal F}(\hat{\omega}))=0$ (blue) and 
	 $\Im({\cal F}(\hat{\omega}))=0$	(yellow) in the complex 	frequency plane. The dots indicate the crossing points where the  form factor ${\cal F}(\hat{\omega}))$ vanishes. 
	 \newline {\bf (b)} The  original charge distribution $\rho(t)$ (red curve) and the  reconstruction (blue dots) using the reconstruction  phase Eq.\,(\ref{reconstr-phase}), i.e.  the difference between Kramers-Kronig phase and  Blaschke phase (computed for $N_B=10$ zeros). As a  reminder,  the Kramers-Kronig reconstruction without Blaschke phase is also shown here by the green dotted curve (labelled $N_B=0$).}}
\label{two-cos2-contour}
	\end{figure}  
To this end, the contour lines 
$\Re({\cal F}(\hat{\omega}))=0$  and $\Im({\cal F}(\hat{\omega}))=0$  are plotted and their intersection points in the upper half of the complex 	$\hat{\omega}$ plane are determined. The numerical results are shown in  Fig.\,\ref{two-cos2-contour}a.
The crossing points give us the complex zeros $\hat{\omega}_n=a_n+ib_n$ which are needed to compute the Blaschke phase $\Phi_{\rm B}(\omega)$. 
In this example we take $N_B=10$ complex zeros $\hat{\omega}_n=a_n+i b_n$  into account which are located in the right upper quarter of the $\hat{\omega}$ plane, and in addition   their mirror images $\hat{\omega}'_n=-a_n+i b_n$ in the left upper quarter.
The total Blaschke phase  and the reconstruction  phase become
$$\Phi_{\rm B}(\omega)=\sum_{n=1}^{N_B} \Phi_{B,n}(\omega)\,,~~~\Phi_{\rm rec}(\omega)=\Phi_{\rm KK}(\omega)-\Phi_{\rm B}(\omega)\,.$$	
Using the phase $\Phi_{\rm rec}(\omega)$ an almost perfect reconstruction  is achieved, see Fig.\,\ref{two-cos2-contour}b.
An interesting question is how many zeros of the form factor have to be taken into consideration. In Fig.\,\ref{two-cos2-nzeros} the Kramers-Kronig-Blaschke (KKB) reconstruction is shown for $N_B=1,2, 5$ zeros. 

\subsection{Three cosine-squared pulses of equal width}
\begin{figure}[htb!]
		\centering
		\includegraphics[width=14cm]{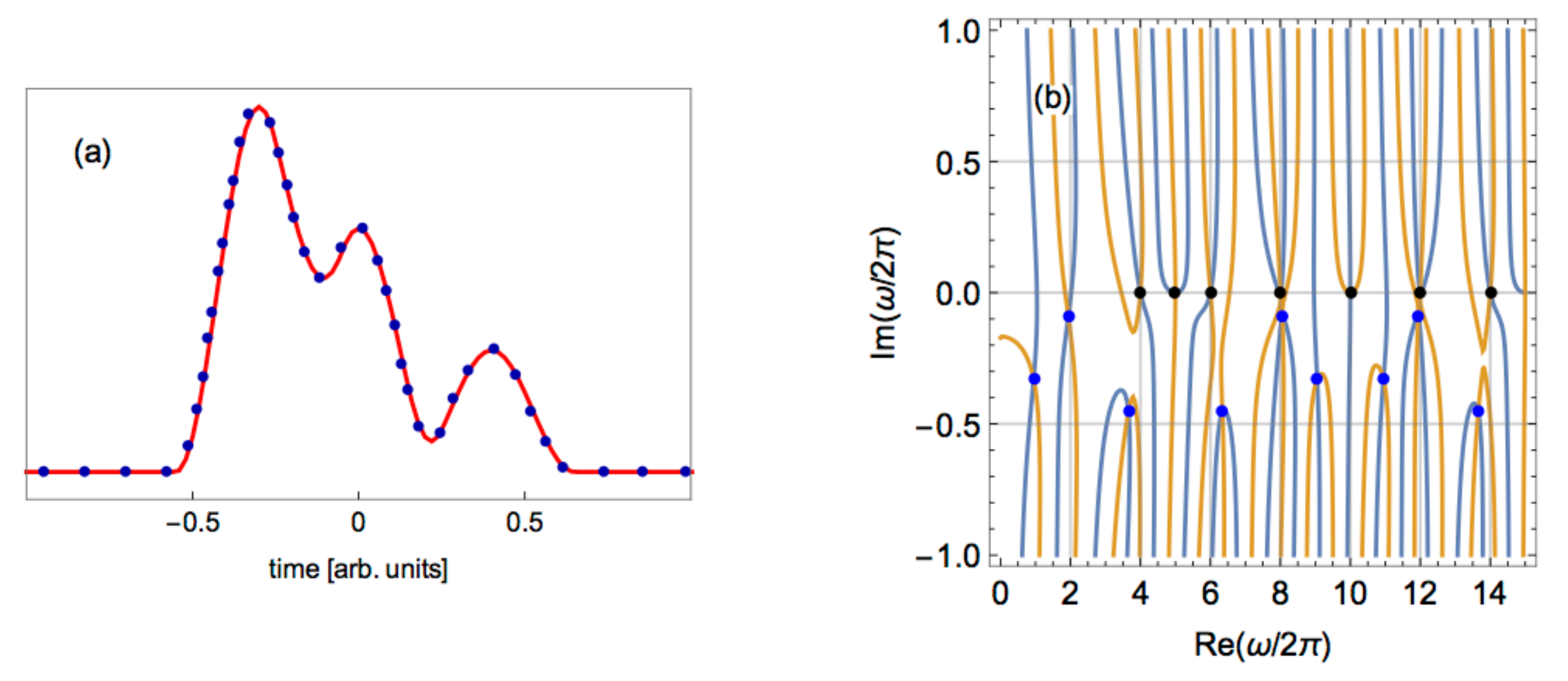} 	
\caption{\small{Three  cosine-squared pulses  of equal width and with non-uniform spacing.  The largest peak is at the front.  {\bf (a)}	The original curve (red) and 
 the KK  reconstruction (blue dots). {\bf (b)} The contour lines $\Re({\cal F}(\hat{\omega}))=0$ (blue) and 
	 $\Im({\cal F}(\hat{\omega}))=0$	(yellow) in the complex 	frequency plane. The complex form factor has zeros on the real axis (black dots) and in the lower half of the complex frequency plane (blue dots), none of which contribute to the Blaschke phase. Hence  the reconstruction phase is equal to the KK phase: $\Phi_{\rm rec}(\omega)\equiv\Phi_{\rm KK}(\omega)$.}}
\label{3Cos2-nonequi-FF-front}
	\end{figure}

We consider the superposition of three  cosine-squared pulses  of equal width and with non-uniform spacing which has already been shown in Fig.\,\ref{3Cos2-KK-nonequi}. First we take the case that the largest peak is at the front.	Then the  KK phase yields a perfect reconstruction, see Figs.\,\ref{3Cos2-KK-nonequi}a and \ref{3Cos2-nonequi-FF-front}a. This can be understood as follows: The complex form factor has no zeros in the upper half of the complex frequency plane, hence the Blaschke phase is zero and the reconstruction phase is identical with the KK phase.

\begin{figure}[htb!]
		\centering
		\includegraphics[width=13cm]{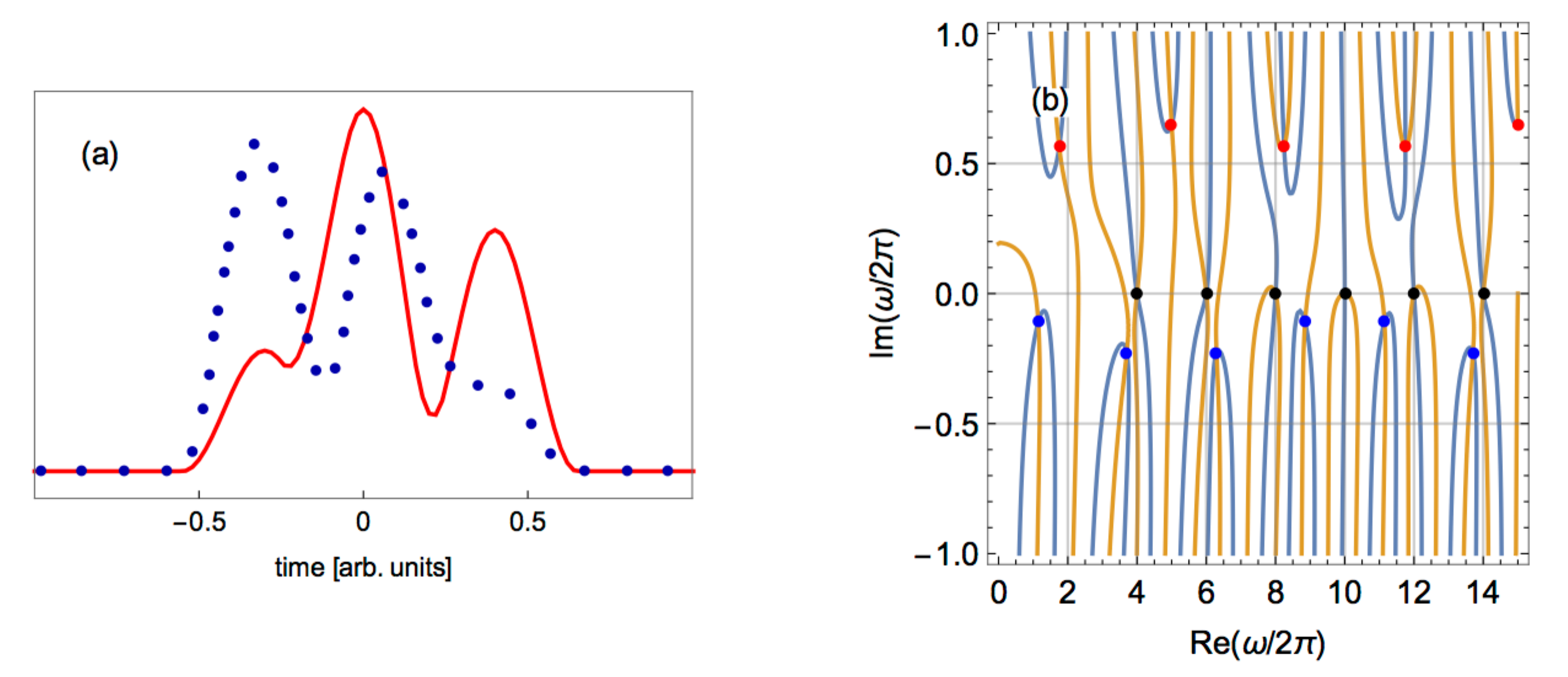} 	
\caption{\small{Three  cosine-squared pulses  of equal width and with non-uniform spacing.  The largest peak is in the center.  {\bf (a)}	The original curve (red) and 
 the KK  reconstruction (blue dots). {\bf (b)}  The complex form factor has zeros  in the upper half of the complex frequency plane (red dots) which contribute to the Blaschke phase. }}
\label{3Cos2-nonequi-FF-center}
	\end{figure}

\begin{figure}[htb!]
		\centering
		\includegraphics[width=6cm]{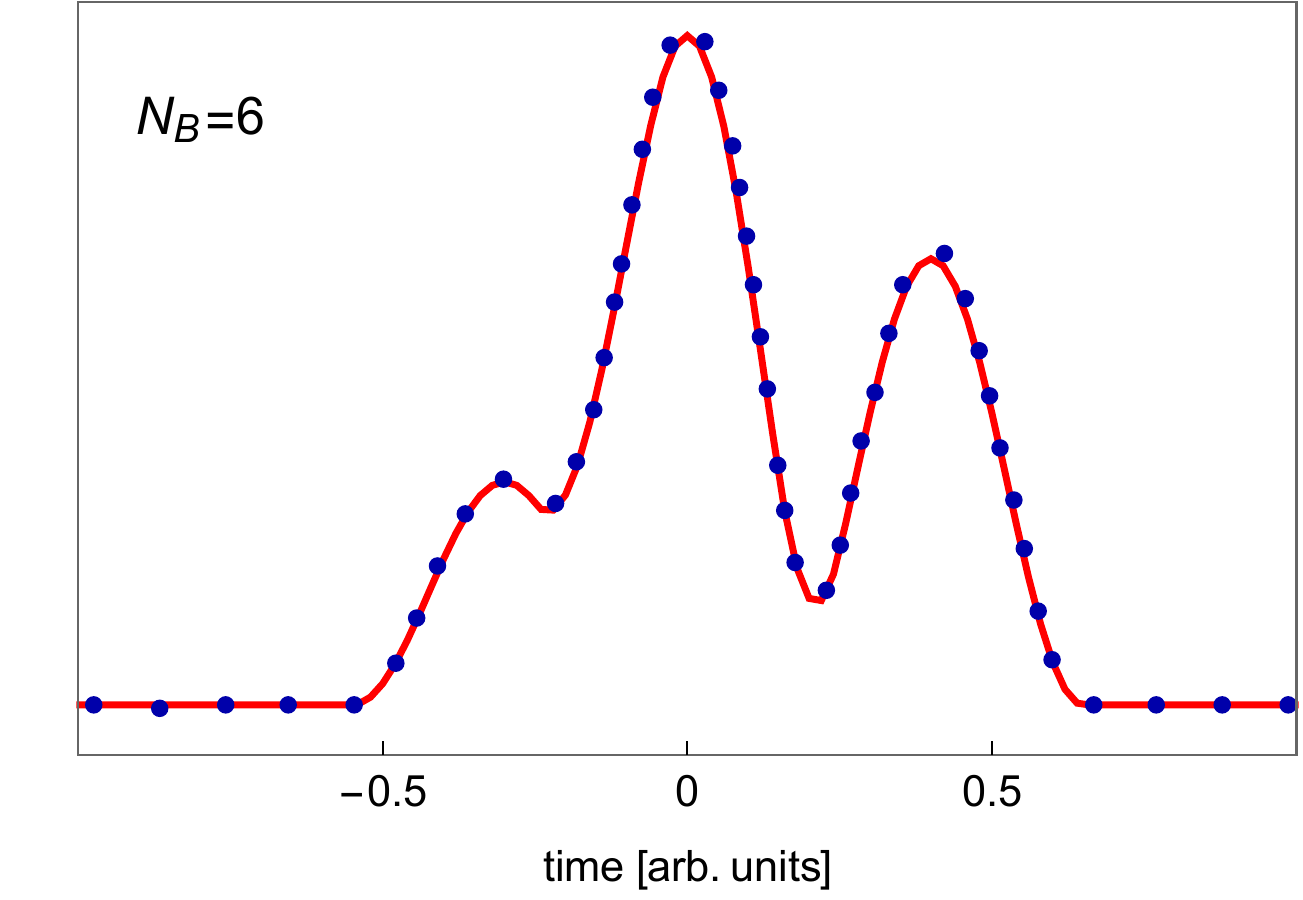} 	
\caption{\small{The Kramers-Kronig-Blaschke (KKB)  reconstruction of the triple cosine-squared structure with the largest peak  in the center, using 
 $\Phi_{\rm rec}(\omega)=\Phi_{\rm KK}(\omega)-\Phi_{\rm B}(\omega)$ with $N_B=6$ pairs of zeros. }}
\label{3Cos2-nonequi-center-KKB}
	\end{figure}

When the largest peak is in the center   the KK reconstruction disagrees with the input distribution, see Fig.\,\ref{3Cos2-KK-nonequi}b. 	The explanation is given  in Fig.\,\ref{3Cos2-nonequi-FF-center}.
 The complex form factor has zeros  in the upper half of the complex frequency plane, so there is a nonvanishing Blaschke phase contributing to the reconstruction phase. The shape reconstruction using the phase $\Phi_{\rm rec}(\omega)=\Phi_{\rm KK}(\omega)-\Phi_{\rm B}(\omega)$ is depicted in Fig.\,\ref{3Cos2-nonequi-center-KKB}. It agrees very well with the original bunch shape.

\subsection{Time reversal and form factor zeros}
Let $\rho_1(t)$ be a causal function and ${\cal F}_1(\hat{\omega})$ its complex form factor. The time-reversed function $\rho_2(t)=\rho_1(-t)$ has the form factor
\begin{equation}
{\cal F}_2(\hat{\omega})={\cal F}_1(-\hat{\omega})\,.
\label{FF1-FF2}
\end{equation}
Proof:\\
\small{
$${\cal F}_1(\hat{\omega})=\int_{t=-\infty}^{t=+\infty}\rho_1(t) \exp(+i\,\hat{\omega}t)dt~~\Rightarrow~~{\cal F}_1(-\hat{\omega})=\int_{t=-\infty}^{t=+\infty}\rho_1(t) \exp(-i\,\hat{\omega}t)dt=\int_{t=-\infty}^{t=+\infty}\rho_2(-t) \exp(-i\,\hat{\omega}t)dt\,.$$
In the last integral  we substitute $t'=-t$, $dt'=-dt$ and get
$${\cal F}_1(-\hat{\omega})=\int_{t'=+\infty}^{t'=-\infty}\rho_2(t') \exp(+i\,\hat{\omega}t')(-dt')
=\int_{t'=-\infty}^{t'=+\infty}\rho_2(t') \exp(+i\,\hat{\omega}t')dt' \equiv {\cal F}_2(\hat{\omega})~~~~~\mathrm{qed.}$$}

\noindent Equation (\ref{FF1-FF2}), combined with Eq.\,(\ref{FF-FF*}), has an important consequence: The complex zeros of  ${\cal F}_1(\hat{\omega})$  and the zeros of ${\cal F}_2(\hat{\omega})$ are mirror-symmetric with respect to the real frequency axis.

\noindent Suppose ${\cal F}_1(\hat{\omega})$ has only  zeros in the upper half of the complex $\hat{\omega}$ plane but no zeros in the lower half. Then ${\cal F}_2(\hat{\omega})$ behaves just in the opposite way, it has   zeros in the lower half  plane  but no zeros in the upper half plane. This implies that the KK reconstruction works perfectly well for $\rho_2(t)$ while it fails for $\rho_1(t)$. In fact, if one uses the KK phase to reconstruct the time profile from the modulus $F_1(\omega)=|{\cal F}_1(\omega)|$, the resulting time profile coincides with $\rho_2(t)$.  \\

\begin{figure}[ht]
		\centering
		\includegraphics[width=17cm]{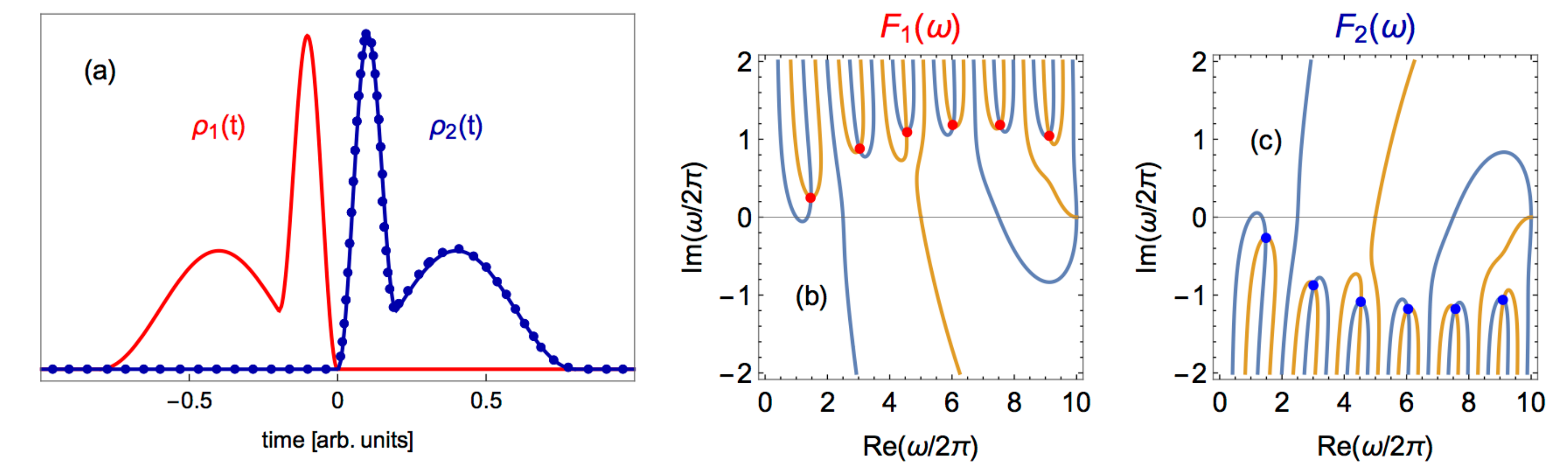} 	
\caption{\small{{\bf (a)} Time profile of a bunch consisting of two cosine-squared pulses of different width,  the narrow peak is either at the rear end 
(red curve $\rho_1(t)$) or at the front (blue curve $\rho_2(t)$). 
In both cases the KK reconstruction, indicated by the blue dots,  yields the same time profile which coincides with the blue curve. 
\newline {\bf (b)} Contour plot for the case that the narrow peak is  at the rear. The red dots are the complex form factor zeros. They are in the upper half of the complex plane. 
 {\bf (c)} Contour plot for the case that the narrow peak is  at the front. The  complex form factor zeros (blue dots)  are in the lower half of the complex plane. }}
\label{2Cos2-front-rear}
	\end{figure}

\noindent This behavior is demonstrated in Fig.\,\ref{2Cos2-front-rear}  for a bunch consisting of two cosine-squared pulses of different width where the narrow peak is either at the rear (time profile $\rho_1(t)$) or at the front (time profile $\rho_2(t)$). The form factor magnitudes are identical on the real axis: 
$F_1(\omega)=|{\cal F}_1(\omega)|=F_2(\omega)=|{\cal F}_2(\omega)|$, and the shape reconstruction with the KK phase (\ref{KK-phase-sect2})  reproduces  the time profile $\rho_2(t)$ in either case.
	\begin{figure}[ht!]
		\centering1
		\includegraphics[width=15cm]{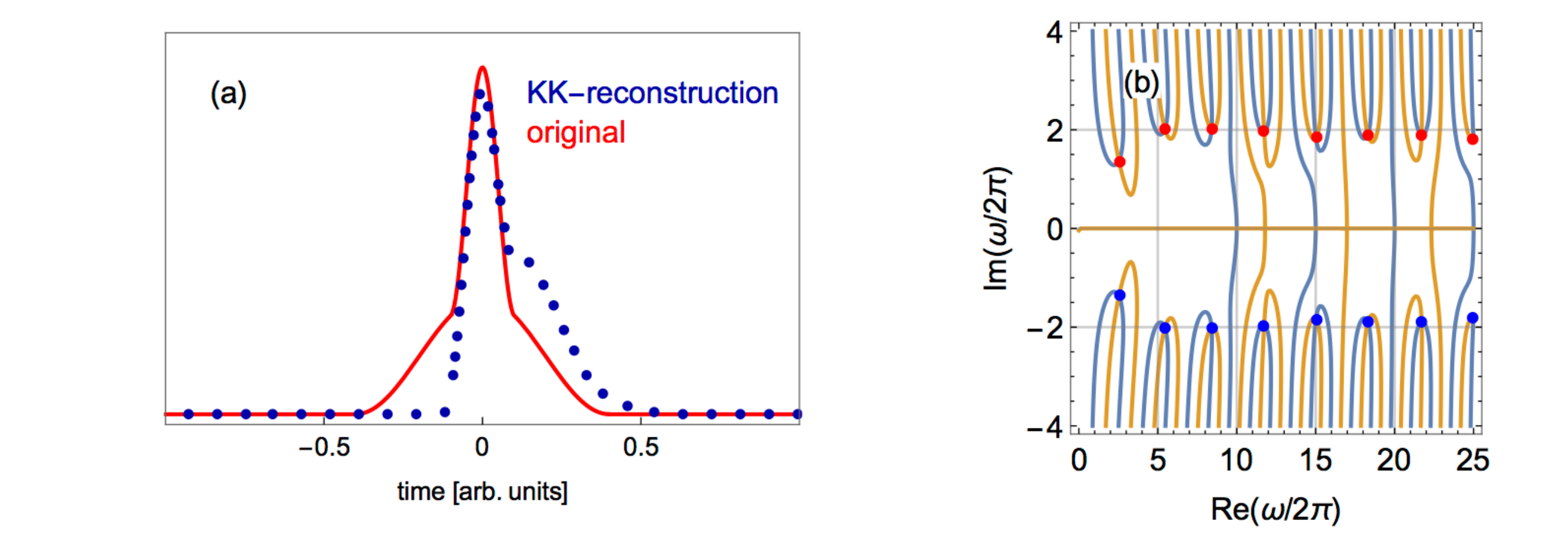} 	
\caption{\small{{\bf (a)} A bunch consisting of two cosine-squared pulses of different width which are both centered at $t=0$. {\bf (b)} Contour plot. 
The complex form factor zeros are both in the upper half (red dots) and in the lower half (blue dots) of the complex plane.}}
\label{2Cos2-centered-FF}
	\end{figure}

\noindent However, if the narrow peak is centered with respect to the wide one (Fig.\,\ref{2Cos2-centered-FF})  then time reversal does not change anything.
As expected, the form factor zeros in the complex $\hat{\omega}$ plane are mirror-symmetric with respect to the real axis. The KK reconstruction must necessarily fail.

\vspace{200mm}

\section{Appendix C:  Gaussians and truncated Gaussians}
Gaussian functions are very convenient since their Fourier transforms are also Gaussians,  and hence they have often been used in model calculations on analytic bunch shape reconstruction. We have shown in Appendix A  that a  Gaussian function  in time-domain violates the fundamental requirement of causality, so the use of genuine Gaussians is mathematically questionable. 
Much preferable are Gaussian-like time profiles obeying causality. These can be generated by truncating a  Gaussian at a sufficiently large  distance from the peak.\vspace{1mm}

\noindent  It is the purpose of this appendix to explore to what extent normal (non-truncated) Gaussians are permitted in analytic model calculations, in spite of their mathematical deficiencies. We will demonstrate that phase retrieval based on the Kramers-Kronig dispersion relation can indeed be done with any superposition of Gaussians provided the integration in the basic KK phase formula (\ref{KK-phase}) is restricted to a finite frequency range. In that case the Gaussians describing  the time  profile of the bunch are in reality truncated Gaussians. Quite a different result is obtained if one wants to determine the complex zeros of the form factor. We  show that there are significant differences between truncated and non-truncated Gaussians. In the computation of the Blaschke phase it is mandatory to use truncated functions.

\subsection{Impact of truncation on the Kramers-Kronig phase}
In  section\,\ref{Gauss-KK-Gamma1} we have proved that the  truncated Gaussian defined in Eq.\,(\ref{trunc-Gauss2})  fulfills all requirements needed in the derivation of the Kramers-Kronig phase formula (\ref{KK-phase}). For completeness we repeat this definition: 
$$
\rho_{\rm trunc}(t)=0~~\mathrm{for}~~~t<0~ \,,~~~~~~~
\rho_{\rm trunc}(t)=\frac{1}{\sqrt{2\pi}\sigma}\exp\left(\frac{-(t-\tau)^2}{2 \sigma^2} \right)  ~~~\mathrm{for}~~~t \ge 0$$
where  $\tau$ is the ``cutoff time''  and $\alpha=\tau/\sigma$ the dimensionless cutoff parameter.
\begin{figure}[hb!]
\begin{center}
\includegraphics[width=17cm]{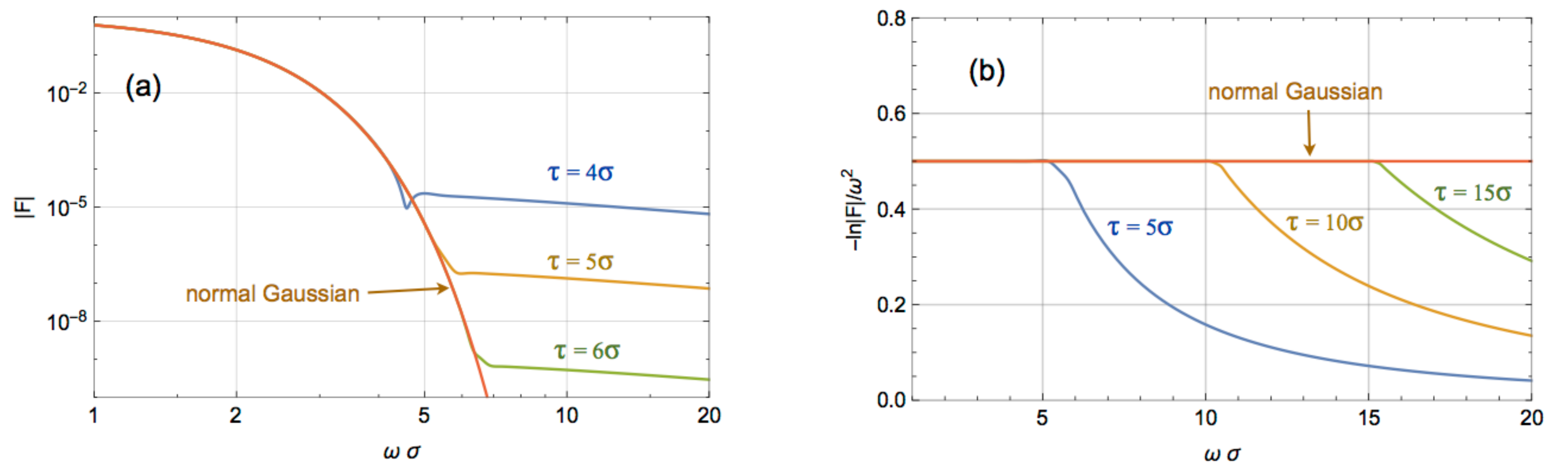}
\end{center}
\caption{\small{  {\bf (a)} The magnitude $|{\cal F}_{\rm trunc}(\omega)|$ of the  form factor  of a truncated Gaussian is plotted as a function of the dimensionless variable $\omega \sigma$ for three  cutoff  times $\tau=\alpha\,\sigma$.   For comparison
the form factor of a normal Gaussian  is also shown. {\bf (b)} The integrand $I_{\rm trunc}(\omega)$ of the KK phase integral (\ref{KKphase-trunc}) for  truncated Gaussians with different cutoff times and the integrand $I_{\rm Gauss}(\omega)=\sigma^2/2$ for a normal Gaussian. In this graph we put $\sigma=1$ in arbitrary units.} 
}
\label{truncatedGauss-C}
\end{figure} 

\noindent The form factor ${\cal F}_{\rm trunc}(\omega)$ is given by  Eq.\,(\ref{Ftrunc-ana}).
In Fig.\,\ref{truncatedGauss-C}a we plot its absolute magnitude  $|{\cal F}_{\rm trunc}(\omega)|$ for several values of the cutoff  time $\tau$ and compare it with the form factor  of a normal Gaussian. Both form factors  agree in the range
$0\le\omega \sigma\le \alpha $,  however at larger frequencies 
the form factor of the normal Gaussian  drops to tiny values while that  of the truncated Gaussian levels off and obeys an inverse power law according to 
Eq.\,(\ref{FF-inverse-power}).
\vspace{3mm}

\noindent To understand the impact of truncation on the KK phase we rewrite the KK phase integral (\ref{KK-phase}) for a truncated Gaussian:
\begin{equation}
\Phi_{\rm KK}^{\rm trunc}(\omega_0)=\frac{2 \omega_0}{\pi} \, \mathcal {P} \int_{0}^{\omega_{\rm cut}} 
\frac{\ln(| {\cal F}_{\rm trunc}(\omega)|)-\ln( |{\cal F}_{\rm trunc}(\omega_0)|) } {\omega_0^2-\omega^2}\,d\omega\,.
\label{KKphase-trunc}
\end{equation}
Since the KK phase integral is usually evaluated by numerical integration over a limited frequency range, the upper integration limit  in formula (\ref{KK-phase})  has been replaced by a suitable cutoff frequency $\omega_{\rm cut}< \infty$. 
For the special case $\omega_0=0$ we have ${\cal F}_{\rm trunc}(\omega_0)=1$, and the integrand takes  the simple form
$$I_{\rm trunc}(\omega)=-\frac{\ln(| {\cal F}_{\rm trunc}(\omega)|) } {\omega^2}\,.$$
In case of a normal Gaussian the integrand is a constant:
$$I_{\rm Gauss}(\omega)=-\frac{\ln(| {\cal F}_{\rm Gauss}(\omega)|) } {\omega^2}=-\frac{-\omega^2 \sigma^2 } {2\omega^2}
=\frac{\sigma^2 } {2}\,.$$
The function $I_{\rm trunc}(\omega)$ is plotted in Fig.\,\ref{truncatedGauss-C}b for three different cutoff times, together with 
$I_{\rm Gauss}(\omega)=\sigma^2/2$. The functions  $I_{\rm trunc}(\omega)$ and $I_{\rm Gauss}(\omega)=\sigma^2/2$ agree in the range
$0\le\omega \sigma\le \alpha $, but at larger frequencies $I_{\rm trunc}(\omega)$ drops rapidly to zero. This implies that  the upper limit of the phase integral 
(\ref{KKphase-trunc}) is $\omega_{\rm cut} \approx \alpha/\sigma = \tau/\sigma^2$. 
\vspace{2mm}

\noindent Now we turn the argumentation around. Suppose we describe the charge distribution inside a bunch by a superposition of Gaussians and want to evaluate the KK phase formula. The numerical integration is necessarily restricted to a finite frequency range $0\le \omega \le  \omega_{\rm cut} $. This however is equivalent to the KK phase computation for the same superposition of truncated Gaussians where the cutoff parameter is given by $\alpha \approx \sigma \,\omega_{\rm cut} $. In other words: The KK phase formula (\ref{KK-phase}) is applicable for any superposition of Gaussian time-domain functions, provided the integration is restricted to a finite frequency range, which happens automatically in a numerical integration.

\subsection{Impact of truncation on the Blaschke zeros}
\begin{figure}[htb]
		\centering
		\includegraphics[width=12cm]{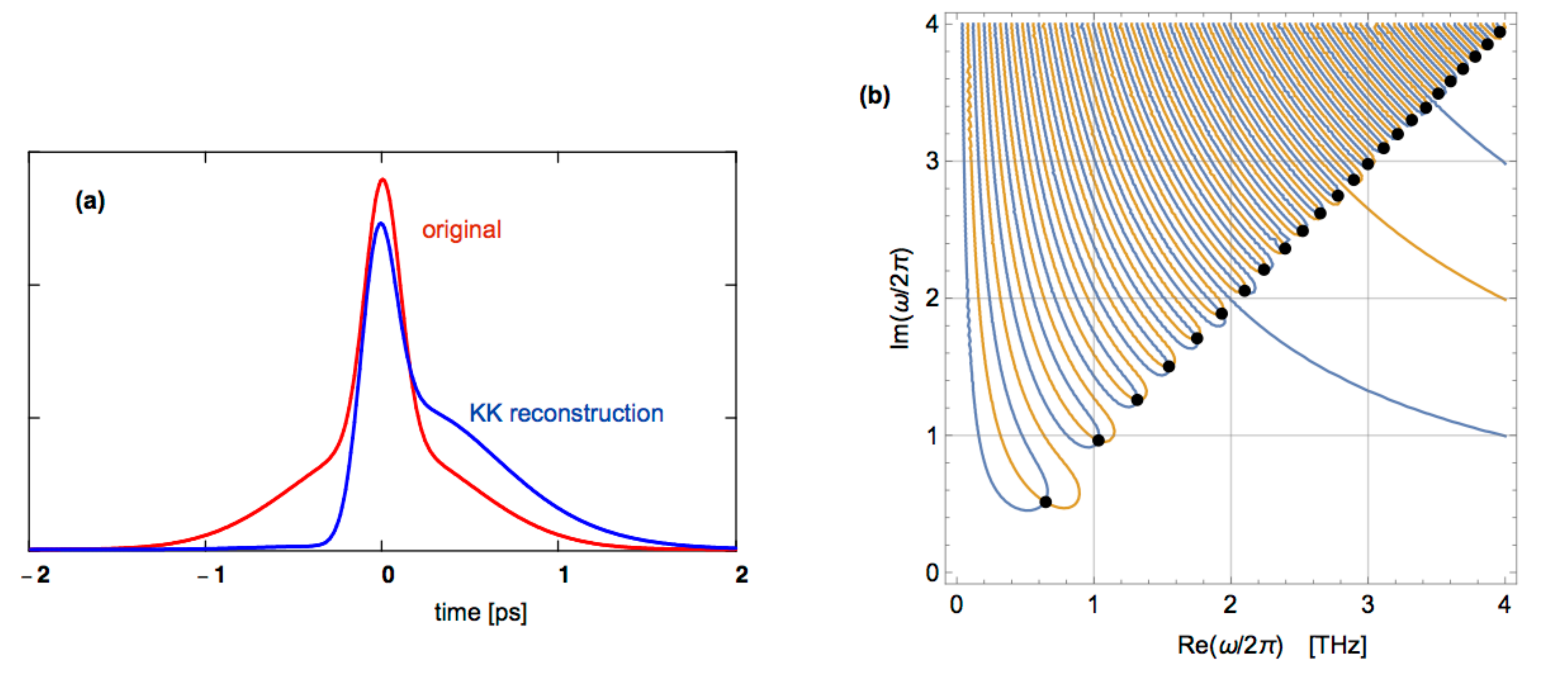} 	
		\caption{\small{{\bf (a)} Superposition of two Gaussians of different width  ($\sigma_1=0.5\,$ps, $\sigma_2=0.1\,$ps) which are centered with respect to each other (red curve) and the KK reconstruction (blue curve).  {\bf (b)} Plot of the contour lines $\Re({\cal F}(\hat{\omega}))=0$ (blue) and 
		$\Im({\cal F}(\hat{\omega}))=0$	(yellow) in the complex 	$\hat{\omega}$ plane.  There are a huge number of densely spaced complex zeros (black dots).}}			\label{2Gauss-difsig}
\end{figure}

To investigate the impact of truncation on the zeros of the complex form factor we study a special case, the superposition of two Gaussian-shaped pulses  of different width.
In section\,\ref{ana-phase-retrieve}  and  Appendix B we have analyzed bunches consisting of two  cosine-squared pulses  of different width, see  Fig.\,\ref{2cos-diffwidth} and Fig.\,\ref{two-cos2-contour}, and we have observed that the KK phase yields a good reconstruction  when the narrow peak  is at the front of the wide one, but fails when the narrow peak  is centered with respect to the wide peak. The same puzzling result is obtained for two Gaussian-shaped pulses  of different width.
This is illustrated in Fig.\,\ref{2Gauss-difsig}a.

In Appendix B we have shown that a perfect reconstruction of the two-cosine-squared bunch is achieved by combining the KK and Blaschke phases. 
For Gaussian pulses this turns out to be quite problematic. As an example we consider the following function
$$\rho(t)=\frac{A_1}{\sqrt{2\pi}\,\sigma_1}\exp\left(-\frac{t^2}{2 \sigma_1^2}\right)+
\frac{A_2}{\sqrt{2\pi}\,\sigma_2}\exp\left(-\frac{t^2}{2 \sigma_2^2}\right)$$
with the parameters $A_1=2/3$, $A_2=1/3$, $\sigma_1=0.5\,$ps and $\sigma_1=0.1\,$ps.
The form factor is
$${\cal F}(\omega)=A_1\exp\left(-\frac{\sigma_1^2 \omega^2}{2 }\right)
+A_2\exp\left(-\frac{\sigma_2^2 \omega^2}{2 }\right).$$
The contour plot of the curves $\Re({\cal F}(\hat{\omega}))=0$ and 
		$\Im({\cal F}(\hat{\omega}))=0$ in Fig.\,\ref{2Gauss-difsig}b reveals there are an enormous number of densely spaced complex zeros which moreover are not close to the real axis but oriented along a diagonal in the complex plane.  As a consequence
		the Blaschke correction becomes very tedious, one has to consider a very large number of zeros to achieve a  reconstruction which might be called ``acceptable'' although it is far inferior to the reconstruction of the two-cosine-squared bunch.
\begin{figure}[htb]
		\centering
		\includegraphics[width=17cm]{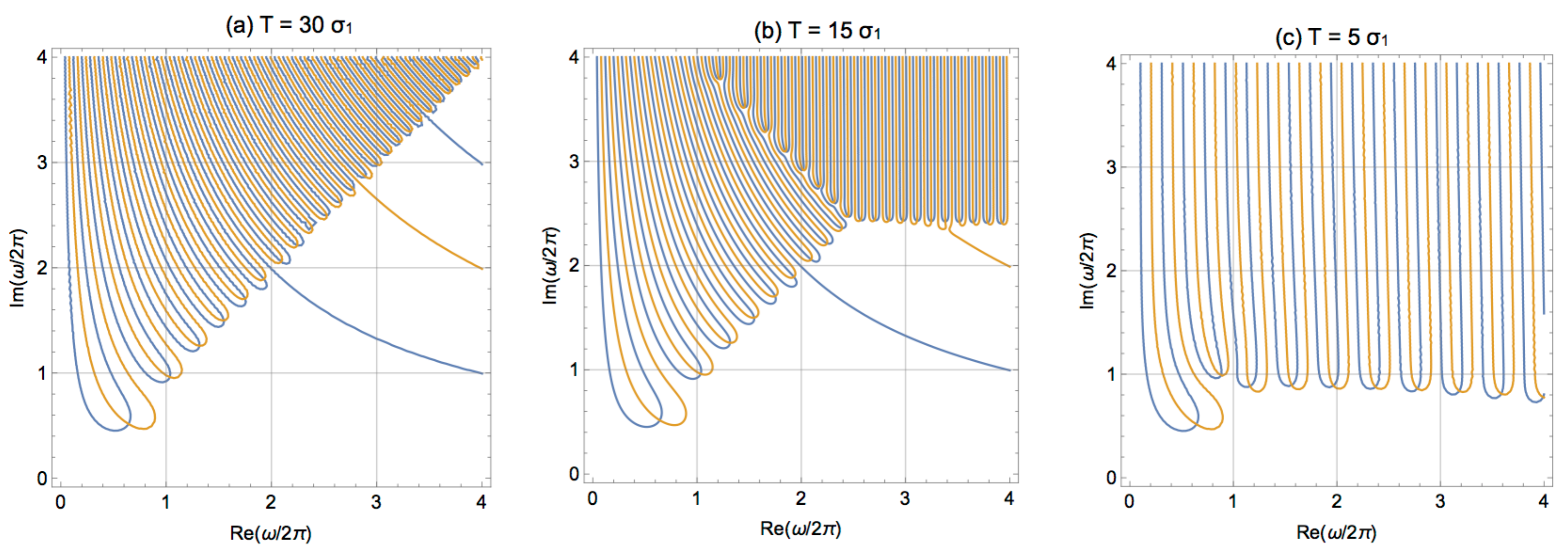} 	
		\caption{\small{Superposition of two truncated Gaussians of different width ($\sigma_1=0.5\,$ps, $\sigma_2=0.1\,$ps), which are centered with respect to each other.  The contour lines $\Re({\cal F}_{\rm tr}(\hat{\omega}))=0$  and 
		$\Im({\cal F}_{\rm tr}(\hat{\omega}))=0$	 in the complex 	$\hat{\omega}$ plane are plotted for three different cutoff times:  
		{\bf (a)} $T=30\,\sigma_1$, {\bf (b)} $T=15\,\sigma_1$ and {\bf (c)} $T=5\,\sigma_1$.}	}		
		\label{2Gauss-trunc}
\end{figure}

\noindent Now we investigate whether this unpleasant property of the complex form factor is related to the non-causality of Gaussian functions. To test  this idea we  truncate the Gaussians in time domain and write the particle density distribution in the form\footnote{To simplify the computation  the Gaussians are truncated symmetrically at $t=-T$ and $t=+T$.} 
\begin{eqnarray}
\rho_{\rm trunc}(t)&=&\frac{A_1}{\sqrt{2\pi}\,\sigma_1}\exp\left(-\frac{t^2}{2 \sigma_1^2}\right)+
\frac{A_2}{\sqrt{2\pi}\,\sigma_2}\exp\left(-\frac{t^2}{2 \sigma_2^2}\right)~~~\mathrm{for}~~|t| \le T \,, \\
\rho_{\rm trunc}(t)&=&0~~~\mathrm{for}~~|t|>T \nonumber
\label{Eq2Gauss-trunc}
\end{eqnarray}
with a sufficiently large cutoff time $T$ such that both Gaussians have dropped to tiny values. The Fourier transform of a single truncated Gaussian can be calculated analytically
\begin{eqnarray}
{\cal F}_j(\omega)&=&\frac{1}{\sqrt{2\pi}\,\sigma_j} \int_{-T}^T\exp\left(-\frac{t^2}{2 \sigma_j^2}\right)
\exp(i\,\omega t)\,dt \nonumber \\
{\cal F}_j(\omega)&=&\frac{\sqrt{\pi}}{2}\exp\left(-\frac{\sigma_j^2 \omega^2}{2 }\right)
\left[\mathrm{erf}\left(\frac{T-i\,\sigma_j^2 \omega}{\sqrt{2}\,\sigma_j}\right)
+\mathrm{erf}\left(\frac{T+i\,\sigma_j^2 \omega}{\sqrt{2}\,\sigma_j}\right)\right].
\end{eqnarray}
The form factor of the  distribution $\rho_{\rm trunc}(t)$
is
$${\cal F}_{\rm trunc}(\omega)=A_1 {\cal F}_1(\omega)+A_2 {\cal F}_2(\omega)\,.$$
It is very instructive to study the contour line pattern  in the complex 	$\hat{\omega}$ plane  for various values of the cutoff time $T$. \vspace{1mm}

\noindent (a) Our first choice is a very large value, $T=30\,\sigma_1$. The  left picture in Fig.\,\ref{2Gauss-trunc} looks exactly like Fig.\,\ref{2Gauss-difsig}b: if the cutoff time is much larger than the standard deviation $\sigma_1$ of the wider  Gaussian,  the form factor  ${\cal F}_{\rm tr}(\omega)$
of the truncated particle density distribution is almost identical with the form factor  ${\cal F}(\omega)$ of the non-truncated distribution. The big theoretical advantage of the truncated function is however that it obeys causality.\\

\noindent  (b) When the cutoff time is reduced  the contour line pattern changes dramatically. For $T=15\,\sigma_1$ the complex zeros follow the diagonal up to a certain frequency, and beyond that they are located a horizontal line. \\

\begin{figure}[ht]
		\centering
		\includegraphics[width=15cm]{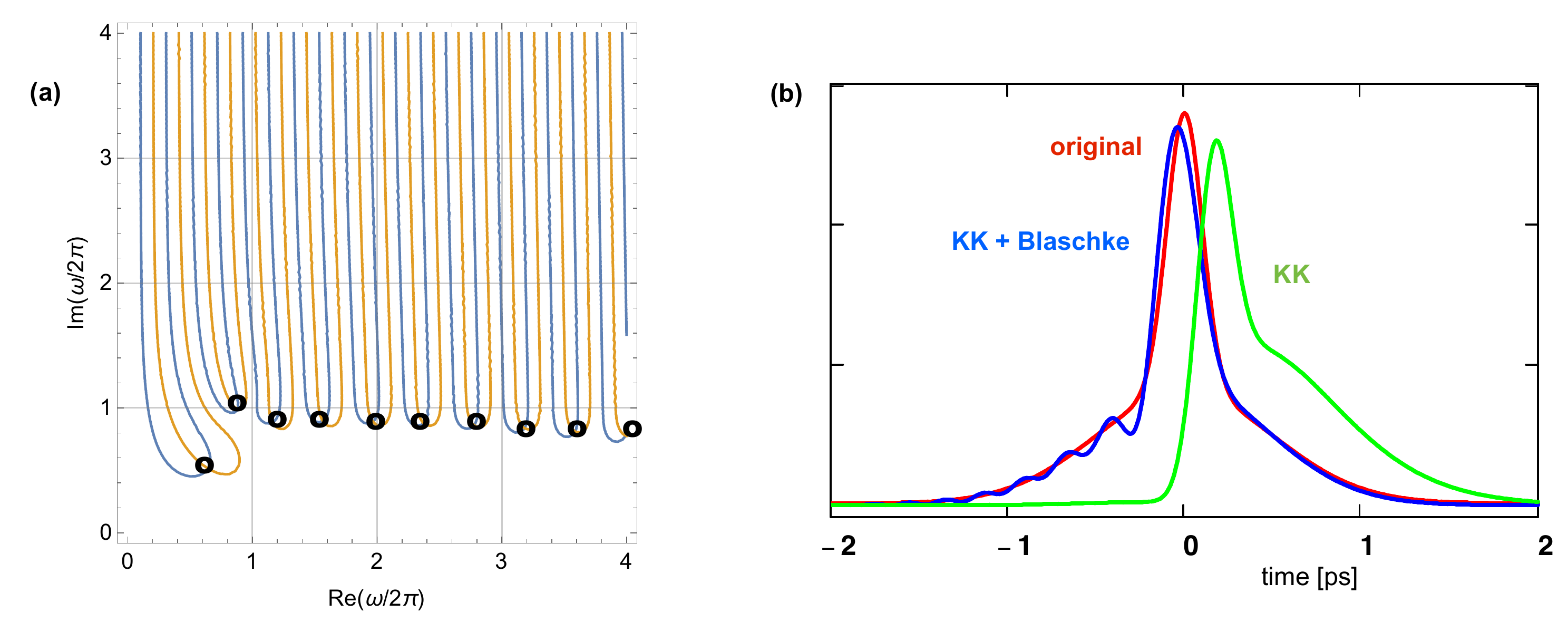} 	
\caption{\small{Reconstruction of a bunch consisting of two truncated Gaussian pulses with $\sigma_1=0.5\,$ps and  $\sigma_2=0.1\,$ps, which are both centered at $t=0$. The cutoff time is $T=5\,\sigma_1$.
{\bf (a)} Plot of the contour lines $\Re({\cal F}(\hat{\omega}))=0$ (blue) and 
	 $\Im({\cal F}(\hat{\omega}))=0$	(yellow) in the complex 	frequency plane. The black circles indicate the crossing points where the  form factor ${\cal F}(\hat{\omega}))$ vanishes. {\bf (b)}    The  reconstruction  based on  the Kramers-Kronig phase and the Blaschke phase (blue curve) is in rather good agreement with the original charge distribution $\rho(t)$ (red curve).  Shown is also the Kramers-Kronig reconstruction without Blaschke phase (green curve) which fails to reproduce the original.}}
\label{doubleGauss-trunc-KKB}
	\end{figure}

\noindent (c) At $T=5\,\sigma_1$ all zeros  all arranged  parallel to the real axis and their spacing is less dense. In fact, Fig.\,\ref{doubleGauss-trunc-KKB}a shows great similarity with contour line pattern of the two cosine-squared pulses in Fig.\,\ref{two-cos2-contour}.
When the reconstruction phase $\Phi_{\rm rec}(\omega)=\Phi_{\rm KK}(\omega)-\Phi_{\rm B}(\omega)$ is used one obtains a good reconstruction of the original time profile, see Fig.\,\ref{doubleGauss-trunc-KKB}b.

\noindent This example shows convincingly that  the Blaschke correction should not be done with normal Gaussians but rather with truncated Gaussians.
\vspace{2mm}

\clearpage

\section{Appendix D:  Propagation of  Radiation by  Fourier Transformation}

The analytical and semi-analytical computations in Section~\ref{TransRad} are based on cylindrical symmetry. This symmetry  is not present in many practical cases, for example when one wants to compute transition radiation from a rectangular or asymmetric screen.  The numerical two-dimensional Fourier transformation permits to deal with such situations and offers the additional advantage that the radiation can be propagated through a whole optical system consisting of drift spaces, apertures and focusing elements such as parabolic or elliptic mirrors. The method is explained at length in Ref.\,\cite{Casalbuoni-I}, see also Refs.\,\cite{Goodman}, \cite{Lawrence}. Here we present the main ideas and formulas. A Mathematica\texttrademark \, code {\it THzTransport} \cite{THzTransport}  was developed for carrying out this analysis.
\vspace{2mm}

\noindent To explain the  principle of electromagnetic wave propagation by Fourier transformation we consider a TR source without cylindrical symmetry. It is then appropriate to work in Cartesian coordinates. Position space is described by the position vectors $\boldsymbol{r}=(x,y,z)$ and Fourier space is described by the wave vectors $\boldsymbol{k}=(k_x,k_y,k_z)$ with $k=\sqrt{k_x^2+k_y^2+k_z^2}=\omega/c$. We choose  general points $Q=(\xi,\eta,0)$ on the source screen  and $P=(x,y,D)$ on the observation screen (see Fig.\,\ref{diff-geom}). 

When the distance $D$ between source screen and observation screen is  larger than the size of the image it is appropriate to apply the Huygens principle: each point  on a wave front acts as origin of a small spherical wave, and the new wave front is formed by the superposition of these spherical waves. Using this principle, 
the horizontal field component at  a general point $P=(x,y,D)$ is computed by a double integral over the source screen:
\begin{equation}
\tilde{E}_x(P,k)= -\frac{i k}{2\pi}\, \underbrace{\iint}_{\rm  source}
\tilde{E}_x(Q,k)  \,  \frac{\exp(i k R')}{R'}\,d\xi d\eta ~~~~\mathrm{with}~~R'=\overline{QP}\,.
\label{ExEy}
\end{equation}
A corresponding expression holds for the vertical field component.
The distance $R'$ between $Q$ and $P$ is expanded up to second order  
\begin{equation}
R'=\overline{QP}=\sqrt{D^2+(x-\xi)^2+(y-\eta)^2}
	\approx D+\frac{x^2+y^2}{2D}- \frac{x \xi+y \eta}{D}+\frac{\xi^2+\eta^2}{2D}
\end{equation}
for $\xi^2+\eta^2 \ll D^2$ and $x^2+y^2 \ll D^2$.
The horizontal component of the Fourier-transformed electric field is thus
\begin{eqnarray}
\tilde{E}_x(P,k)&=& -\frac{i k}{2\pi}\, \,\frac{\exp(i k D)}{D} 
 \exp\left(\frac{ik(x^2+y^2)}{2D}\right) \nonumber \\
 && \cdot  \underbrace{\iint}_{\rm source}
\tilde{E}_x(Q,k)  \,  \exp\left( \frac{i k (\xi^2+\eta^2)}{2D}\right)
\exp\left(-i(k_x \xi+k_y \eta)\right)\,d\xi d\eta  \nonumber	
\end{eqnarray}
with the transverse components of the wave vector $k_x=k\,x/D$, $k_y=k\,y/D$.
The double integral can be written as a two-dimensional Fourier transformation
\begin{equation}\label{2DFourier}
F_x(k_x,k_y) = \frac{1}{2\pi}\underbrace{\iint}_{\rm source} G_x(\xi,\eta)	\exp[-i(k_x \xi+k_y \eta)]  \,d\xi d\eta 
\end{equation}
of the function
\begin{equation}
G_x(\xi,\eta) = \tilde{E}_x(\xi,\eta,k)  \,  \exp\left( \frac{i k (\xi^2+\eta^2)}{2D}\right) \,.
\end{equation}
Corresponding expressions are obtained for the vertical component.\vspace{1mm}

\noindent  It is important to note that the Fourier integral (\ref{2DFourier}) can describe both far-field and near-field diffraction since it contains  the second-order phase factor 
$\exp\left( i k (\xi^2+\eta^2)/(2D) \right)$.\vspace{3mm}

\noindent The Fourier  transform method is quite general and not restricted to the propagation of transition radiation from the TR source to an observation screen. In fact, the method  permits the propagation of an arbitrary radiation field through a whole beamline in a stepwise procedure, going  from one screen to the next. In each step, the complex electric field vector (amplitude, direction and phase) must be known in a fine grid on the ``source screen''. Using the above algorithm,  the complex electric field vector (amplitude, direction and phase) on the ``observation screen'' is then computed by FFT, again in a fine grid. In the next step, the previous ``observation screen'' is treated as the new ``source screen''. A screen can  for example be an aperture, but it may also represent a lens or a focusing mirror of a given size and shape. In that case, position-dependent phase factors are applied to account for the focusing action. This is explained in Ref.\,\cite{Casalbuoni-I}.

\end{document}